\documentclass[astrosymb]{aastex631}

\usepackage{scalerel}
\graphicspath{{./}{figures/}}
\usepackage{graphicx}
\usepackage{enumitem}

\newcommand{\totobs}{350}
\newcommand{\totdet}{295}
\newcommand{\perdet}{84.3}
\newcommand{\unconfdet}{253}
\newcommand{\unconfdetperc}{72.3\%}
\newcommand{\HLobs}{241}

\newcommand{\UGCobs}{119}

\newcommand{\GBTobs}{161}
\newcommand{\NRTobs}{239}
\newcommand{\AOobs}{66}
\newcommand{\onetelobs}{255}
\newcommand{\onetelobsperc}{72.9\%}
\newcommand{\twoplustelobs}{95}
\newcommand{\GBTNRTobs}{63}
\newcommand{\GBTAOobs}{6}
\newcommand{\NRTAOobs}{20}
\newcommand{\NRTNRTobs}{1}
\newcommand{\GBTGBTobs}{8}
\newcommand{\doublesample}{14}
\newcommand{\threetelobs}{5}
\newcommand{\twoplustelunconf}{54}

\newcommand{\veldispall}{1 $\pm$ 4}
\newcommand{\veldispNRTGBT}{-1 $\pm$ 8}
\newcommand{\veldispGBTAO}{16 $\pm$ 9}
\newcommand{\veldispNRTAO}{18 $\pm$ 4}
\newcommand{\veldispNRTNRT}{-2 $\pm$ 4}
\newcommand{\veldispGBTGBT}{0 $\pm$ 3}
\newcommand{\Wfiftydispall}{-8 $\pm$ 9}
\newcommand{\WfiftydispNRTGBT}{-19 $\pm$ 16}
\newcommand{\WfiftydispGBTAO}{16 $\pm$ 21}
\newcommand{\WfiftydispNRTAO}{-9 $\pm$ 9}
\newcommand{\WfiftydispNRTNRT}{7 $\pm$ 14}
\newcommand{\WfiftydispGBTGBT}{2 $\pm$ 15}

\newcommand{\fluxdispNRTGBT}{0.01 $\pm$ 0.08}
\newcommand{\fluxdispGBTAO}{0.36 $\pm$ 0.15}
\newcommand{\fluxdispNRTAO}{0.22 $\pm$ 0.07}
\newcommand{\fluxdispNRTNRT}{0.04 $\pm$ 0.07}
\newcommand{\fluxdispGBTGBT}{0.08 $\pm$ 0.07}
\newcommand{\litredshifts}{177}
\newcommand{\litoptvel}{135}
\newcommand{\litHIvel}{80}
\newcommand{\litoptHIvel}{38}

\newcommand{\litoptveldet}{124}

\newcommand{\litoptvelunconf}{101}
\newcommand{\litHIvelunconf}{66}

\newcommand{\litoptveldist}{-2 $\pm$ 2}

\newcommand{\litHIveldist}{-7 $\pm$ 2}

\newcommand{\litWfifty}{0 $\pm$ 3}
\newcommand{\litflux}{-0.4 $\pm$  0.1}
\newcommand{\MHIgeTenunconf}{35}
\newcommand{\MHIgeTenunconfperc}{11.9\%}

\newcommand{\MHIgeTenunconfmuLo}{31}
\newcommand{\MHIgeNineunconf}{189}
\newcommand{\MHIgeNineunconfperc}{64.1\%}

\newcommand{\WfiftygeFourUnconf}{30}

\newcommand{\WfiftygeFourUnconfmuLo}{30}
\newcommand{\MLSBGposunconf}{45}
\newcommand{\MLSBGposunconfmuLo}{41}

\newcommand{\MLSBGunconfmuLo}{20}

\newcommand{\MHIall}{$10^{9.35 \pm 0.04}$}

\newcommand{\Wfiftyunconf}{199 $\pm$ 7}

\newcommand{\KSHIUGC}{15\% $\pm$ 13\%}
\newcommand{\KSHIHL}{46\% $\pm$ 6\%}

\newcommand{\KSWfiftyUGC}{72\% $\pm$  8.4\%}
\newcommand{\KSWfiftyHL}{ 0\% $\pm$ 19\%}

\newcommand{\as}[2]{$#1''\,\hspace{-1.7mm}.\hspace{.0mm}#2$}
\newcommand{\HI}{\mbox{H\,{\sc i}}}
\newcommand{\HIbf}{\mbox{H\hspace{0.155 em}{\footnotesize \bf I}}}
\newcommand{\HIit}{\mbox{H\hspace{0.155 em}{\footnotesize \it I}}}

\newcommand{\masq}{\mbox{mag~arcsec$^{-2}$}}
\newcommand{\MHI}{$M_{\rm HI}$}

\newcommand{\Msun}{$M_\odot$}
\newcommand{\kms}{\mbox{km\,s$^{-1}$}}
\newcommand{\nan}{Nan\c{c}ay}

\newcommand{\FHI}{\mbox{$F_{\rm HI}$}}

\newcommand{\Jykms}{\mbox{Jy~km~s$^{-1}$}}
\newcommand{\kmsMpc}{\mbox{km~s$^{-1}$~Mpc$^{-1}$}}

\newcommand{\LsunB}{$L_{\odot,\rm B}$}

\newcommand{\MHIMsun}{\mbox{$M_{HI}$/${M}_{\odot}$}}
\newcommand{\MHILB}{\mbox{$M_{\rm HI}/L_{\rm B}$}}

\newcommand{\BT}{\mbox{$B_{\rm T}$}}
\newcommand{\muB}{\mbox{$\langle$$\mu_{\rm B}$$\rangle$}}

\newcommand{\DHL}{\mbox{$D_{\mathrm 25}$}}

\newcommand{\Vhel}{\mbox{$V_{\rm hel}$}}
\newcommand{\VHI}{\mbox{$V_{\rm HI}$}}
\newcommand{\VHIb}{\mbox{$\bf V_{\rm HI}$}}
\newcommand{\Vopt}{\mbox{$V_{\rm opt}$}}

\newcommand{\Wfifty}{\mbox{$W_{\mathrm 50}$}}
\newcommand{\Wfiftycor}{\mbox{$W_{\mathrm 50,\rm cor}$}}
\newcommand{\Wfiftyb}{\mbox{$\bf W_{\mathrm 50}$}}
\newcommand{\Wfiftycorb}{\mbox{$\bf W_{\mathrm 50,\rm cor}$}}
\newcommand{\Wtwenty}{\mbox{$W_{\mathrm 20}$}}
\newcommand{\Wtwentyb}{\mbox{$\bf W_{\mathrm 20}$}}
\newcommand{\LB}{\mbox{$L_{\rm B}$}}

\newcommand{\SNR}{{\em SNR}}

\shorttitle{MLSBGs}
\shortauthors{O'Neil, et al.}

\begin{document}


\title{Searching in \HI{} for Massive Low Surface Brightness Galaxies: \\
 Samples from HyperLeda and the UGC}

\author[0000-0002-2502-5808]{K. O'Neil}
\affiliation{Green Bank Observatory, 155 Observatory Rd, Green Bank, WV 24934, USA}

\author{Stephan E. Schneider}
\affiliation{University of Massachusetts, Astronomy Program, 536 LGRC, Amherst, MA 01003, U.S.A.} 
\author[0000-0003-4770-9829]{W. van Driel}
\affiliation{GEPI, Observatoire de Paris, Universit\'e PSL, CNRS, 5 place Jules Janssen, 92195 Meudon, France} 
\affiliation{Observatoire Radioastronomique de \nan, Observatoire de Paris, Universit\'e PSL, Universit\'e d'Orl\'eans, 18330 \nan, France}
\author{G. Liu}
\affiliation{University of Massachusetts, Astronomy Program, 536 LGRC, Amherst, MA 01003, U.S.A.} 
\author{T. Joseph}
\affiliation{Astrophysics, Cosmology and Gravity Centre (ACGC), Department of Astronomy, University of Cape Town, Private Bag X3, Rondebosch 7701, South Africa}
\affiliation{Anton Pannekoek Institute for Astronomy, Faculty of Science, University of Amsterdam, Science Park 904, 1098 XH Amsterdam, The Netherlands} 
\author{A. C. Schwortz}
\affiliation{University of Massachusetts, Astronomy Program, 536 LGRC, Amherst, MA 01003, U.S.A.}
\affiliation{Quinsigamond Community College, 670 W Boylston St, Box 224, Worcester, MA 01606-2092, U.S.A.} 
\author{Z. Butcher}
\affiliation{University of Massachusetts, Astronomy Program, 536 LGRC, Amherst, MA 01003, U.S.A.}
\affiliation{National Radio Astronomy Observatory, 520 Edgemont Rd, Charlottesville, VA22903 , U.S.A.}
\date{Received 24 April 2023; Accepted: 30 April 2023 (AJ)}

\begin{abstract} 
A search has been made for 21 cm \HI{} line emission in a total of \totobs{} unique 
galaxies from two samples whose optical properties indicate they may be massive 
The first consists of \HLobs{} low surface brightness (LSB) galaxies of morphological type Sb and later 
selected from the HyperLeda database and the the second consists of \UGCobs{} 
LSB galaxies from the UGC with morphological types Sd-m and later.
Of the \totobs{} unique galaxies, \NRTobs{} were observed at the \nan{} Radio Telescope,
\GBTobs{} at the Green Bank Telescope, and \AOobs{} at the Arecibo telescope. 
A total of \totdet{} (\perdet{}\%) were detected, of which \unconfdet{} (\unconfdetperc{}) 
appear to be uncontaminated by any other galaxies within the telescope beam. 
Finally, of the total detected, uncontaminated galaxies, at least \MHIgeTenunconfmuLo{} 
appear to be massive LSB galaxies, with a total \HI\ mass $\ge$10$^{10}$ \Msun, 
for $H_0 = 70\; \kmsMpc$. If we expand the definition to also include galaxies
with significant total (rather than just gas) mass, i.e., those
with inclination-corrected \HI\ line width \Wfiftycor{} $>$ 500 \kms{}, 
this bring the total number of massive LSB galaxies to \MLSBGposunconfmuLo{}.
There are no obvious trends between the various measured global galaxy properties, 
particularly between mean surface brightness and galaxy mass.
\end{abstract}

      
\section{INTRODUCTION}\label{sec:Intro} 

Low Surface Brightness (LSB) galaxies -- spiral galaxies with a central surface brightness 
at least one magnitude arcsec$^{-2}$ fainter than the night sky -- are now well established as a 
class of galaxies with properties distinct from the High Surface Brightness (HSB) objects 
that define the Hubble sequence. However, considerable uncertainty still exists about the range 
of LSB galaxy properties and their number density in the z$\le$0.1 Universe.
As LSB galaxies encompass many of the extremes in galaxy properties, gaining a firm understanding 
of their properties and number counts is vital for testing galaxy formation and evolution 
theories, as well as for determining the relative amounts of baryons that are contained in 
galaxy potentials, compared to those that may comprise the intergalactic medium.
As we will show, we have not yet fully sampled the LSB galaxy parameter space. 
In addition, it should be emphasized that there may still be large numbers of LSB galaxies 
with properties beyond our present detection limits.

The traditional (albeit erroneous) perception of LSB galaxies is that they are young dwarf galaxies
which have undergone little star formation: 
low mass, late-type, fairly blue systems with relatively high \MHILB{} ratios and low metallicities. 
In practice, however, LSB disk galaxies are known to have a remarkable diversity in properties, 
including very red objects with near-solar metallicity \citep[e.g.][]{2007AJ....134..547O}, 
as well as massive (\MHI$\ge$10$^{10}$ \Msun, for $H_0 = 70$ \kmsMpc), systems 
such as Malin 1 (the largest disk galaxy found to date), [SII93] 1226+010, Malin 2 and others 
\citep[e.g.][]{bothun87, sprayberry95}.
Note, too, that the galaxies' observed \HI-richness may be biased by the fact that their redshifts 
are often determined through \HI{} observations -- an optical redshift survey of LSB galaxies 
observed at 21 cm \citep{burkholder01} also contains objects with very low \MHILB{}
ratios \citep{bell00}.

In principle, massive, or giant, LSB galaxies can be defined on different criteria. 
Based on surface photometry, \citet{sprayberry95} defined a ``diffuseness index'' to 
distinguish massive LSB galaxies, which is based on the deprojected blue central surface 
brightness $\mu_{\rm B}(0)$ and the scale length of the disk $h_{\rm r}$ -- the 7 giants in 
their paper have \muB\ = 23.2 \masq\ and 
$\langle h_{r} \rangle$ = 13.0 kpc.
Other selection criteria can be used as well, such as:
deprojected central blue disk surface brightness $\mu_{\rm B}$(0)$\geq$ 23 \masq{} and
\HI{} mass \MHI$\geq$10$^{9.5-10}$ \Msun{} or optical diameter $\geq$50 kpc 
\citep[e.g.][]{2020MNRAS.496.3996K,2017MNRAS.464.2741M, 1997AJ....114.1858P}.
In the present study we use the criterion \MHI$\ge$10$^{10}$ \Msun to identify massive galaxies,
although we also consider the cases with high dynamical mass as defined by 
an inclination-corrected line width \Wfiftycor{} $\ge$ 500 \kms{}.

Massive LSB galaxies are interesting for a number of reasons:
\begin{enumerate}[noitemsep,nolistsep,topsep = 0pt]
\item The current rate and history of star formation in massive LSB galaxies is puzzling. Optical
photometry shows that most have blue colors, and this appears to be
due in part to lower metallicities, as well as to a young population of stars \citep[e.g.][]{oneil97}. 
What has delayed star formation in such massive galaxies? 

\item How the massive LSB galaxies have evolved into large disk systems without converting 
much of their gas into stars poses an interesting problem for galaxy evolution models. 
Are they in environments where the disks have remained undisturbed, are they all 
in the late stage of mergers, have they only recently been assembled, or does their high 
dark matter content play a r\^{o}le in their stellar evolution 
\citep{2015AJ....149..199D, 2011ApJ...728...74G, 2000A&A...357..397V, 2010A&A...516A..11L, 1997AJ....114.1858P}?

\item Much of the \HI{} in the local universe is tied up in LSBs, 
either in dwarfs (which constitute the great majority of LSBs) or in higher-mass galaxies that 
have large \HI-to-optical flux ratios \citep{rosenberg02}. Massive LSBs are rare, but can contain 
very large \HI{} masses. What is their overall contribution to the \HI{} mass function? 

\item Finally, massive LSB galaxies often have a significant bulge component, 
and frequently an active nucleus raising the question of how the extremely large and 
extremely diffuse stellar disks of many of these objects continue to exist 
\citep[e.g.][]{gallagher83,knezek93,knezek99, schombert98}
\end{enumerate}


The origin and evolutionary histories of massive LSB galaxies remain unclear. 
\citet{hoffman92} proposed a scenario with giants forming from rare, 
low density fluctuations in very low density regions, which should give rise to
quiescent, gas-rich disks, with flat rotation curves with $v_{max}\sim$300~\kms. 
\citet{knezek93} suggested an alternative scenario, based on \citet{kormendy89},
whereby LSB Giants may have dissipatively formed from massive, metal-poor dark 
matter halos. More recent studies have expounded on these, theorizing that
LSB galaxies start as a high surface brightness disk galaxy \citep{2021MNRAS.503..830S} 
or spheroid \citep{2018MNRAS.478.3994C} with the LSB disk forming later through 
accretion of external gas.
Another possibility is that they form through the merger of more than one galaxy 
\citep[e.g.][]{2019MNRAS.485..796M} and/or that the disk itself is the result of a 
recent merger and will eventually coalesce into a higher surface brightness disk 
\citep[e.g.][]{2018MNRAS.478.3994C, 2016MNRAS.460.4466Z}. 
Finally, of course, it is possible that all of these scenarios are correct in different 
circumstances, and that the evolutionary history of massive LSB galaxies is as complex 
as of their HSB counterparts \citep[e.g.][]{2020MNRAS.496.3996K}. However, the limited 
number of objects available for study makes testing these theories challenging, at best.

We have previously undertaken surveys to explore the number density and \HI{} properties 
of massive LSB galaxies in the nearby Universe. These earlier projects focused on two 
different samples -- late-type LSB galaxies found in the UGC 
\citep[Uppsala General Catalogue of Galaxies, ][]{UGC,oneil04HI2}
and near-infrared LSB galaxies found in the 2MASS catalog \citep{monnier03b, monnier03c,monnier03a}. 
Both samples were observed in \HI{} using the \nan{} and Arecibo radio telescopes with good success: 
out of 231 UGC galaxies observed, 146 were detected, including 47 (32\% of the detected objects) 
massive LSB galaxies, and out of the 701 2MASS galaxies observed, 278 were detected, of which 31 (11\%) 
were also found to be massive. These results established that massive LSB galaxies are 
surprisingly abundant, but the surveys were incomplete and a more systematic set of \HI{} 
observations of a homogeneous sample were required. 

To search for additional massive LSB galaxies we undertook a second survey, using a combination of 
(1) \HLobs{} LSB galaxies selected from the online Lyon Extragalactic Database 
(HyperLeda - \hyperlink{http://leda.univ-lyon1.fr}{http://leda.univ-lyon1.fr}) 
on their mean surface brightness in the $B$ band (referred to as the HyperLeda sample),
and (2) \UGCobs{} LSB galaxies from the UGC \citep{UGC} with morphological types Sd-m (the UGC sample). 
Initially the HyperLeda sample was observed in \HI{} using the Arecibo, Green Bank, and \nan{} 
radio telescopes, while the UGC sample was observed at \nan{} only.
At a later stage, the GBT was used for follow-up \HI{} observations of both samples, 
to clarify any potential (RFI, or other) confusion regarding prior detections and to search 
for prior non-detections.

The selection of the two samples of LSB galaxies we observed in the 21 cm \HI{} line
is described in Section~\ref{sec:Gal_Selection}, the observations and the data reduction are presented 
in Section~\ref{sec:observations}, and the results in Section~\ref{sec:results}. A discussion 
of the results is given in Sections~\ref{sec:HIprops} and \ref{sec:mlsbgs}, and our conclusions 
are presented in Section~\ref{sec:conclusion}.
Appendix~\ref{Sec:IndGals} provides details of many of the individual galaxies observed.

Please note that throughout this paper a Hubble constant value of $H_0 = 70 \kms{} \kmsMpc$ was used, 
and that all radial velocities are heliocentric and calculated according to the conventional 
optical definition ($V = $cz$ = c\left({\lambda - \lambda_0}\over{\lambda_0}\right)$).

\section{GALAXY SAMPLE SELECTION} \label{sec:Gal_Selection}  

The Hyperleda and UGC galaxy samples were defined in 2004 and 2002, respectively,
before the stream of galaxy redshifts from the Sloan Digital Sky Survey 
(SDSS; see, e.g., \citealt{2000AJ....120.1579Y}) became available. All values used here 
reflect the most recently available information, but this results in a few galaxies 
that are included in the sample which would not have been selected based on the more 
recently updated spectroscopic velocities and photometric values.

\subsection{HyperLeda Sample}  \label{subsec:Hyperleda} 
The HyperLeda galaxies were selected from the HyperLeda database (see e.g. \citealt{2014AA...570A..13M}), 
based on the following criteria, as available in 2004:
\begin{itemize}[noitemsep,topsep = 0pt]
\item average blue surface brightness $\langle\mu_{\rm B}\rangle \ge$25 \masq{}, 
defined following \citet{1985AJ.....90.2487B}, to select for galaxies with a low surface brightness disk: 
$\langle\mu_{\rm B}\rangle$ = \BT{} + 2.5 log(\DHL{}) + 8.63. Here \BT{} is the apparent total 
blue magnitude and \DHL\ the blue major axis diameter, both at the 25 \masq{} isophotal level, in arcmin;
\item morphological type later than Sb, to insure galaxies have a disk and to increase the 
likelihood of detection in \HI; 
\item optical diameter $\DHL>$30\arcsec, to insure against selecting dwarf galaxies;
\item no known redshift (at the time of observation, in 2004); 
\item inclination $<$70$^\circ$; 
\item declination -38$^\circ$$<\delta<$80$^\circ$ to allow for observation by the radio telescopes.
\end{itemize}

The complete list of galaxies meeting these criteria comprised $\sim$290 objects. 
Of these, \HLobs{} were randomly selected for observation. (The complete 
sample could not to be observed due to limitations in telescope availability.)

\subsection{UGC Sample} 

The UGC galaxies were selected from the complete set of UGC dwarf
and LSB galaxies that were originally observed by \citet{schneider90}
and \citet{schneider92} in \HI{}, which comprises all listed dwarf,
irregular, and magellanic galaxies in the UGC -- a total of 1845 galaxies, or
14\% of the total of 12,940 galaxies in the catalog. 
The original observations were made with the 305m Arecibo telescope 
and at the Green Bank 300 foot telescope otherwise.
Most of the Green Bank detections were of nearby, \HI-rich dwarfs only, due to
the telescope's narrow bandpasses (3000 \kms) and low sensitivity. 

By 1992, 85\% were detected and by 2002, when we defined our present sample, the total 
detection rate had increased to $>90\%$, but more than 180 sources remained undetected. 
We selected all \UGCobs{} remaining objects which lie outside the Arecibo
declination range for observation at \nan{}, excluding the following six objects 
whose declination is too high ($>$82$^{\circ}$) for the telescope's geometry
(see below) to allow tracking: UGC 3461, UGC 5298, UGC 9478, UGC 10263, UGC 10581, and 
UGC 10605. The non-detections from the Arecibo sample were not re-observed with \nan{}, 
as their non-detection limit is below what is practical with the \nan{}.
The GBT was later used to confirm a number of the \nan{} detections and to 
look for non-detections from this subsample, as described in Section~\ref{sec:observations}.

\section{OBSERVATIONS AND DATA REDUCTION} \label{sec:observations}  

\subsection{The Arecibo Telescope} %
The Arecibo observations of the Hyperleda sample were carried out between November 26, 2004 
and May 14, 2005 for a total of 32 hours of observing time. 
To avoid baseline ripples caused by the Sun, all the observations were made at night. 
The position-switching technique was applied, using an ON/OFF pattern with equal time 
given to the on-source and off-source position. The on-source integration time was 
5 minutes for each galaxy. The backend was configured using 9-level sampling 
with the Wide band Arecibo Pulsar Processors (WAPPs) and 5 \kms{} resolution. 
This allowed for a velocity coverage of either a low velocity range of $\sim$-2000 to 19,000 \kms\ 
(1330 -- 1430 MHz), or a high velocity range of $\sim$19,000 to 36,000 \kms\ (1250--1330 MHz) 
for the two correlator boards separately. It should be noted that large sections of 
the high velocity range were filled with radio frequency interference (RFI) and that the r.m.s. 
noise levels in this band listed in the tables were measured in RFI-free sections. 
The original 5.3 \kms{} velocity resolution was kept throughout the data processing phase.

The data were calibrated using the on telescope noise diodes, from the original WAPP files including 
the two different velocity search ranges. Data analysis was performed using Supermongo routines 
developed by our team. The two polarizations were first averaged for each observation of a given 
galaxy, from which the overall r.m.s. noise level was measured by choosing a broad range without 
apparent unusual signatures. Obviously bad data were abandoned at this stage. After that, a 
polynomial baseline was fitted to the data within the vicinity of the galactic \HI{} profile, 
excluding those velocity ranges with \HI{} line emission (both from the Galaxy and the target 
galaxy) or RFI (e.g., the GPS L3 signal around 8300 \kms). With a view toward optimization and 
faithfulness of the fitting, F-tests were performed on all the orders (0 -- 9) of the polynomial 
fit and those that gave a result greater than 10 were considerably preferred. Fits of order 
higher than 4 were rarely adopted, and only the first order fit was used for profiles with 
low signal to noise ratios. Once the baseline was subtracted, the velocities were corrected 
to the optical, heliocentric system.

At 21 cm the Arecibo beam is $3\farcm4$$\times$$3\farcm6$.

\subsection{The Robert C. Byrd Green Bank Telescope} %
The first observations of the Hyperleda sample galaxies taken with the Robert C. Byrd Green Bank 
Telescope (GBT) were taken in October, 2004 over a total of 9 nights for a total of 65 hours. 
All these observations used the Gregorian L-band receiver with the (now retired) auto-correlation 
spectrometer \citep{1998SPIE.3357..424E}. 
As the redshifts for the observed galaxies were unknown, the spectrometer was set up to observe 
with a 9-level sampling, from -2000 to 25,600 \kms\ (1300 - 1430 MHz) with 2.57 \kms{} resolution, 
which was smoothed to 5.2 \kms{} during the data processing phase.
Standard position switching techniques were used, with an ON/OFF source pattern and 300s for to each 
on-souce and off-source observation. A calibration noise diode was fired on blank sky at the end of 
each ON+OFF observing pair. 

Follow-up observation of a number of the previously non-detected galaxies from the HyperLeda sample
were taken with the GBT in early 2012. Again the autocorrelation spectrometer was used for the 
observations, with the same set-up as for the earlier observations. Position switching was used, 
with typically 3-5 on-source observations were taken for every off (the total number depending 
on available telescope time). 

A second round of follow-up observations was made in the fall of 2022 to examine 
discrepancies among our prior detections, between our detections and those from the literature, 
and to try to detect the previously undetected galaxies from both the UGC and HyperLeda samples. 
Again the GBT's L-band receiver was used, but now with the VEGAS spectrometer \citep{2015ursi.confE...4P}. 
To maximize coverage, the backend was set-up to observe from -2000 to 30,250 \kms\ (1290 to 1430 MHz) 
with 0.15 \kms{} velocity resolution, which was smoothed to 15 \kms{} during the data processing phase. 
Please note though, that the presence of RFI at the lower frequencies limited the usable observing 
range to -2000 to 25,600 \kms, i.e., the same as for the earlier GBT observations.
Position switching was again used, with 300s scans. To maximize our science output from the telescope
and taking advantage of the very stable baseline on the GBT, the observing pattern consisted of 
one ON+OFF source pair of scans, typically followed by 8 ON source observations, and a final ON+OFF pair.
This pattern was used for all observations, if time allowed. 

All data were calibrated using the engineering noise diode values measured at the GBT and checked by 
observing a minimum of one (and typically 2-3) standard flux calibrators each night.

Data were reduced using standard GBTIDL (http://gbtidl.nrao.edu) routines modified for our 
observing pattern. Any individual spectra showing baseline ripples due to RFI were removed from 
the data reduction package.
Frequencies were converted to radial velocities in the optical, heliocentric velocity frame.
Data obtained in 2022 were boxcar smoothed to 15 \kms{} resolution, cf. the 5.2 \kms{} resolution 
for the first sets of observations.

It should be noted that changes in instrumentation between observations make the averagine of the data together impractical.

At 21 cm, the GBT HPBW is $8\farcm7$$\times$$8\farcm7$.

\subsection{The \nan{} Radio Telescope} 
The 100-m class \nan{} decimetric radio telescope (NRT), a meridian transit-type instrument of the 
Kraus/Ohio State design, consists of a fixed spherical mirror (300m long and 35m high), a tiltable 
flat mirror (200$\times$40m), and a focal carriage moving along a curved rail track; for further details
on the instrument and data reduction, see \citet[][]{vandriel16} and references therein. Sources on the 
celestial equator can be tracked for about 60 minutes. Its collecting area is about 
7000~m$^{2}$ (equivalent to a 94-m diameter parabolic dish). Due to the E-W elongated shape of the 
mirrors, some of the instrument's characteristics depend on the observed declination. 
At 21 cm, the telescope's HPBW is $3\farcm5$ in right ascension, independent of 
declination, while in the North-South direction it is 23$'$ for declinations up to $\sim$20$^{\circ}$, 
rising to 25$'$ at $\delta$ = 50$^{\circ}$ and 38$'$ at $\delta$ = 79$^{\circ}$, 
the northern limit of the survey. The instrument's effective collecting area and, consequently, 
its gain follow the same geometric effect, decreasing correspondingly with declination. 
Flux calibration is determined through regular measurements 
of a cold load calibrator and periodic monitoring of strong continuum sources.

All data were taken in a standard ON/OFF position-switching mode, with an on-source integration 
time step of 40 seconds. An auto-correlator set-up of 4096 channels was used in a 50~MHz bandpass, 
with a velocity resolution of 2.6~\kms. For most of the Hyperleda objects the default search range 
was $\sim$-250 to 10,600 \kms, whereas for the UGC sample it was $\sim$ 325 to 11,825 \kms.
For a number of objects with known higher redshifts, a search was also made at
$\sim$6875 to 18,330 \kms{} (for 4 objects from the HyperLeda sample and 25 from the UGC sample).
The bulk of the observations of the HyperLeda sample galaxies were made in the period 
July - November 2004, for a total of about 410 hours of telescope time, and the 
high-velocity observations were obtained in 2007, whereas the UGC sample observations 
were made in the period January - December 2003, using a total of about 240 hours of 
telescope time. 

Averaging the two receiver polarizations and applying a declination-dependent factor to convert 
from units of system temperature to flux density in Jy was done using standard NRT software.
In order to reduce the effect of relatively strong RFI in our observations, we used the 
RFI flagging and mitigation routine described in  \citet{monnier03b} for further details). 
The RFI signal trigger was usually set to a level of 10$\sigma$ for each 40 second integration.
Subsequent smoothing in velocity and baseline fitting was performed for the HyperLeda sample 
with standard \nan{} software and for the UGC sample with Supermongo routines developed by 
one of our team which were based on the standard ANALYZ routines then in use at Arecibo.

Spectra with an \HI{} line peak signal-to-noise ratio larger than 5 were boxcar smoothed to a 
velocity resolution of 10~\kms{} during the data reduction, whereas spectra with fainter lines 
and non-detections were smoothed to a resolution of 18~\kms. 
Radial velocities were ultimately converted to a heliocentric, optical $cz$ scale.

\section{RESULTS}  \label{sec:results}  
\subsection{Literature Properties} \label{subsec:optical} 

Basic optical properties, including spectroscopic radial velocities, of the sample galaxies are 
listed for both \HI{} detections and non-detections in Table~\ref{tab:opticalprops}. 
Details on the individual galaxies, when given, can be found in Appendix~\ref{Sec:IndGals}.
From the optical parameters (see list hereafter), the total blue magnitude, \BT, and the 
axial ratio, $b/a$, were taken from HyperLeda. 

At present, redshifts are available for \litredshifts{} of the observed galaxies, with
\litoptvel{} objects having optical spectroscopic redshifts (only), \litHIvel{} having previously published \HI{} redshifts, 
and \litoptHIvel{} having both optical and \HI{} redshifts in the literature (see Tables~\ref{tab:opticalprops} 
and ~\ref{tab:resultsnondets}). References for all redshifts found are given in Table~\ref{tab:opticalprops}, 
and are divided into two categories -- optical spectral line and \HI{} velocities.

Listed in Table~\ref{tab:opticalprops} are the following elements, in alphabetical order of galaxy name, 
and divided into four categories: galaxies detected in \HI{}, spurious detections, galaxies not
detected in \HI{} and galaxies whose optical velocity measurements (taken after our observations) 
place them outside the \HI{} search range.
Please note that we have listed the current HyperLeda values in our Tables and used them for calculations.
\begin{enumerate}
\item {\bf Name:} Common name of the object;
\item {\bf RA:} Right ascension in J2000.0 coordinates;
\item {\bf Dec:} Declination in J2000.0 coordinates;
\item {\bf Type:} Morphological type, preferably from the HyperLeda database.  
If not available, the classification retrieved from NED, the NASA/IPAC Extragalactic Database
\hyperlink{https://ned.ipac.caltech.edu}, is given in brackets;
\item {\bf \BT{}:} Total apparent magnitude in the $B$ band, from HyperLeda;
\item {\bf D$_{25}$:} Length of the projected major axis of a galaxy at the isophotal level 25 \masq{} in the $B$-band, from HyperLeda;
\item {\bf $b/a$:} Axial ratio measured at a surface brightness level of 25 \masq{} in the $B$-band, from HyperLeda;
\item {\bf \Vopt:} Heliocentric velocity, from optical spectra;
\item {\bf Opt. ref.:} Literature reference for \Vopt\ (the full references to the abbreviated notations
used in the Table are given in the Notes to the Table);
\item {\bf \Vopt:} Heliocentric velocity, from \HI{} spectra;
\item {\bf \HI\ ref.:} Literature reference for \VHI{}.
\end{enumerate}

\subsection{The H{\small I} Data} \label{subsec:Data} 

\begin{figure*} 
  \centering
  \includegraphics[width = 14.4cm]{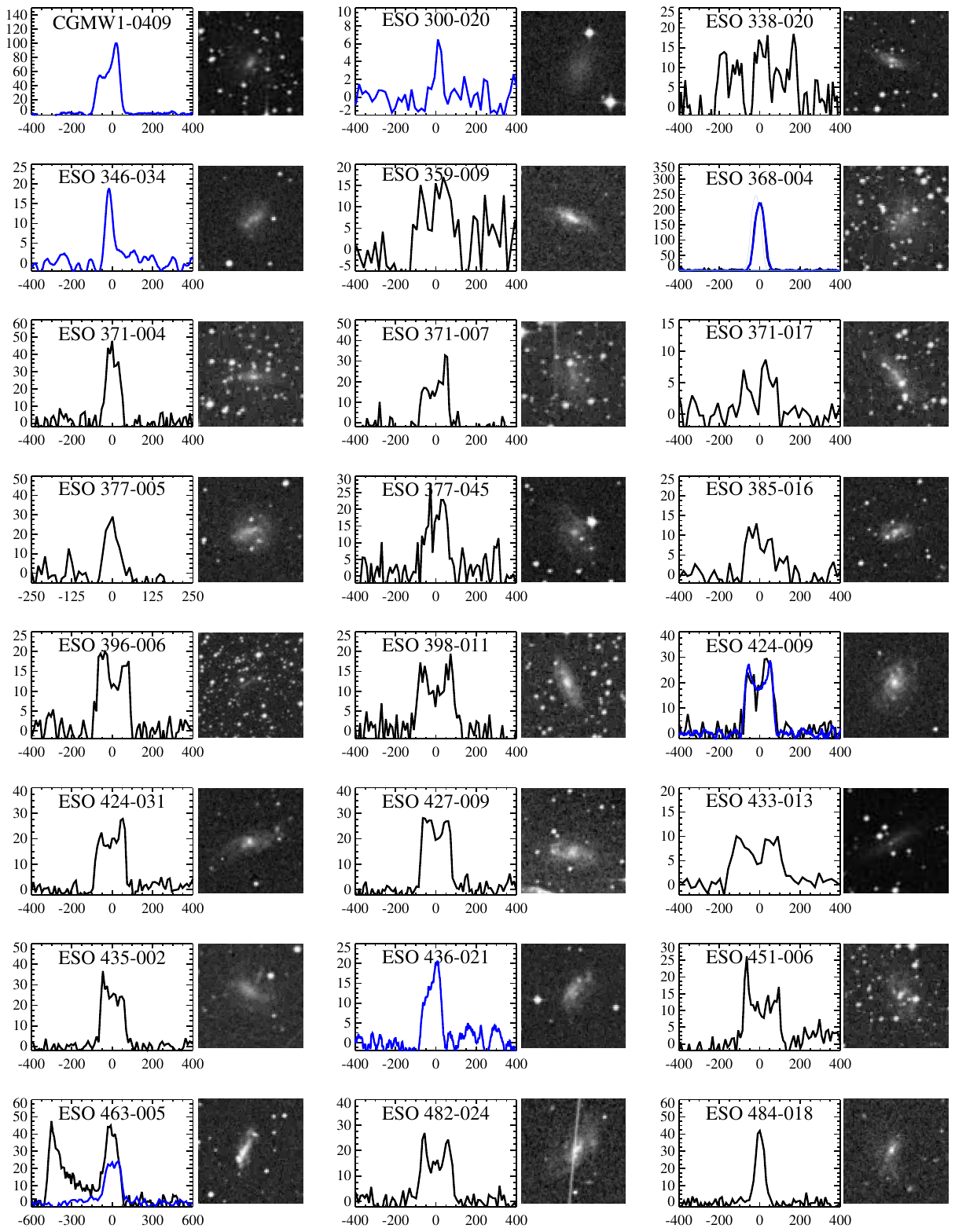}
  \caption{{\bf a.} 21 cm \HI{} line spectra and optical images of the detected galaxies. 
Black lines in the spectra represent \nan{} data, blue lines GBT data, and red 
lines Arecibo data. In the case where an object was observed twice by the same telescope, 
the earliest observation is shown in by a dashed line.
Optical images are 2$\times$2 arcmin in size. False color ($i$, $r$, and $g$ filter) images are 
from the SDSS DR12 \citep{2015ApJS..219...12A}, and black and white images are from the 
2$^{nd}$ Digitized Sky Survey (DSS2) Blue plates \citep{2004AJ....128..502A}, 
used when SDSS images are not available. Objects are arranged in alphabetical order.
The complete figure set (13 images) is available in the online journal.}
  \label{fig:HLdets}
\end{figure*}

\begin{figure*}  
  \centering
  \includegraphics[width = 14.9cm]{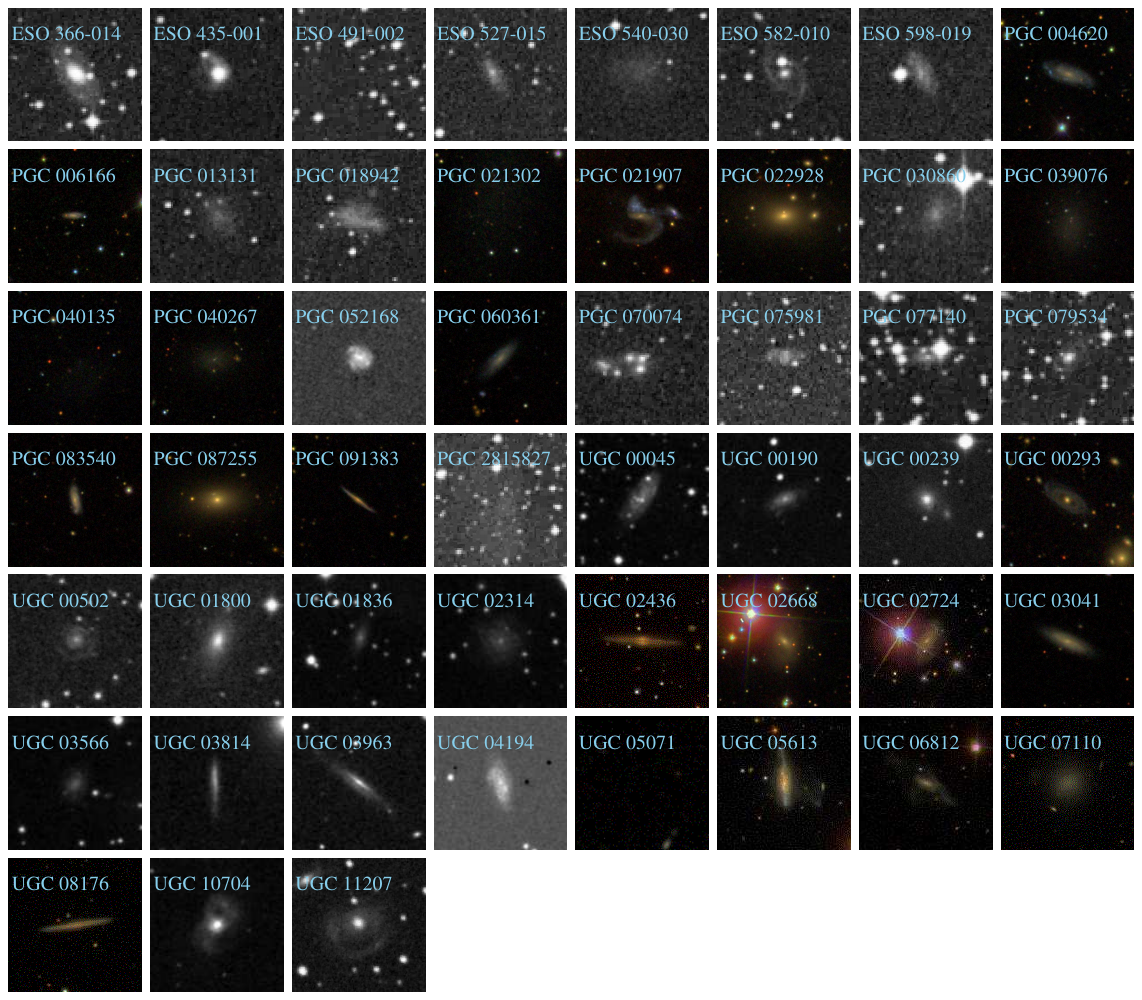}
\caption{Optical images of target galaxies that were not detected by us in \HI{}. 
Images are 2$\times$2 arcmin in size, unless otherwise indicated in the lower right corner. 
False color ($i$, $r$, and $g$ filter) images are from the SDSS DR16 \citep{2020ApJS..249....3A}
and black and white images are from the 2$^{nd}$ Digitized Sky Survey (DSS2) Blue plates 
\citep{2004AJ....128..502A}, used when DR16 images are not available. 
Objects are arranged in alphabetical order.}
  \label{fig:nondets}
\end{figure*}

\begin{figure*} 
\centering
\gridline{
\fig{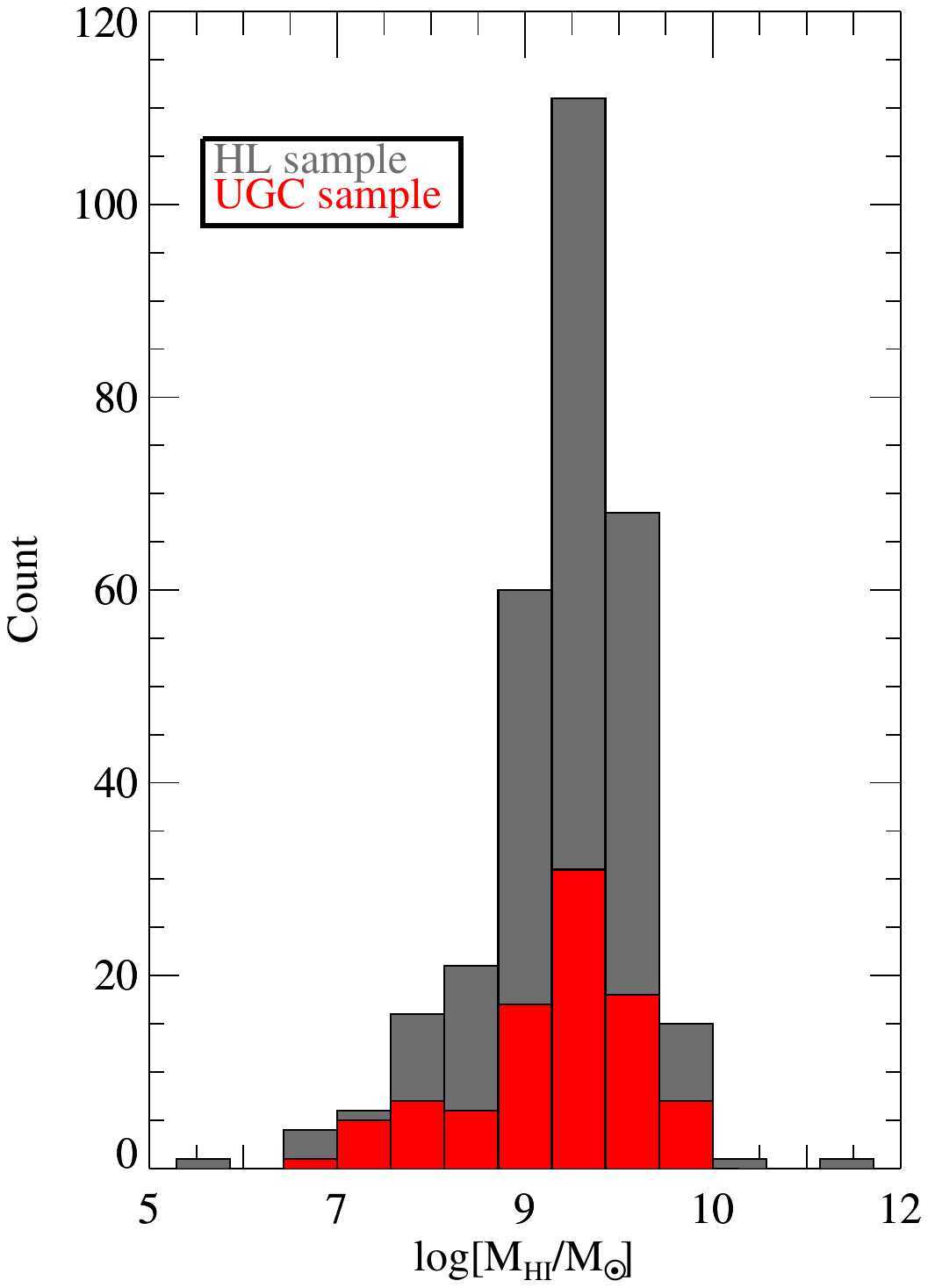}{0.4\textwidth}{(a)}
\fig{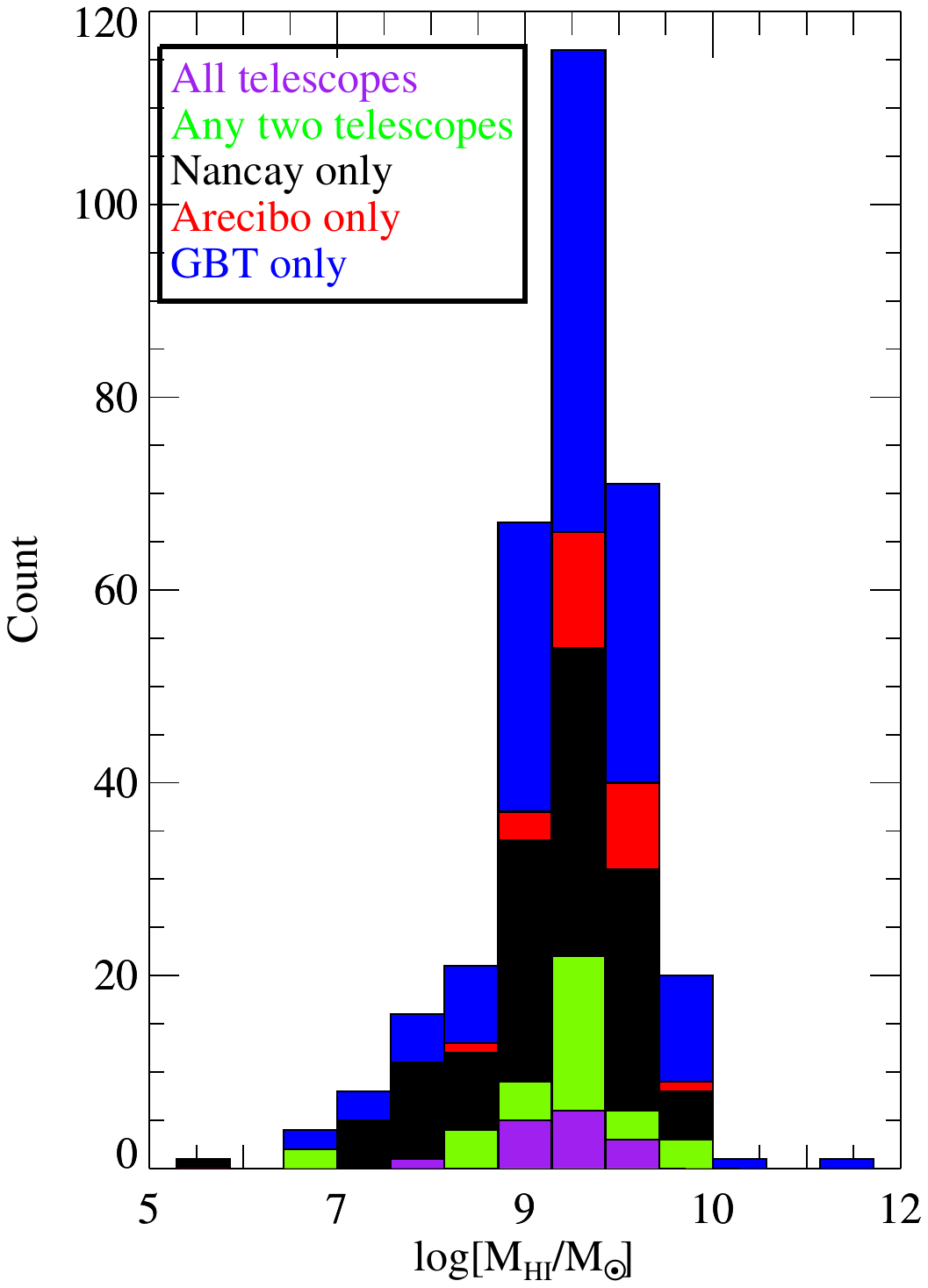}{0.4\textwidth}{(b)}
}
\caption{Distributions of total \HI{} masses \MHI{}, in \Msun, for all detected galaxies 
in our survey: (a) comparison between the UGC sample (red) and the HyperLeda sample (grey), 
and (b) as a function of the telescope(s) which detected the galaxy. 
\label{fig:MHI_Hist_all}}
\end{figure*}

The primary results from the observations are given in Tables~\ref{tab:avgresultsdets}, \ref{tab:indresultsdets}, 
and \ref{tab:resultsnondets}. \HI{} spectra and optical images of the detections are shown in Figure~\ref{fig:HLdets}
and optical images of the non-detections are shown in Figure~\ref{fig:nondets}.
The \HI{} mass distributions for the samples can be seen in Figure~\ref{fig:MHI_Hist_all}.

Tables ~\ref{tab:avgresultsdets} and ~\ref{tab:indresultsdets} give the measured \HI{} line properties 
of all detected galaxies. Table~\ref{tab:avgresultsdets} provides the median and therefore highest 
quality values, while Table~\ref{tab:indresultsdets} provides the values as measured by each telescope, 
for all galaxies which were observed by more than one telescope.

Finally, Table~\ref{tab:resultsnondets} lists the r.m.s.
noise level values of all non-detections for each telescope.

In Table ~\ref{tab:avgresultsdets}, if an object was observed with more than one telescope within the HyperLeda survey, non-weighted 
averages were used of the measured values, unless one or more of the individual measurements 
showed a significant discrepancy with the others, such as a much larger 
estimated uncertainty, lower line flux or difference in central velocity, in which case they were not 
used here (see also the notes on individual galaxies in the Appendix). 
 
Listed in Tables~\ref{tab:avgresultsdets} and ~\ref{tab:indresultsdets} are the following elements
for all detected galaxies, in alphabetical order of their name:
\begin{enumerate}
\item {\bf Name:} The common name of the object;
\item {\bf Sample:} The survey(s) for which the observation was made. 
The options are: HyperLeda (HL), UGC (UGC), both HyperLeda and UGC (HL/UGC), 
HyperLeda spurious dectection (HLS) and UGC spurious detection (UGCS). 
Please note that this column only appears in Table ~\ref{tab:avgresultsdets};
\item {\bf Notes:} Indication if the galaxy has notes in Appendix~\ref{Sec:IndGals}; 
given as Y/N for Yes/No;
\item {\bf Conf:} Contamination (confusion) status for detections, indicating the level of possibility 
that another galaxy within the telescope beam contributed some or all of the detected \HI{} flux, 
graded as 1 (unlikely), 2 (possible) and 3 (very likely). For details on the assessment of the contamination status,
see the description of potentially confusing sources given in Appendix  for all galaxies with a
confusion status of 3 or 2;
\begin{itemize}
\item For the HyperLeda sample, a search was made (using HyperLeda and NED) within respectively
12$'$ radius and 1000 \kms{} around the position and systemic velocity of the detected galaxy. 
If no known source was found within that region, the confusion flag was 
set to 1, indicating the detection is likely not confused. If a known galaxy 
lay within that region but was unlikely to be contributing to the detected \HI{} flux 
(e.g., the known galaxy is of type S0 or earlier), the confusion status flag was set to 2. 
If at least one galaxy lay within the region and was very likely to contribute to some 
or all of the detected \HI{} gas, the confusion status was set to 3. 
\item For the UGC sample, we first inspected Digitized Sky Survey (DSS) images centered on 
the position of each clearly or marginally detected source, over an area of
12$'$ $\times$ 36$'$ ($\alpha$ $\times$ $\delta$). In case galaxies were
noted that might give rise to confusion with the \HI{} profile of the
target galaxy, we then queried the NED and HyperLeda databases for
information on the objects. 
\end{itemize}
All galaxies with possible contamination have notes in Section~\ref{Sec:IndGals};
\item {\bf \VHIb{}:} Central \HI{} line (optical, heliocentric) velocity as measured in our profiles, in \kms{}; 
\item {\bf error:} Error for \VHI{}, determined following \citep{schneider1986,schneider1990}, as
\begin{equation} \sigma_{V_{HI}} = 1.5(W_{20}-W_{50})S\!NR^{-1}\, (\kms)
\end{equation} 
where \SNR{} is the peak signal-to-noise ratio of a spectrum, which we define as the ratio of 
the peak flux density $S_{\rm max}$ and the r.m.s. dispersion in the baseline (both in Jy);
\item {\bf \Wfiftyb{}:} \HI{} line velocity width measured at 50\% of the \HI{} profile peak value, in \kms;
\item {\bf error:} Error for \Wfifty{}, in \kms{} (Table ~\ref{tab:indresultsdets} only.). 
For GBT and \nan{} data, these errors are expected to be 2 times those in \VHI, following \citep{schneider1990}. 
For the Arecibo data, an upper limit of 15 \kms{} is given for the errors, i.e., the width of one (smoothed) 
velocity channel. 
\item {\Wfiftycorb{}:}  \HI{} line velocity width measured at 50\% of the \HI{} profile peak value, in \kms,
corrected for random motion effects, using \citep{1985ApJS...58...67T}, and inclination:

\begin{equation}
W_{50,cor}\; = \;
\sqrt{
W_{50,i}^2 \;+\; {W_t}^2\;+\;
2 W_{50,i} W_t \left[ 1-e^{-{(W_{50,i}/W_{t})}^2} \right]  
\;-\;2 {W_t}^2 
e^{-({W_{50,i}/W_t})^2 } 
} 
\end{equation}
Here, w$_t$ is the assumed turbulent motion, which is 14 \kms{} for LSB galaxies \citep{2000AJ....119..136O} 
and $W_{50,i} = W_{50}/\left[{sin(i)}\right]$, where the inclination is based on the $b/a$ axial ratio listed 
in Table~\ref{tab:opticalprops}.
\item {\bf \Wtwentyb{}:} \HI{} line velocity width measured at 20 \% of the \HI{} profile peak value, in \kms;
\item {\bf error:} Error for \Wtwenty{}, in \kms.
For GBT and \nan{} data, these errors are expected to be 3.1 times those in \VHI, following \citep{schneider1990}. 
For the Arecibo data, the listed errors are 15 \kms{} (see \Wfifty{} error). 
(Table ~\ref{tab:indresultsdets} only.)
\item {\bf r.m.s.:} r.m.s. noise level of the \HI{} spectrum, in mJy, measured around the detected
\HI{} line. For Table ~\ref{tab:avgresultsdets} if the detection was made with more than one telescope 
only the r.m.s. for the most sensitive observation is given; 
\item {\bf \FHI{}:} Measured integrated \HI{} line flux, in Jy \kms;
\item {\bf error:} Error for \FHI{},  in Jy \kms{}, determined following \citep{schneider1986,schneider1990}, as
\begin{equation} \sigma_{F_{HI}} = 2(1.2W_{20}R)^{0.5}r.m.s.\, (\kms)
\end{equation} 
where $R$ is the instrumental resolution in \kms{} (see Section~\ref{sec:observations});
\item {\bf \MHI{}:} Log of the total \HI{} mass in units of \Msun{}, where 
\MHI{} = 2.356$\times$10$^5$$\cdot$D$^2$$\cdot$\FHI{}, and $D = V/70$ is the galaxy's distance (in Mpc);
\item {\bf error:} Error for log(\MHI{}/\Msun{});
 \item {\bf Telescope:} Telescope(s) used for observation: A = Arecibo, G = GBT, N = \nan{}. 
The number of times a telescope is listed denotes the number of separate observations 
made with that telescope over the duration of our survey.
 \end{enumerate}

Listed in Table~\ref{tab:resultsnondets} are the following elements for all galaxies not detected by any telescope, 
in alphabetical order of their name. 
\begin{enumerate}
\item {\bf Name:} The common name of the object;
\item {\bf Sample:} Which survey resulted in this detection. The options are: HyperLeda (HL), UGC (UGC) and both (HL/UGC);
\item {\bf Notes:} Indicating if a galaxy has notes in Section~\ref{Sec:IndGals}  
(see Table~\ref{tab:avgresultsdets} for a description);
\item {\bf \nan{}:} r.m.s. noise level across the observed spectra, in mJy, for all \nan{} observations. 
For the HyperLeda sample, the search range is -250 to 10,600 \kms{}. For the UGC sample a single value indicates 
a search was made only in the ``lo'' range, 325 to 11,825 \kms, and a second value an additional search in the ``hi''range,
$\sim$6875 to 18,330 \kms{};
\item {\bf GBT:} r.m.s noise level across the observed spectra, in mJy, for all GBT observations, 
in the range 2000 to 25,600 \kms{};
\item {\bf Arecibo low:} r.m.s noise level across the observed spectra, in mJy, for all Arecibo observations in the 
low velocity range, -2000 to 19000 \kms{};
\item {\bf Arecibo high:} r.m.s noise level across the observed spectra, in mJy, for all Arecibo observations in the 
high velocity range, 19000 to 36000 \kms{}.
\end{enumerate}

\subsection{Inter-telescope Comparisons} \label{subsec:intertel} 

\subsubsection{Telescope-Telescope Comparison}  
Of the \totobs{} galaxies observed for this paper, the majority (\onetelobs{}, or 
\onetelobsperc{}) was observed with only one telescope. The remaining \twoplustelobs{} 
galaxies, though, allow for a comparison of \HI{} profiles detected with the different 
instruments. Included within these statistics are the \doublesample{} galaxies detected 
in both the UGC and HyperLeda surveys. 

In all, then, the breakdown between the samples is:
\begin{itemize}
\item \GBTNRTobs{} galaxies observed by both the GBT and \nan{} telescopes (\GBTGBTobs{} were observed twice by the GBT) ;
\item \NRTAOobs{} galaxies observed by both the Arecibo and \nan{} telescopes;
\item \GBTAOobs{} galaxies observed by both the Arecibo and GBT telescopes;
\item \NRTNRTobs{} galaxy observed only at \nan{} but separately for the UGC and HyperLeda samples;
\item \threetelobs{} galaxies observed by all three telescopes.
\end{itemize}

\begin{figure*} 
\centering
\gridline{
\fig{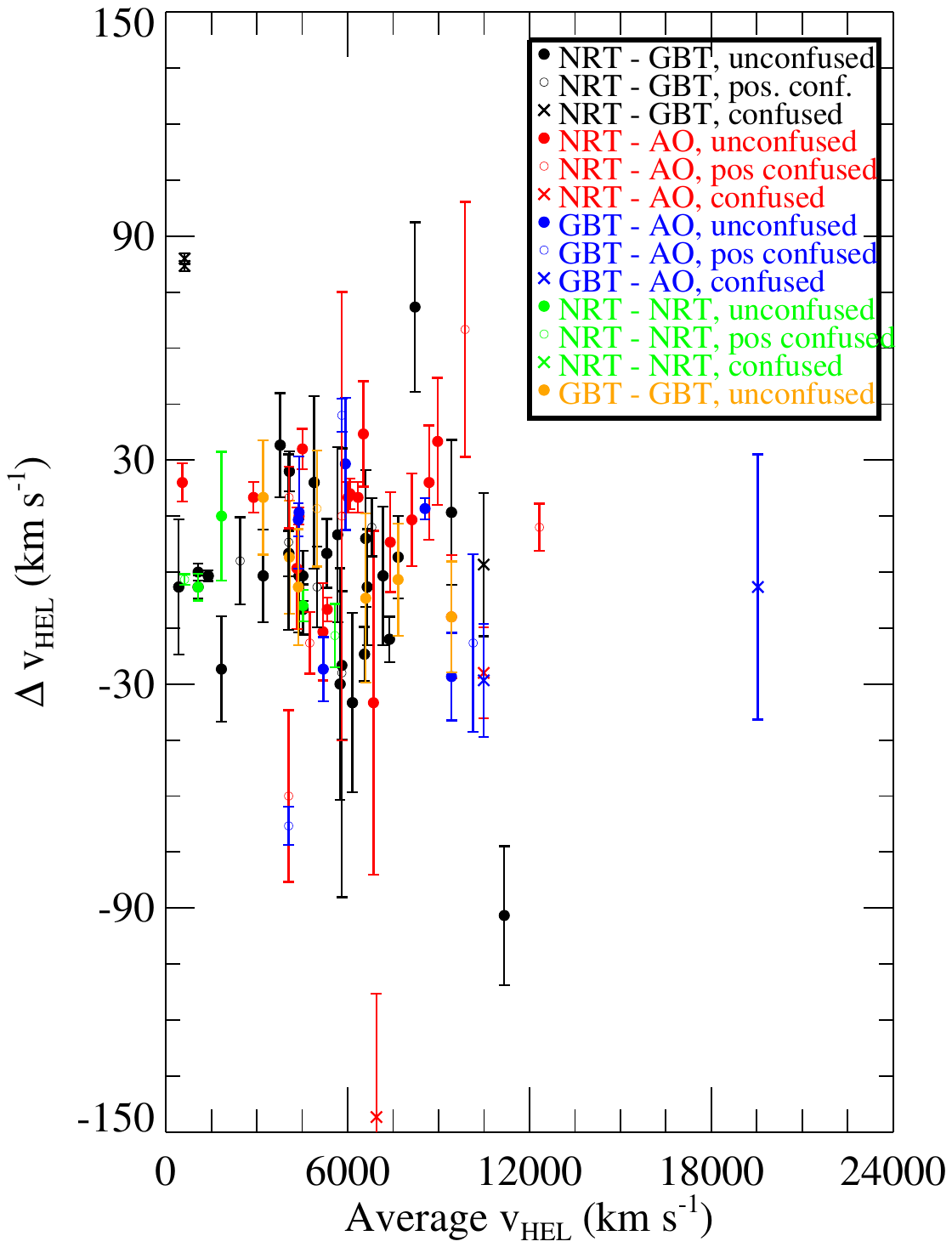}{0.3\textwidth}{(a)}
\fig{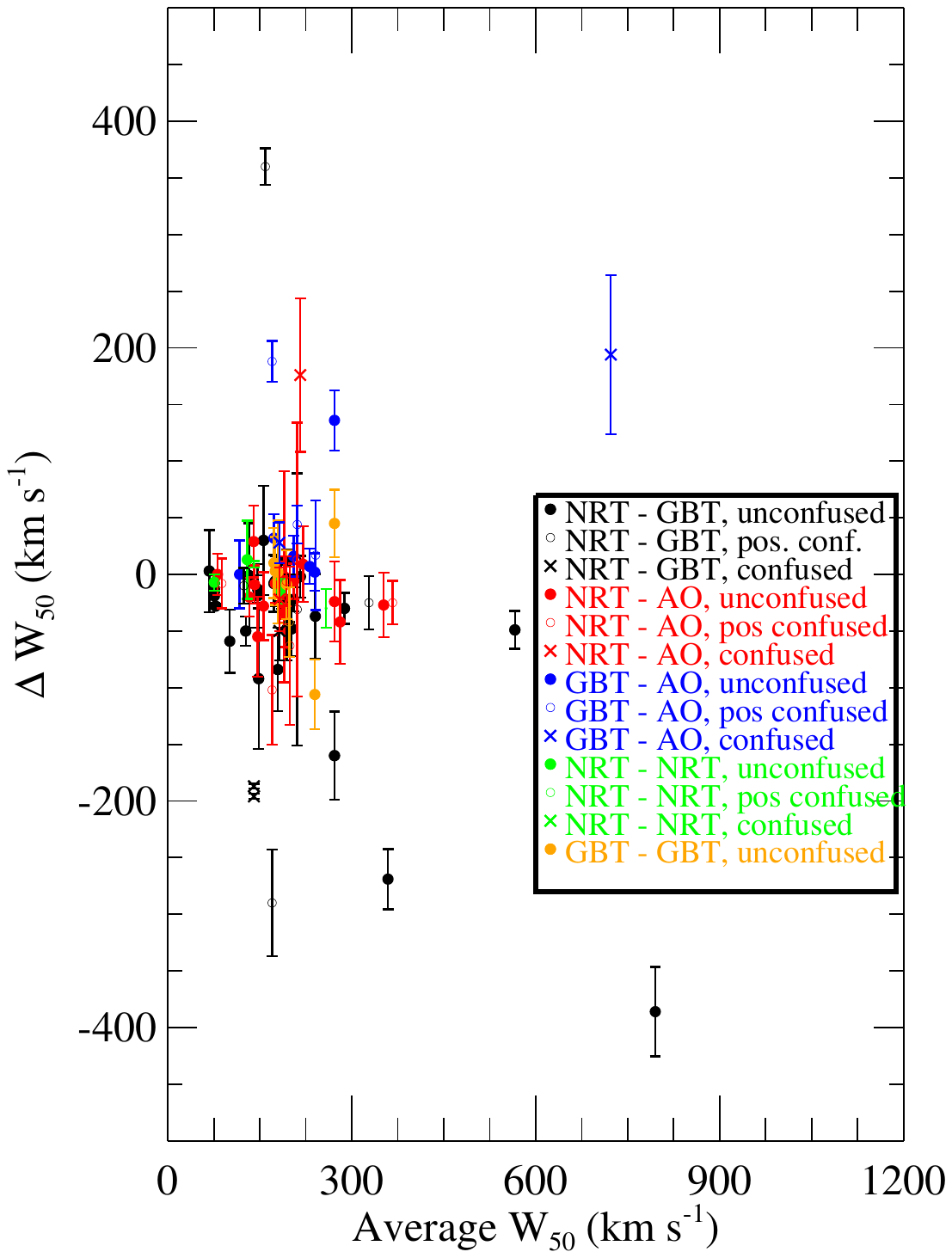}{0.3\textwidth}{(b)}
\fig{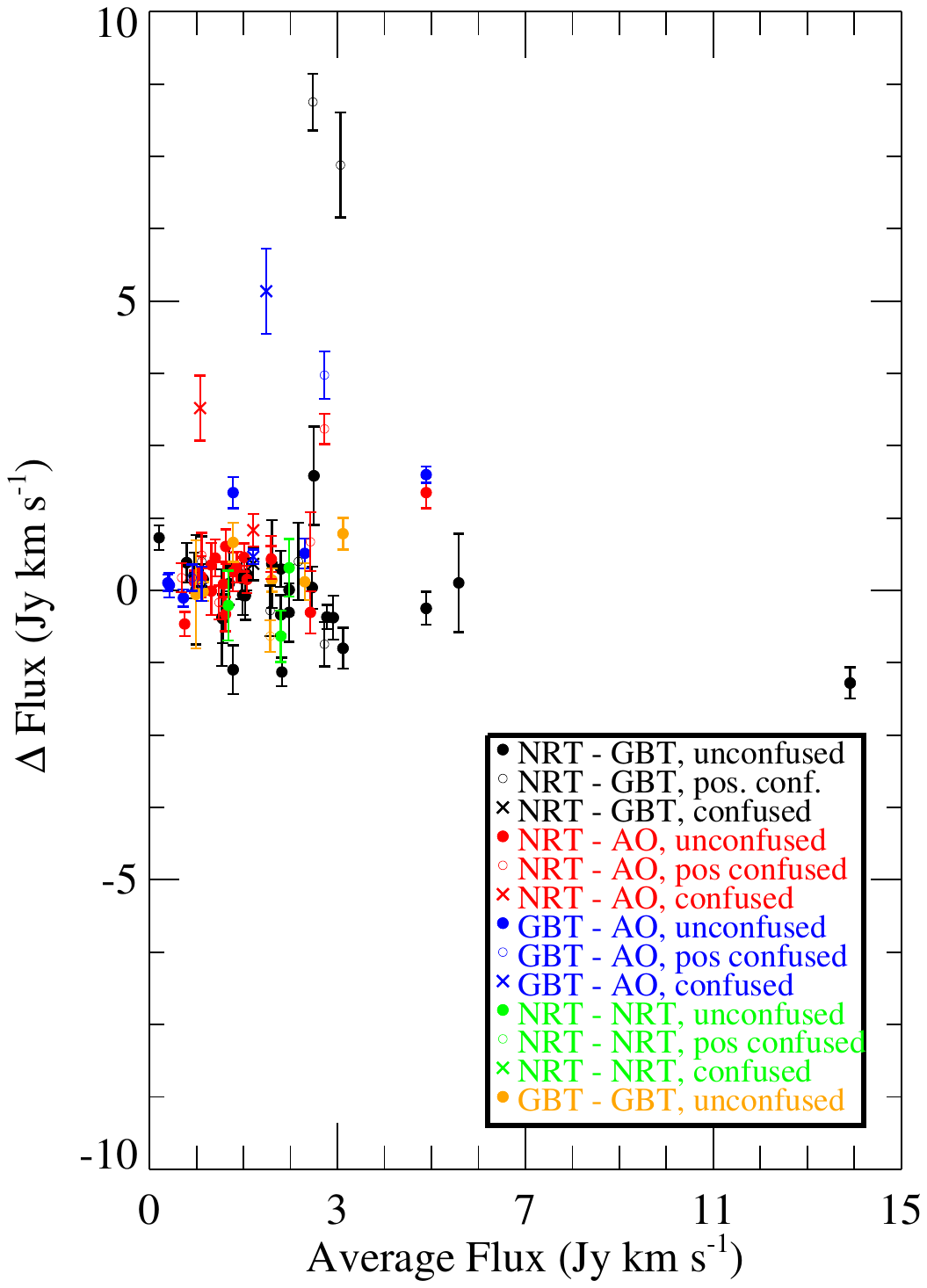}{0.3\textwidth}{(c)}
}
\caption{Comparison of differences in \HI{} line profile parameters of galaxies in our 
two samples that were detected by us with more than one telescope.
Plotted as a function of their average central \HI{} velocity, \VHI{}, are the differences in:
(a) central \HI{} velocity, $\Delta$\VHI\ in \kms, (b) in \Wfifty\ line width, 
$\Delta$\Wfifty\ in \kms, and (c) in integrated \HI{} line flux, $\Delta$\FHI\ in Jy \kms. 
As indicated in the legend, colors indicate which telescopes' results are compared: 
Green Bank (GBT) vs \nan{} (NRT) in black, Arecibo (AO) vs NRT in blue,  
AO vs GBT in red, and NRT (HyperLeda sample) vs NRT (UGC sample) in green. 
Different symbols indicate the probability of a detection being
confused by another galaxy within a telescope beam: large filled dots for 
uncontaminated, crosses for possibly contaminated, and small open dots for clearly contaminated. 
Outliers in all plots are described in Section~\ref{subsec:intertel} and \ref{Sec:IndGals}. 
\label{fig:HLtelcomp} }
\end{figure*}

These inter-telescope comparisons are shown in Figure~\ref{fig:HLtelcomp} for 
\HI{} profile central velocities, \Wfifty\ velocity widths and integrated line fluxes. 
Overall, they are consistent.

Comparison between the central velocities measured by the different telescopes/surveys 
shows a spread of differences centered around \veldispall{} \kms{} for the \twoplustelunconf{} 
galaxies not contaminated by a nearby companion and observed by more than one telescope 
(indicated by filled dots in Figure~\ref{fig:HLtelcomp}). 
The comparisons between the individual telescope are consistent, with the \nan{} and GBT telescopes
having the smallest differences, but small number statistics makes 
drawing any conclusions difficult. The difference between telescopes and their standard error are:
v$_{NRT}$-v$_{GBT}$ = \veldispNRTGBT{} \kms;
v$_{GBT}$-v$_{AO}$ = \veldispGBTAO{} \kms; 
v$_{NRT}$-v$_{AO}$ = \veldispNRTAO{} \kms; 
v$_{NRT, UGC}$-v$_{NRT, HL}$ = \veldispNRTNRT{} \kms; and
v$_{GBT}$-v$_{GBT}$ = \veldispGBTGBT{} \kms{}.

The \Wfifty{} profile widths measured for non-contaminated detections are also similar. 
The measured \Wfifty{} values for the \twoplustelunconf{} uncontaminated galaxies averaged 
$\Delta$\Wfifty{} = \Wfiftydispall{} \kms{}. Here the NRT and Arecibo values are
the most similar.  The difference between telescopes and their standard error are:
$W _{\rm 50, NRT}$ - $W _{\rm 50, GBT}$ = \WfiftydispNRTGBT{} \kms;
$W _{\rm 50, GBT}$ - $W _{\rm 50, AO}$ = \WfiftydispGBTAO{} \kms;
$W _{\rm 50, NRT}$ - $W _{\rm 50, AO}$ = \WfiftydispNRTAO{} \kms; 
$W _{\rm 50, NRT, UGC}$-$W _{\rm 50, NRT, HL}$ = \WfiftydispNRTNRT{} \kms; and
$W _{\rm 50, GBT}$ - $W _{\rm 50, GBT}$ = \WfiftydispGBTGBT{} \kms.

The integrated \HI{} line fluxes are comparable as well. 
Listed here are differences in measured flux values for the uncontaminated galaxies, 
given as a percentage of the galaxies' flux.
Here, the GBT values readily agree with those of \nan{}, but Arecibo values
do not agree as well with either telescope.
The difference between telescopes and their standard error are:
f$_{NRT}$-f$_{GBT}$ = \fluxdispNRTGBT{}\%;
f$_{GBT}$-f$_{AO}$ = \fluxdispGBTAO{}\%;
f$_{NRT}$-f$_{AO}$ = \fluxdispNRTAO{}\%;
f$_{NRT, UGC}$-f$_{NRT, HL}$ = \fluxdispNRTNRT{}; and
f$_{GBT}$-f$_{GBT}$ = \fluxdispGBTGBT{}\%.  
Also in this case, the differences are small compared to the noise
for small number statistics.

\subsubsection{The Outliers} 

Table~\ref{tab:outliers} lists all galaxies which were observed by more than one telescope 
and which show differences in one or more measured \HI{} property that lie outside the 
ranges listed above. 
Each of these galaxies is also described in detail in Section~\ref{Sec:IndGals}.

\subsection{Comparisons with the Literature}  \label{subsec:lit_comp} 

\subsubsection{Optical Velocity Comparison}  

Of the \totobs{} galaxies observed, \litredshifts{} have previously published spectral line 
velocities, with \litoptvel{} having optical (spectroscopic) velocities in the literature. 
Of these \litoptvel{}, \litoptveldet{} were detected in our survey, and of these \litoptvelunconf{}  
are unlikely to be confused with any other object within the telescope beam.

\begin{figure*} 
\centering
\includegraphics[width = 12cm]{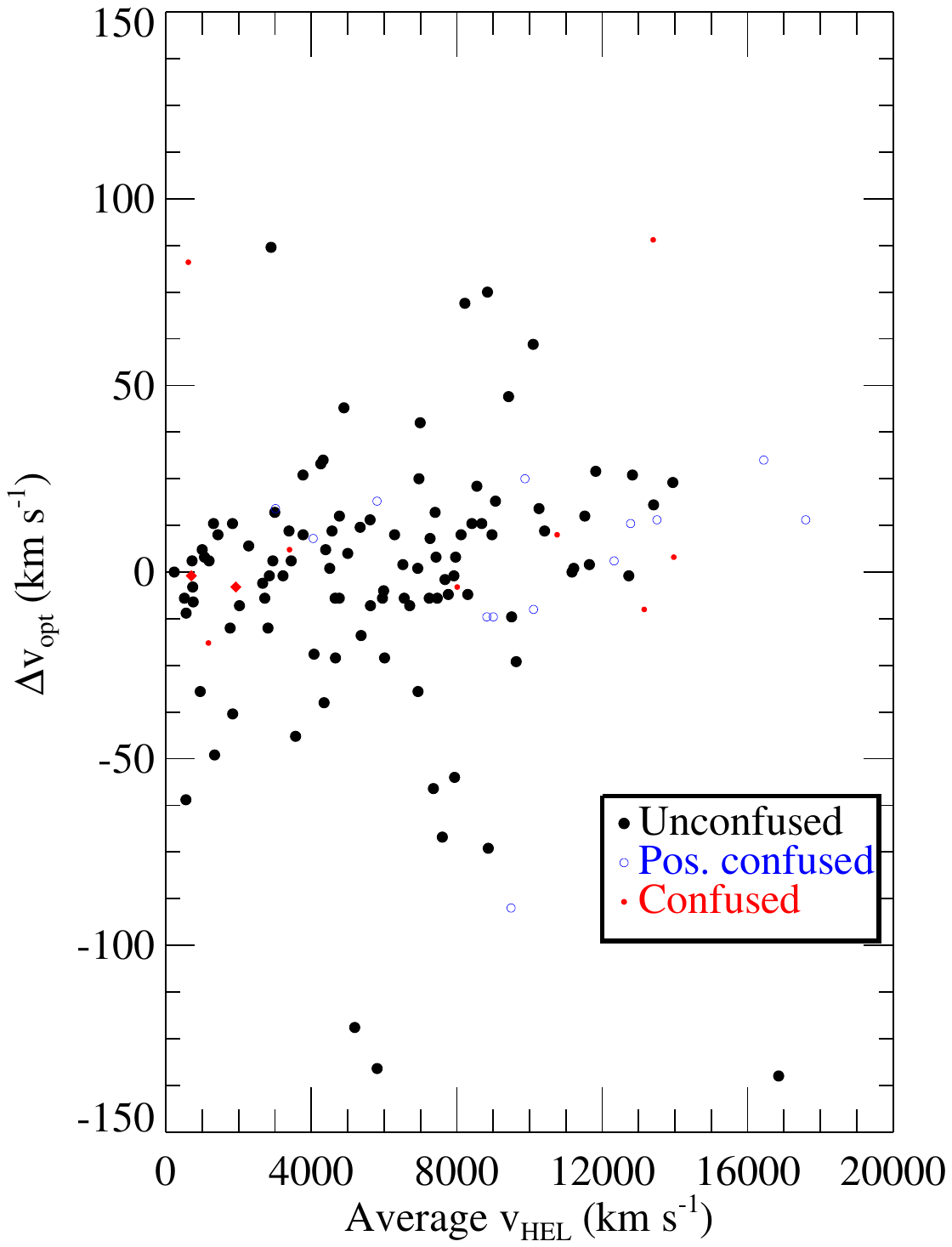}
\caption{Difference between our central \HI{} line velocities and averaged values of 
optical velocities listed in the literature, $\Delta$\Vopt\ in \kms, as a function of 
optical velocities from the literature, \Vhel{} in \kms, for galaxies in our combined sample.
As indicated in the legend, colors indicate the probability of a detection being
confused by another galaxy within a telescope beam: black for uncontaminated, 
blue for possibly contaminated, and red for likely contaminated.}
\label{fig:vHI_vopt} 
\end{figure*}

Figure~\ref{fig:vHI_vopt} shows the difference between our \VHI\ central \HI{} velocities 
(see Table~\ref{tab:avgresultsdets}) and the optical spectroscopic velocities 
(hereafter referred to as "optical velocities") found in the literature, as a function of \VHI. 
Almost all optical velocities are within $\pm$100 \kms{} of our \HI\ measurements, with a median 
velocity difference of \litoptveldist{} \kms{}. Those galaxies with velocity differences $>\pm$100\kms{} 
are discussed in Table~\ref{tab:voptdif} and Appendix~\ref{Sec:IndGals}.

The \HI{} velocities of the three galaxies listed in Table~\ref{tab:voptdif} are all likely 
to be accurate and reliable values.
None of their \HI{} profiles are likely to be confused with \HI{} from another galaxy 
within the beam. Two of the galaxies, UGC 263 and PGC 7225, have rather uncertain 
optical velocities, which are both only $<$1.5$\sigma$ different from our \HI{} values,
and the peak \HI{} velocity of UGC 1145 was confirmed by two telescopes.
%

\subsubsection{\HIit{} Literature Values} 

\begin{figure*} 
\centering
\gridline{
\fig{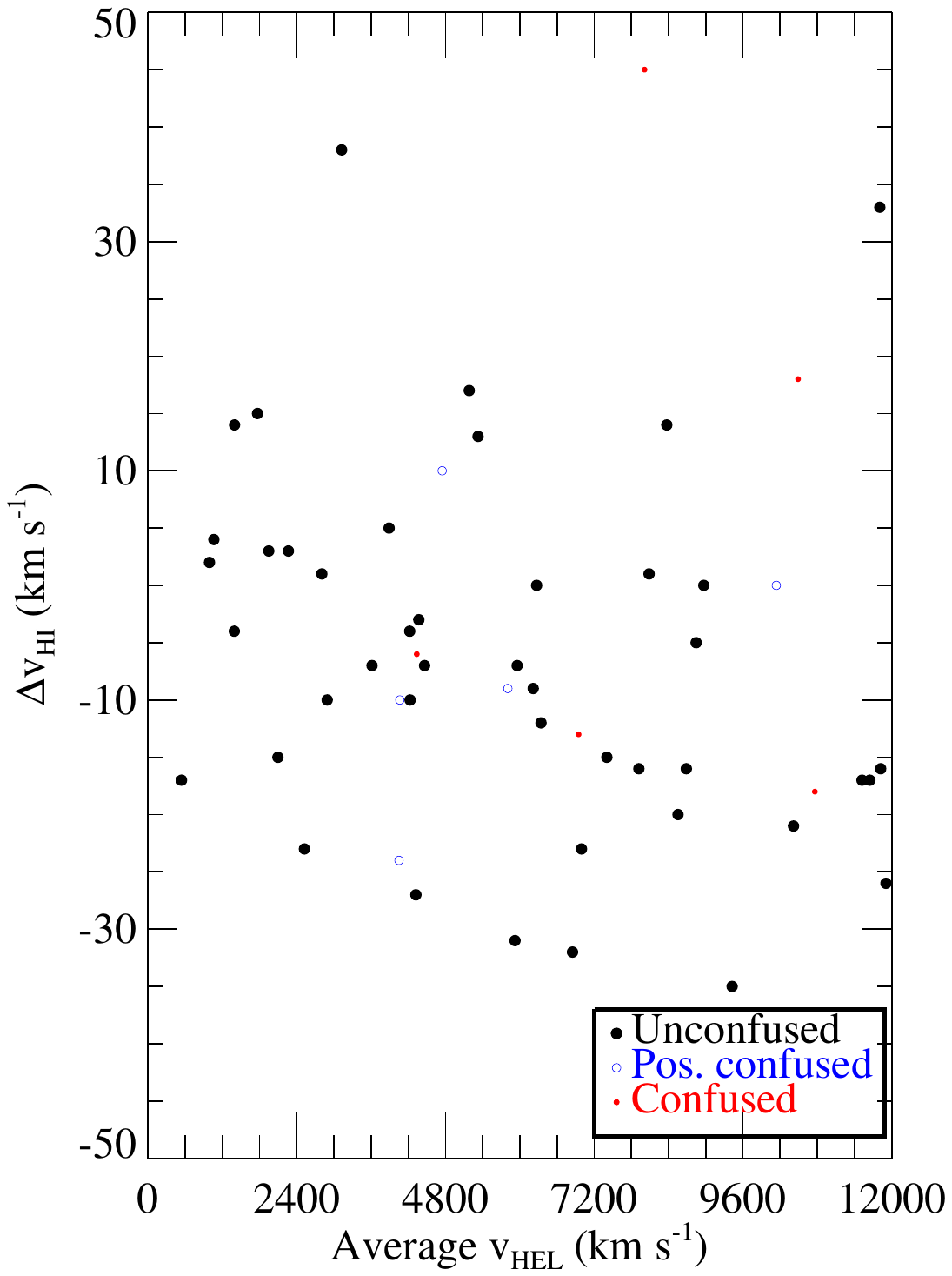}{0.3\textwidth}{(a)}
\fig{Figures/W50_comp_LIT.pdf}{0.3\textwidth}{(b)}
\fig{Figures/flux_comp_Lit.pdf}{0.3\textwidth}{(c)}
}
\caption{Comparison of differences in average \HI{} line profile parameters as measured by us with 
averages of values listed in the literature, for galaxies in our combined sample. Plotted as a function of our average \HI{} line parameters are the differences in: 
(a) central velocity, $\Delta$\VHI\ in \kms, (b) $\Delta$\Wfifty\ line width, in \kms, and 
(c) integrated line flux, $\Delta$\FHI\ in Jy \kms. 
As indicated in the legend, colors indicate the probability of a detection being
confused by another galaxy within a telescope beam: black for uncontaminated, 
blue for possibly contaminated, and red for likely contaminated.
Please note that figure (a) does not include PGC 38958 (see Appendix \ref{Sec:IndGals} for explanation).
}
\label{fig:HI_comp_lit}
\end{figure*}

Published \HI{} measurements are now available for \litHIvel{} of our detected galaxies. 
Of these, \litHIvelunconf{} are of uncontaminated detections. 
A comparison between literature values and our measurements is shown in Figure~\ref{fig:HI_comp_lit} 
for \VHI, \Wfifty, and \FHI. On average the values match extremely well, with
$\Delta$ \Vhel{} = \litHIveldist{}, 
$\Delta$\Wfifty{} = \litWfifty{}, and $\Delta$\FHI{} = \litflux{} Jy \kms.

Only one galaxy, PGC 2815809, has significant differences between our measured \HI{} 
values and those found in the literature, but this may be due to nearby \HI{} source HIPASS J0756-26.
Further observations are required to determine if PGC 2815809 and HIPASS J0756-26 are in fact the 
same object and, if so, where its center lies (see Appendix~\ref{Sec:IndGals} for further details). 

\section{THE \HIbf{} PROPERTIES OF THE SAMPLE} \label{sec:HIprops}

\subsection{Sample Properties}  

\begin{figure*} 
\centering
\includegraphics[width = 12cm]{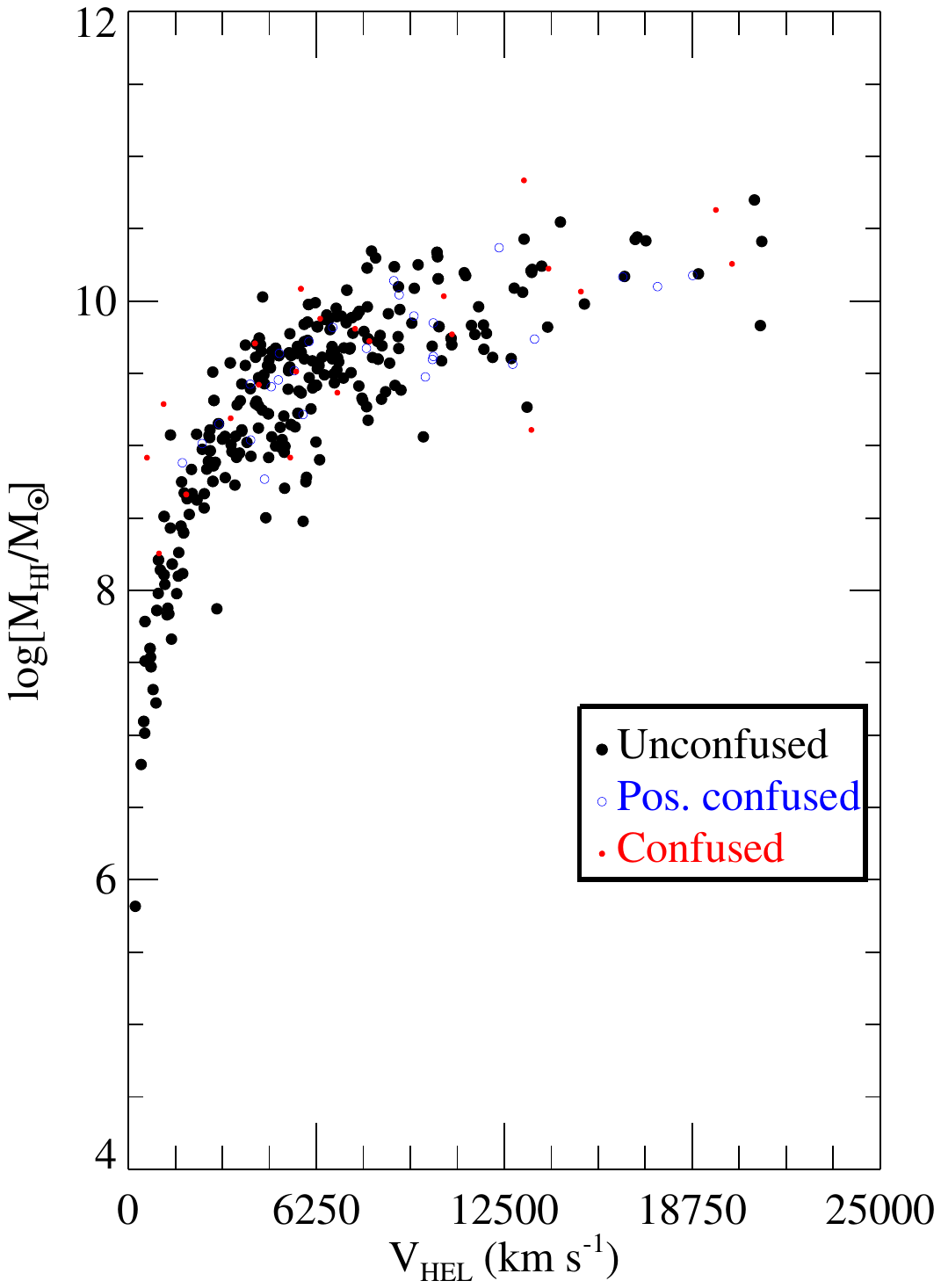}
\caption{Total \HI{} mass, log(\MHI) in \Msun, as a function of radial velocity, 
$V_{\rm hel}$ in \kms, for detected galaxies in our combined sample. 
As indicated in the legend, symbols indicate the probability of a detection being
confused by another galaxy within a telescope beam: solid for uncontaminated, 
open for possibly contaminated, and small for likely contaminated. 
}
\label{fig:allMHIV}
\end{figure*}

Figure~\ref{fig:allMHIV} shows the total \HI{} masses of our \totdet{} detected galaxies as a 
function of radial velocity (\VHI). The plot shows the expected increase of \MHI{} with velocity, 
with most of the lower-mass detections at a given velocity obtained with the much larger Arecibo telescope. 

\begin{figure*} 
\centering
\includegraphics[width = 12cm]{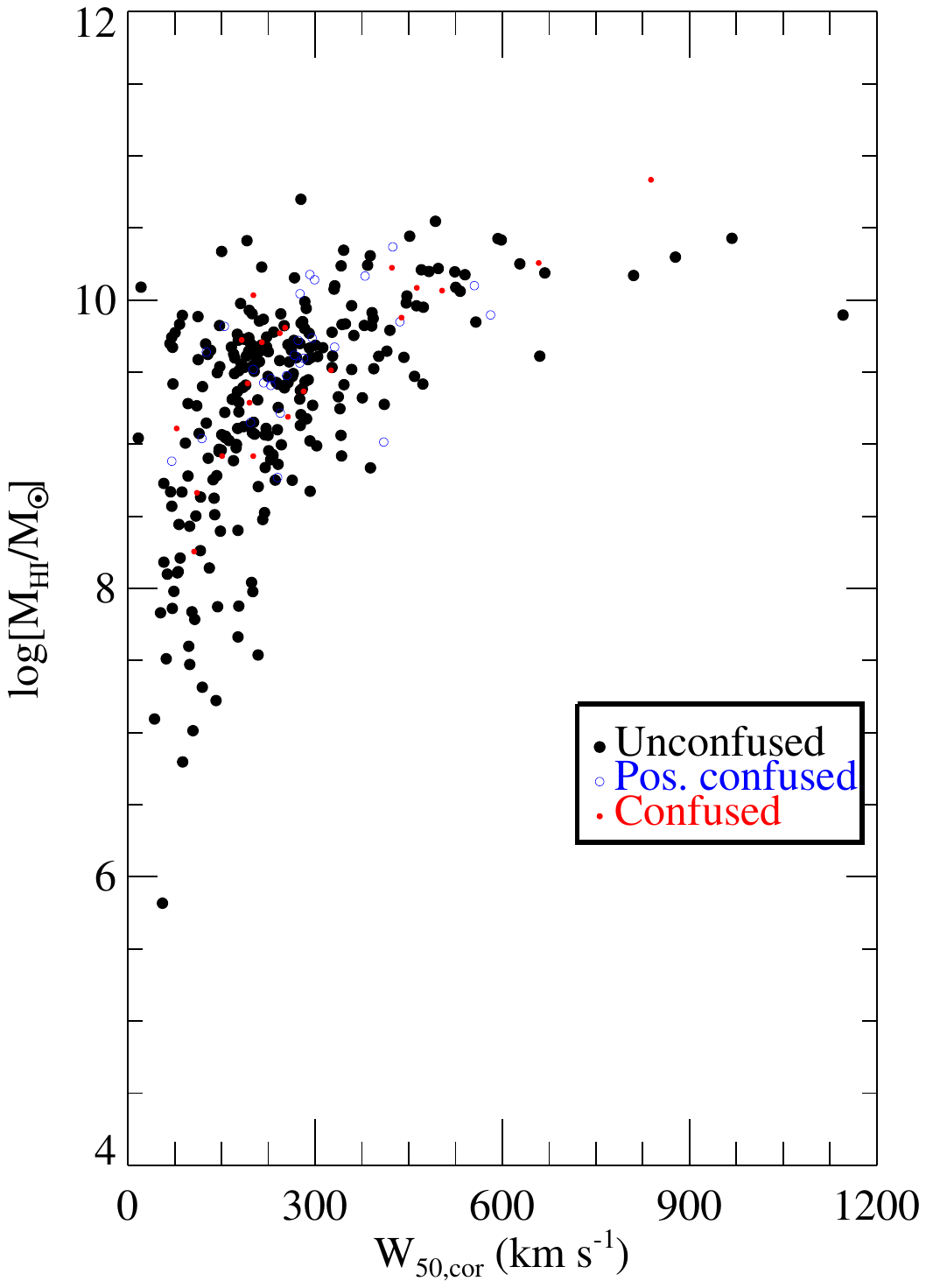}
\caption{Total \HI{} mass, log(\MHI) in \Msun, as a function of total mass, as represented
by the inclination-corrected \Wfiftycor{} \HI\ line velocity width (in \kms) for detected galaxies 
in our combined sample. 
As indicated in the legend, colors indicate the 
probability of a detection being confused by another galaxy within a telescope beam: 
black for uncontaminated, blue for possibly contaminated, and red for likely contaminated. 
\label{fig:MHI_W50} }
\end{figure*}

Plotted in Figure~\ref{fig:MHI_W50} is the total \HI{} mass as a function of the inclination-corrected 
\Wfiftycor{} line width, a parameter indicative of the total galaxy mass, again showing the expected increase with 
\Wfiftycor{}  and \HI{} mass. 

\begin{figure*} 
\centering
\gridline{
\fig{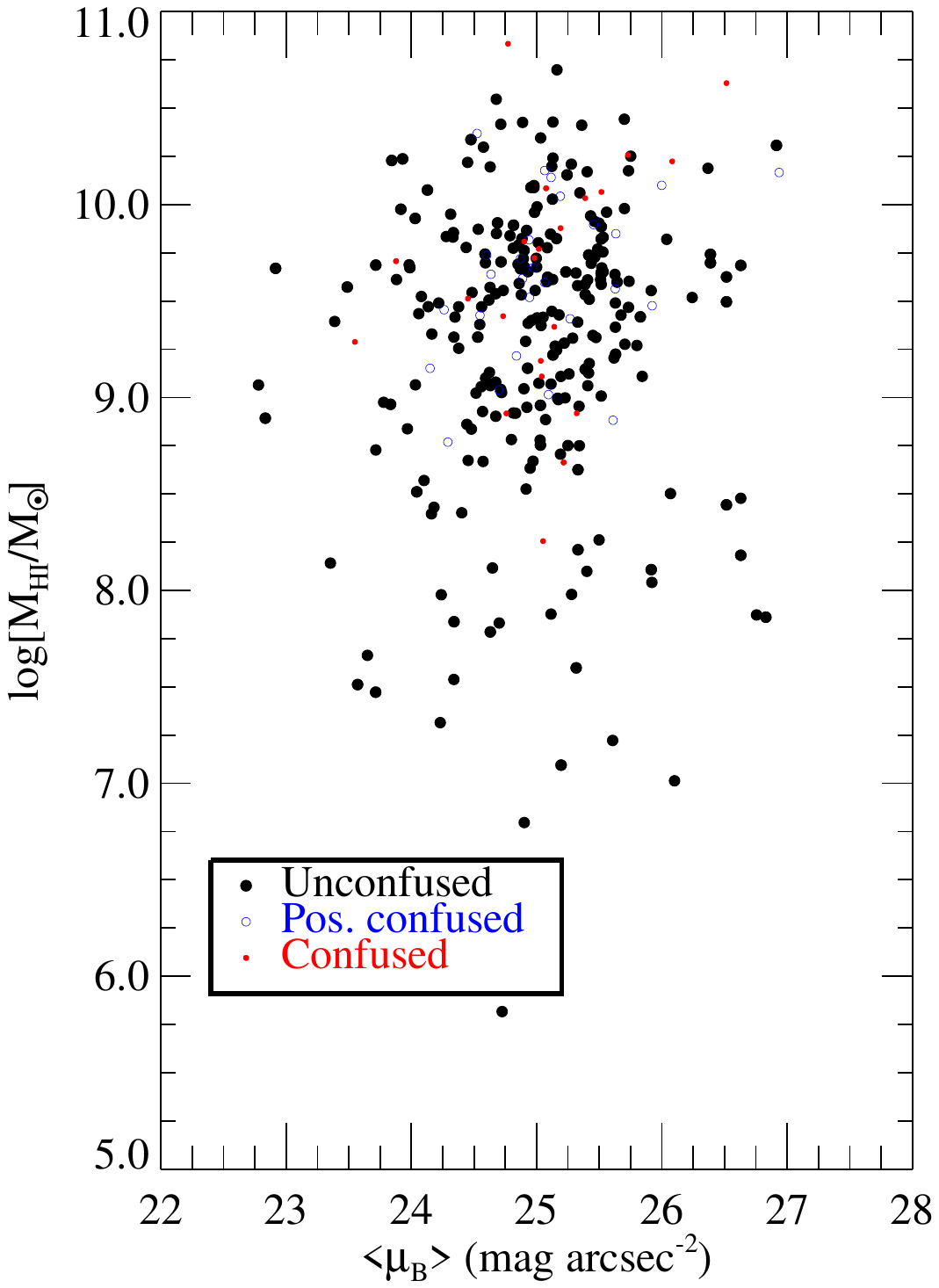}{0.3\textwidth}{(a)}
\fig{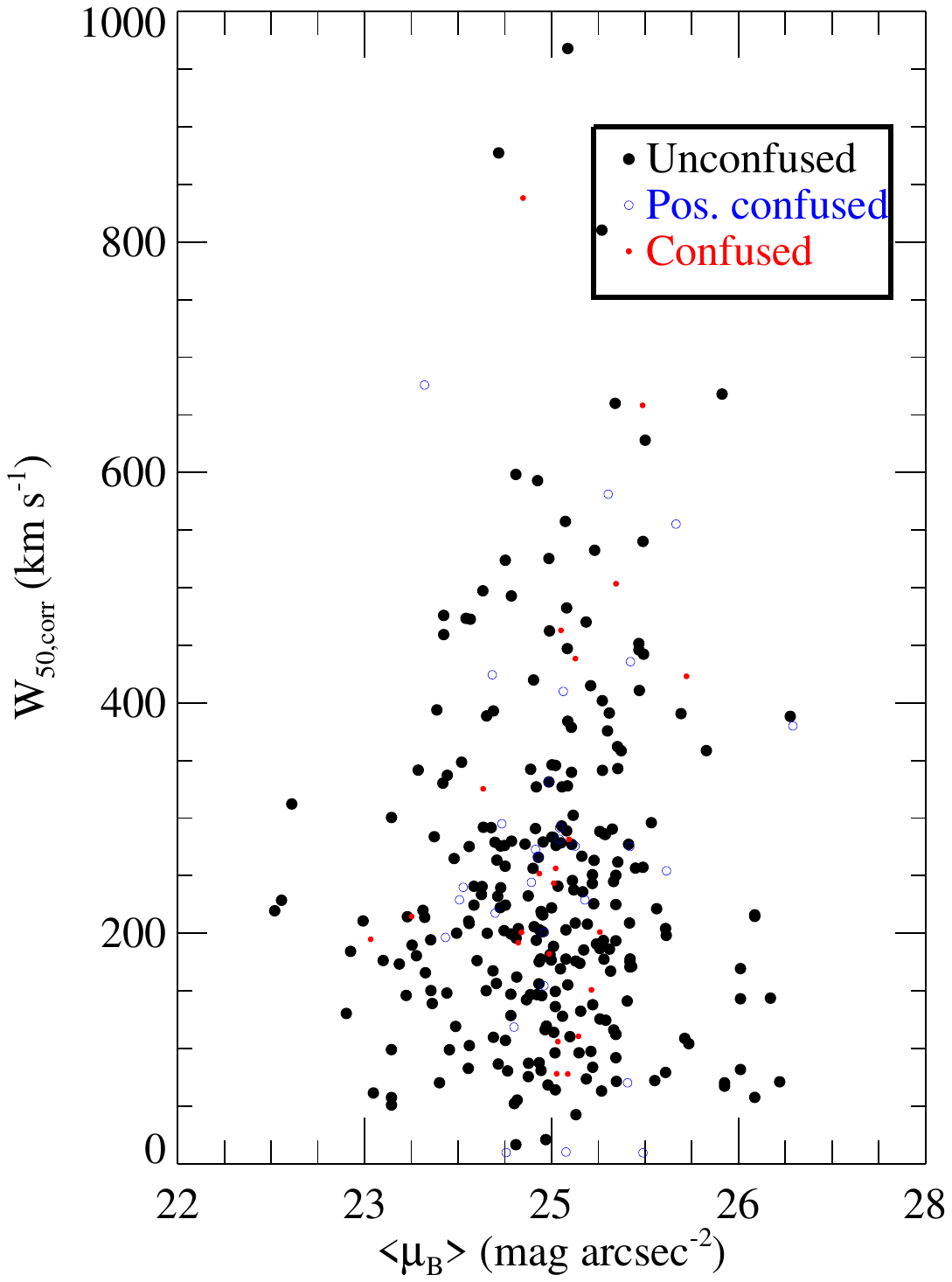}{0.3\textwidth}{(b)}
}
\caption{
Plotted as a function of mean blue surface brightness, \muB{} in \masq,
are (a) the total \HI{} mass, log(\MHI) in \Msun, and (b) the inclination-corrected \Wfiftycor{} 
\HI{} line width in \kms, for detected galaxies in our combined sample. 
As indicated in the legend, colors indicate the probability of a detection 
being confused by another galaxy within a telescope beam: black for uncontaminated, 
blue for possibly contaminated, and red for likely contaminated. 
\label{fig:allmuMHIW50} 
}
\end{figure*}

\begin{figure*} 
\centering
\gridline{
\fig{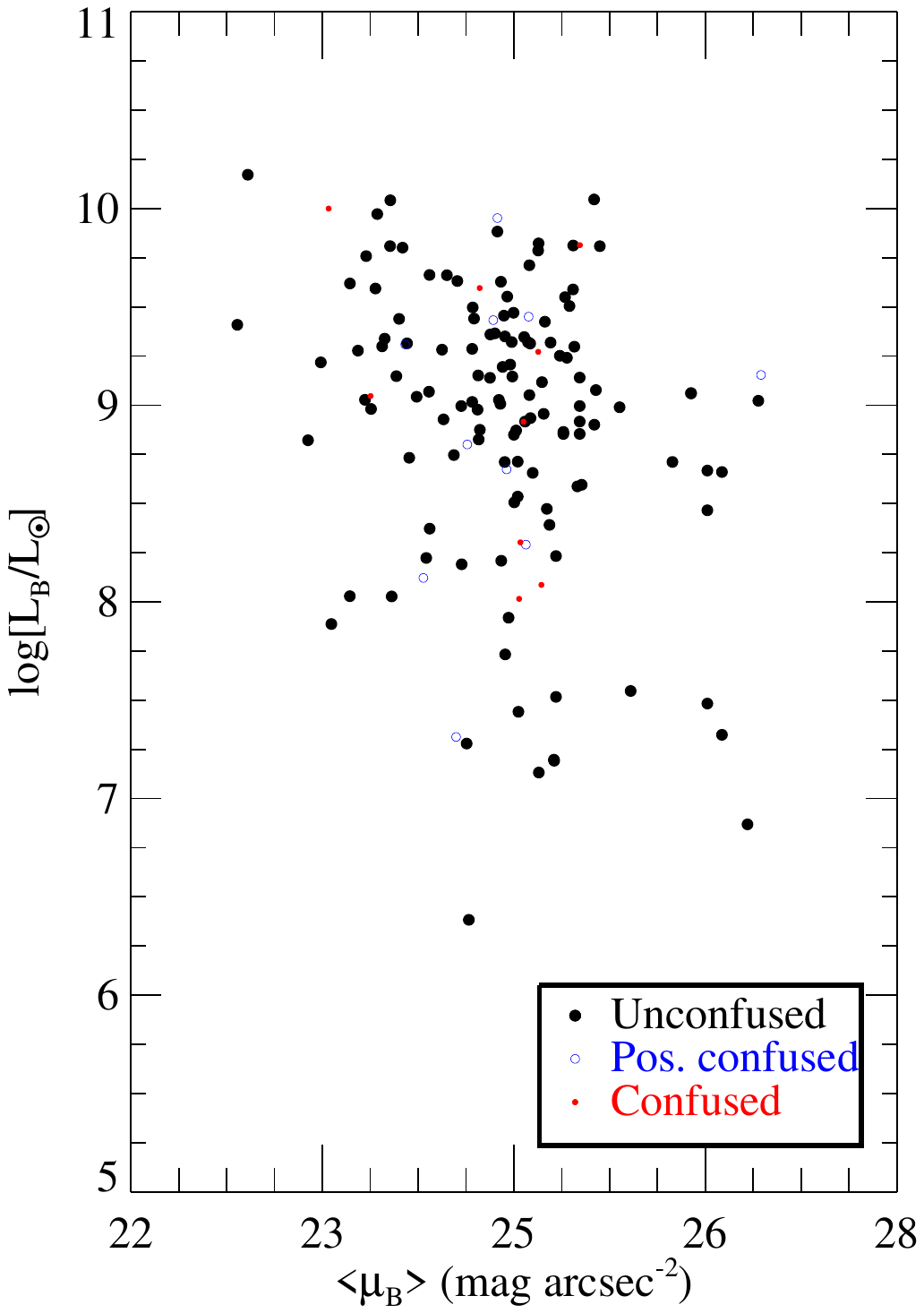}{0.3\textwidth}{(a)}
\fig{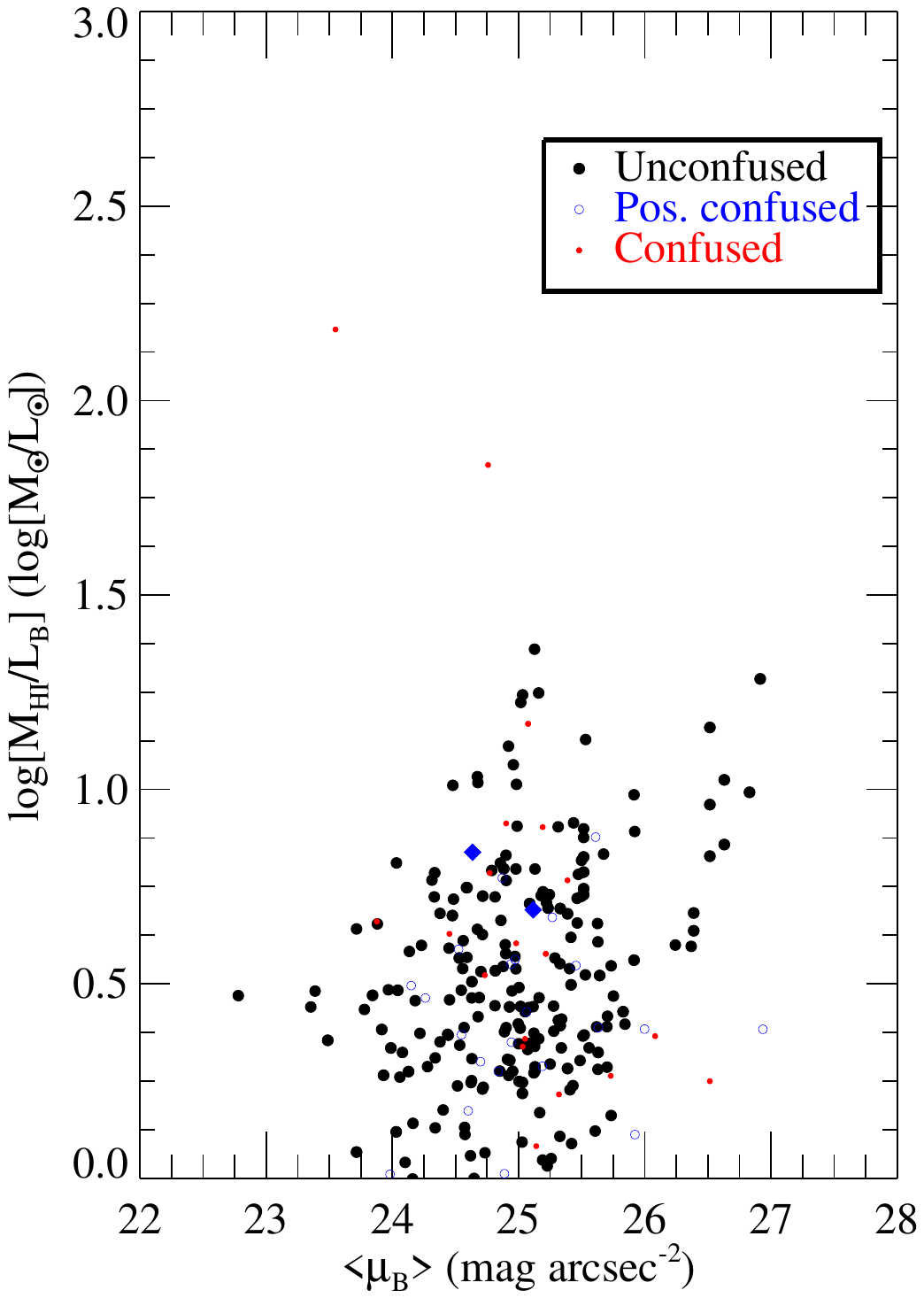}{0.3\textwidth}{(b)}
}
\caption{
Plotted as a function of mean blue surface brightness,  \muB{} (in \masq),
are (a) the blue luminosity, log(\LB) in \LsunB, and (b) the log(\MHI{}/\LB) \HI\ mass-to-light ratio ratio, 
in solar units, for galaxies in our combined sample. 
As indicated in the legend, colors indicate the probability of a detection 
being confused by another galaxy within a telescope beam: black for uncontaminated, 
blue for possibly contaminated, and red for likely contaminated. 
\label{fig:allMHILBLBmuB}
}
\end{figure*}

Looking at the data from this survey, there is no trend between average blue surface brightness \muB{}
and either \HI{} mass, \HI{} mass-to-luminosity ratio, blue luminosity \LB, or total mass as indicated by 
\Wfiftycor{} (see Figures ~\ref{fig:allmuMHIW50} and \ref{fig:allMHILBLBmuB}). That is, we do not see 
any trend toward galaxies becoming less (or more) massive, gas-rich or luminous as their surface brightness decreases. 
There is, however, a small trend for the upper limit to the \MHILB{} ratio to {\it increase} as surface brightness 
{\it decreases}, which is consistent with the idea that the average LSB galaxy is more gas rich than its HSB 
counterpart, for a given luminosity.

\subsection{Comparison with the HIPASS 1000 Brightest Galaxies Sample}  

\begin{figure*} 
\centering
\gridline{
\fig{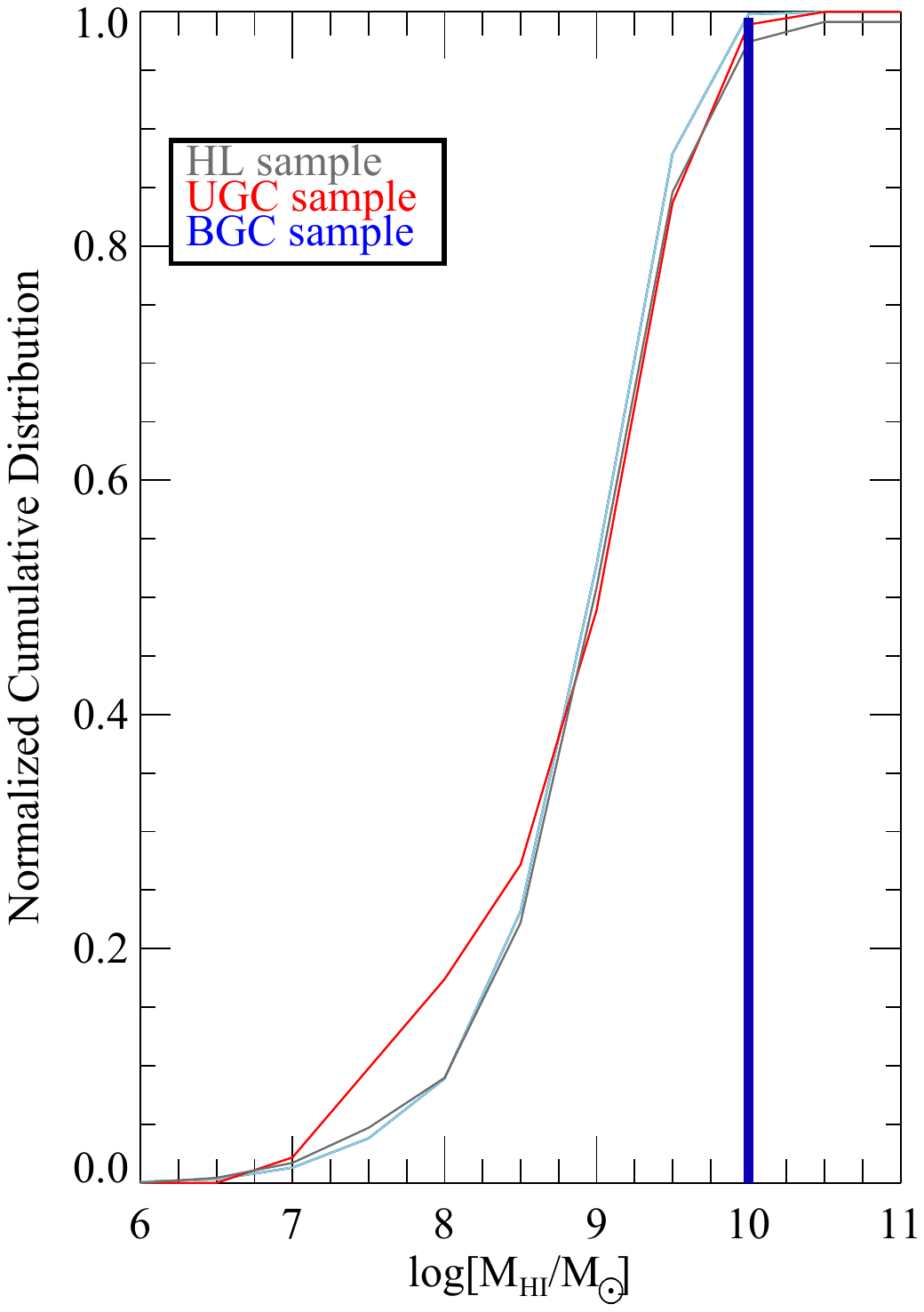}{0.3\textwidth}{(a)}
\fig{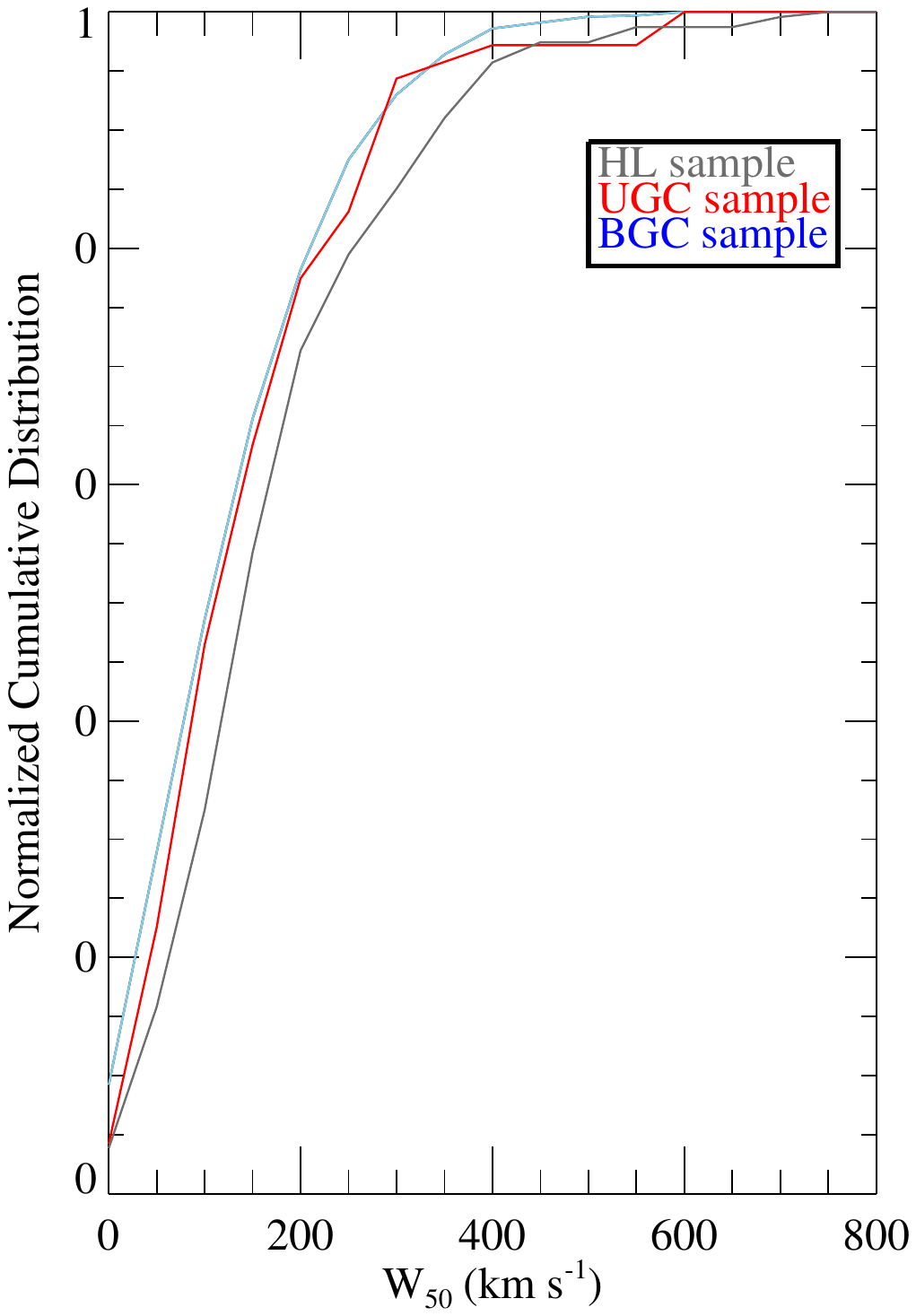}{0.3\textwidth}{(b)}
}

\caption{Comparison of cumulative distributions of our two galaxy samples 
(HyperLeda in black, UGC in red) and the HIPASS Bright Galaxies Catalog  (blue) with  
(a) total \HI{} mass, log(\MHI) in \Msun, and (b) total dynamical mass, as represented 
by the measured \Wfifty{} \HI\ line width, uncorrected for inclination.  The  blue vertical line in (a) shows the massive LSB galaxy limit.
}
\label{fig:cumul}
\end{figure*}

It is an interesting idea to compare the \HI{} mass distribution of the galaxies in this survey 
with those found in other \HI{} surveys. It is difficult to find an identical sample however, 
as this survey has intentionally avoided selecting galaxies which are likely of low \HI{} mass and
catalogs such as \citet{2014ApJ...793...40H} and \citet{2017MNRAS.467.1083L} do not contain enough
galaxies for a meaningful statistical comparison.
The best comparison sample is likely the HIPASS 1000 \HI{} brightest galaxies sample (BGC)
\citep{koribalski04}. Figure~\ref{fig:cumul} shows a comparison of the cumulative distribution 
of the \HI{} masses of the three galaxy samples (BGC, and our HyperLeda and UGC), and it is 
clear that the samples appear similar, but certainly not the same,
in spite of attempting to choose {\it a posteriori} a closely matched galaxy sample.
Confirming this, a two-sample Kolmogorov-Smirnov test on the probability for \HI{} masses 
coming from the same distribution indicates a \KSHIHL{} probability for the HyperLeda and BGC samples, 
and only a \KSHIUGC{} probability for the UGC and BGC samples. 

Also in terms of the total galaxy mass indicator \Wfifty{} the distributions of the three samples 
show little similarity (Figure~\ref{fig:cumul}), like for their \HI{} mass distributions. 
Here, the HIPASS bright galaxies distribution contains fewer lower total mass galaxies than either
the HyperLeda or UGC sample. Again the two sample Kolmogorov-Smirnov test bear this out, giving only a
\KSWfiftyHL{} chance the HyperLeda and BGC samples are from the same pool, and a \KSWfiftyUGC{} 
chance for the UGC and BGC samples. Please note that \Wfifty{}, and not the inclination-corrected line width
\Wfiftycor{} was used as the HIPASS bright galaxies catalog does not include inclination information.

The noted differences in total \HI{} mass between the three samples are unsurprising, as the 
UGC and HyperLeda samples consist of galaxies that were {\bf not} detected by previous \HI{} surveys. 
That is, our samples are not unbiased and certainly do not consist of \HI{} bright sources.

The differences in dynamical mass indicator \Wfifty{} between the three samples should also not be surprising:
for a given \HI{} mass a source with a smaller dynamic mass will tend to have a  narrower \HI{} line profile 
and thus a higher \HI{} peak flux density, and will therefore be more likely to be included in the 
\HI{} Bright Galaxy Catalog of the HIPASS blind \HI{} line survey. 


\section{MASSIVE LSB GALAXIES?} \label{sec:mlsbgs}

The impetus behind this project was to look for massive low surface brightness galaxies 
and to study the properties of these systems. The obvious question, then is -- how many 
massive LSB galaxies were detected in this search, and can 
any conclusions be drawn regarding this sample from our observations?

The initial question of how many massive LSB galaxies were found must be answered in two 
parts. First is the question of how many truly massive galaxies were found, and second is 
where or not these massive systems contain significant LSB disks. 

Looking at the mass of the galaxies, in terms of their total \HI{} gas mass we find that
of the \unconfdet{} uncontaminated galaxy detections in the survey, 
\MHIgeNineunconf{} (\MHIgeNineunconfperc{}) have \MHI$>$10$^9$ \Msun, while \MHIgeTenunconf{} 
({\MHIgeTenunconfperc}) of these are massive in \HI{}, with \MHI$>$10$^{10}$ \Msun.
We can also expand the definition of massive galaxy to include those with high dynamical masses,  
based on their inclination-corrected \HI{} line velocity widths \Wfiftycor{}. 
We find that of the uncontaminated detections, \WfiftygeFourUnconf{} have 
\Wfiftycor{} $\ge$ 500 \kms. 
Eliminating the overlapping galaxies in these two lists gives a total of \MLSBGposunconf{} massive galaxies.

Determining the central surface brightness of these galaxies from literature data is not currently possible, 
but we can check \muB.
Of the \MHIgeTenunconf{} uncontaminated galaxies with \MHI$>$10$^{10}$ \Msun, 
\MHIgeTenunconfmuLo{} also have \muB{}$\ge$ 24 \masq. 
Of the \WfiftygeFourUnconf{} uncontaminated galaxies with \Wfiftycor{} $\ge$ 400 \kms, 
\WfiftygeFourUnconfmuLo{} have \muB{}$\ge$ 24 \masq.
Finally, \MLSBGunconfmuLo{} galaxies have  \MHI$>$10$^{10}$ \Msun, \Wfiftycor{} $\ge$ 500 \kms, and
and \muB{}$\ge$ 24 \masq. 
All of these are clearly worthy of follow-up observations in the optical and radio 
to better understand their properties.

\section{CONCLUSIONS} \label{sec:conclusion} 

While the intent of this survey was to identify massive low surface brightness galaxies, it
is very instructive to look at the overall mass distribution of the sample. 
The average \HI{} mass of the (uncontaminated) sample is \MHIall{} \Msun, and their average 
velocity width $\langle$\Wfiftycor$\rangle$ is \Wfiftyunconf{} \kms{}, 
reinforcing the fact that low surface brightness galaxies have the same mass distribution as their 
higher surface brightness counterparts and are {\bf not} preferentially dwarf systems.

Of more interest will be follow-up \HI{} radio synthesis and optical surface photometry observations 
of these galaxies. 
\cite{mapelli08} have shown that at least some of the massive LSB galaxies are formed through 
the interaction and merger of smaller galaxy systems. This sample will provide an excellent 
test case to determine if there is indeed a mass beyond which all LSB galaxies are formed through
interactions and mergers or if, like their higher surface brightness counterparts, these
systems follow a variety of paths to reach their current state.


\begin{acknowledgements} 
This material is based upon work supported by the Green Bank Observatory which is a major facility 
funded by the National Science Foundation operated by Associated Universities, Inc. 
This research has made use of the NASA/IPAC Extragalactic Database (NED), which is funded 
by the National Aeronautics and Space Administration and operated by the California Institute 
of Technology; the HyperLeda database (http://leda.univ-lyon1.fr); SDSS-II data (http://www.sdss3.org) 
which has been provided by the Alfred P. Sloan Foundation, the Participating Institutions, 
the National Science Foundation, and the U.S. Department of Energy Office of Science; 
and POSS-II data which were produced at the Space Telescope Science Institute under 
U.S. Government grant NAG W-2166 based on photographic data obtained using the 
Oschin Schmidt Telescope on Palomar Mountain and the UK Schmidt Telescope. 
The Arecibo Observatory is operated by SRI International under a cooperative agreement 
with the National Science Foundation (AST-1100968), and in alliance with Ana G. 
M\'endez-Universidad Metropolitana, and the Universities Space Research Association. 
The \nan{} Radio Observatory is operated by the Paris Observatory, 
associated with the French Centre National de la Recherche Scientifique.

\end{acknowledgements}

\bibliography{oneil_combilsb.bib}
\bibliographystyle{aasjournal}


\appendix
\section{NOTES ON INDIVIDUAL GALAXIES}\label{Sec:IndGals}

\subsection{HyperLeda Galaxy Sample} \label{subsec:hyperleda_ind_Gals} 

\subsubsection{CGMW 1-0409}  
This galaxy was not in the list of planned targets. 
It was however detected at the GBT in an OFF source (blank sky) spectrum 
used to calibrate another galaxy. Although our detection matches the Parkes' HIPASS profile 
\citep{2004MNRAS.350.1195M} in \VHI\ (2805 \kms) and \Wfifty\ (131 \kms), our \HI{} flux (8.6$\pm$0.2 Jy \kms)
is much lower than the HIPASS value of 16.1 Jy \kms, indicating the source may be quite extended
given the difference in the telescopes' HPBW (GBT $8\farcm7$, Parkes $15\farcm5$ although the Parkes positional 
accuracy is typically better than $7\farcm5$; \citet{2004MNRAS.350.1210Z}). 

\subsubsection{ESO 338-020}  
It is likely that some of the detected flux for ESO 338-020 is from the nearby galaxy 
2MASX J19412558-3820561, whose optical velocity is only 12 \kms{} 
higher and is located $8\farcm3$ away from our target, ESO 338-020, on the edge of the \nan{} HPBW. 

\subsubsection{ESO 368-004} 
ESO 368-004 has no known galaxies within 15\arcmin{} and 9,000 \kms{}, yet the \nan{} and GBT flux measurements 
for this galaxy differ by 1.6 Jy \kms\ (12\%). Follow-up GBT observations confirmed the original GBT values.
Yet, as all three spectra have high signal-to-noise, the average of the three results was reported in Table~\ref{tab:avgresultsdets}.
Please note, too that the HIPASS detection of this object has even higher flux than our three measurements \citep{2004MNRAS.350.1195M}:
$flux_{NAN} = 12.91$, $flux_{GBT1,GBT2} = 14.51, 13.45$, $flux_{HIPASS} = 15.08$ \Jykms.



\subsubsection{ESO 377-045} 
Three marginal \HI{} detections of ESO 377-045 with peak \SNR\ values of 3.2 to 4.3 were reported by 
\citet{1996AJ....111.1098M} based on Green Bank 42m telescope observations with a 21$'$ HPBW, 
at 1323, 2057, and 3087 \kms; the latter has \Wtwenty{} = 167 \kms{} and \FHI{} = 2.5 Jy \kms, 
similar to our detection with \VHI{} = 3125 \kms, \Wtwenty{} = 171 \kms, and \FHI{} = 2.4 Jy \kms. 

\subsubsection{ESO 398-011} 
Although the galaxy ESO 398-010 is only 12 \kms{} away from ESO 398-011. The galaxies' E-W separation is 8$'$, 
or five times the E-W \nan{} beam radius. It is therefore not likely that the measured spectrum contains 
some contribution from ESO 398-010.

\subsubsection{ESO 463-005} 
The reported optical velocity for this galaxy \citep{jones09} is 3031$\pm$45 \kms{}, significantly
(5$\sigma$) different from the 2809$\pm$3 \kms{} \HI{} measurement originally made with the \nan{} telescope.
Follow-up GBT observations, though, showed our \nan{} spectrum to be likely contaminated by a nearby galaxy, 
NGC 6925, which was probably caught in the larger \nan{} beam. The GBT observations give
an \HI{} velocity of 3014 \kms{}, much closer to the \citep{jones09} result.
Only the GBT results were used for the final values (Table~\ref{tab:avgresultsdets}).

\subsubsection{ESO 482-024} 
It is highly likely that our \nan{} \HI{} spectrum is contaminated by the galaxy NGC 1403 
which is 56 \kms{} and $12\farcm0$ away from ESO 482-024, on the edge of the \nan{} beam. 
NGC 1403 was detected at Parkes in HIPASS \citep{2004MNRAS.350.1195M}.

\subsubsection{ESO 491-002} 
\begin{figure}[ht] 
\centering\includegraphics[width = 4.5in]{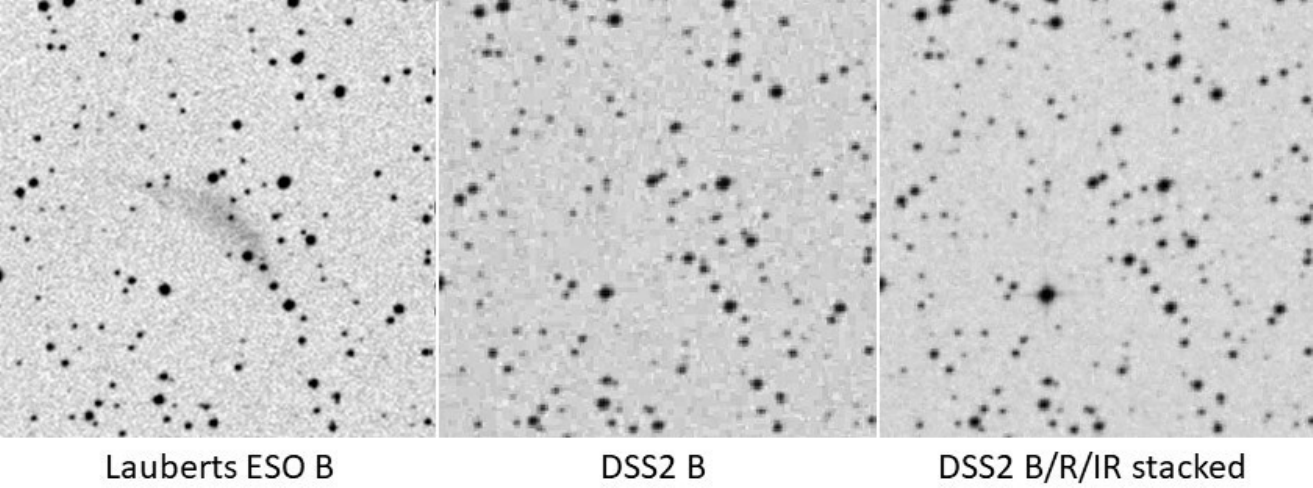}
\caption{Optical images of ESO 491-002. A $B$-band image (left) from \citet{lauberts89} 
taken at the ESO 1m Schmidt telescope shows an LSB object, which could be a distorted galaxy, 
but no galaxy is discernible on the DSS2 $B$, $R$ or $IR$ images \citep{2004AJ....128..502A} 
taken with the 1.2m Palomar Schmidt telescope. Images are $4\farcm3$$\times$$4\farcm3$ in size.  \label{fig:ESO491_002}}
\end{figure}

No galaxy is discernible on the DSS2 images, but a $B$-band image from 
\citet[][as displayed on NED]{lauberts89} taken at the ESO 1m Schmidt 
telescope shows an LSB object, which could be a distorted galaxy (Figure~\ref{fig:ESO491_002}).
We did not detect it in \HI.

\subsubsection{ESO 491-003} 
As the original \nan{} detection of this source had a nearby (in frequency) RFI source, 
follow-up observations were taken. Only the more recent GBT results are reported in 
Table~\ref{tab:avgresultsdets}.


\subsubsection{ESO 492-001} 
The GBT profile has a 84 \kms\ (40\%, or 6$\sigma$) larger \Wfifty\ line width than our \nan\ profile, 
and a 22\% (3.6$\sigma$) higher integrated flux. As there are no neighbors within either telescope beam, 
it is likely the higher signal-to-noise GBT measurements are more reliable than the \nan\ values. 
Subsequent GBT observations agreed with the initial GBT values, and we therefore used only the GBT 
values used for Table~\ref{tab:avgresultsdets}.

\subsubsection{ESO 501-029} 
It appears possible that nearby galaxy UGCA 212 has contaminated our GBT \HI{ }detection of our target, ESO 501-029. 
UGCA 212 is 28 \kms{} and $11\farcm2$ (2.6 times the GBT beam radius) away from our target and its published 
Parkes (HPBW $15\farcm5$) HIPASS profile \citep{2004MNRAS.350.1195M} shows \VHI\ = 1042 \kms, \Wfifty\ = 63 \kms, 
and \FHI\ = 29.4 Jy \kms, or 7.5 times our flux measured towards ESO 501-029. 


\subsubsection{ESO 540-030} %
A weak \HI{} detection of ESO 540-030 was reported by \citep{bouchard05} in Parkes single-dish spectra and 
Australia Telescope Compact Array (ATCA) images at \VHI{} = 233 \kms, with \Wfifty{} = 26 \kms{} and \FHI{} = 0.33 \Jykms.
This low velocity is consistent with the distance of 3.2 Mpc estimated from Surface Brightness Fluctuations
by \citet{jerjen98}. Our \nan\ data show a 7 mJy, \SNR\ = 3.3 peak near the Parkes velocity, which is too weak 
for a confirmation. Follow-up observations with the GBT indicate no signal, with an r.m.s. of 1.8 mJy, 
compared to the 6.5 times higher $\sim$12mJy peak reported by \citep{bouchard05}. 
The GBT data confirm that the \citep{bouchard05} detection was spurious. Similarly, we also consider the 
low signal-to-noise (\SNR = 3.3) signal seen by \nan{} at 8093 \kms{} to be spurious, as it was also not seen by the GBT. 



\subsubsection{IC 3852} 
There are no known galaxies within 15\arcmin{} and 8,000 \kms{} of IC 3852, yet the difference 
between the fluxes measured at Arecibo and the two GBT measurements is large (2.7 and 3.2/3.4 Jy \kms, respectively) 
as is the difference in velocity widths (\Wfifty{} 203 and 205/222 \kms, respectively). 
Published Arecibo data (\citealt{2005ApJS..160..149S,2018ApJ...861...49H}) show 
\VHI{} = 4375 \kms, \Wfifty{} = 198 \kms{} and \FHI{} = 3.75 \Jykms.
With no clear reason for the differences, the average of our observations are listed in Table~\ref{tab:avgresultsdets}.


\subsubsection{PGC 2582} 
The velocity of our GBT detection for PGC 2582 at 15,050 \kms\ does not match those of the two galaxies within the 
telescope beam with measured optical velocities, 2MASXJ00430067-0913463 at 29,310 \kms\ \citep{christlein03} and 
WISEA J004314.24-091247.0 at 22,892 \kms\ \citep{moretti17}. 
It is likely our detection of \HI{} is actually from the Abell 85 cluster in the region which has a mean velocity of
16,500 \kms\ and a velocity dispersion of 1100 \kms\ \citep{oegerle01}. 
Unfortunately, radio frequency interference prevented the GBT observations from reliably detecting any \HI{} 
at or near the optical velocities measured for the two galaxies within the telescope beam.

\subsubsection{PGC 3843}\label{subsec:P3843}
\begin{figure}[ht] 
\centering
\includegraphics[width = 2.8in]{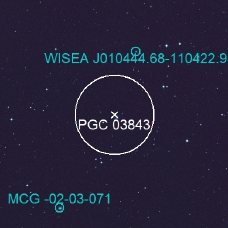}
\caption{DSS multi-color (IR, red, and blue plates) image of PGC 3843 (center, white cross). 
The large white circle shows the $8\farcm7$ GBT beam, and the cyan circles denote the 
two galaxies which have likely recently interacted with PGC 3843, 
WISEA J010444.68-110422.9 and MCG -02-03-071. Image is $25\farcm$ across. 
\label{fig:P3842_neighbors}}
\end{figure}

The \HI{} profile measured at the GBT for PGC 3843 is quite broad (\Wfifty\ = 403  \kms).
No previous redshift has been reported for PGC 3843, and there are no other galaxies within the GBT beam with 
velocities within $\pm$500 \kms{} of 14,367 \kms, the central velocity of PGC 3843's \HI{} detection. 
However, two galaxies, WISEA J010444.68-110422.9 and MCG -02-03-071, lie on either side of PGC 3843, 
7\arcmin{} and 12\arcmin{} away, respectively, and at 14,243 and 14,084 \kms{} (Figure~\ref{fig:P3842_neighbors}). 
It is possible that PGC 3483 has had a recent encounter with (one of) its neighbors, resulting 
in its disturbed optical morphology and its exceptionally large \HI{} line width.  Follow-up observations
will be required to determine if this is the case.

\subsubsection{PGC 7225} 
\citet{1998ApJ...496...39Z} gives an optical velocity of 5071$\pm$80 \kms{}, which is 123 \kms{} lower than
that found by us in \HI{}, but its high uncertainty, combined with our independent detections at \nan{} 
and GBT, makes it likely our measurement is a more accurate value for the average velocity of the galaxy.

\subsubsection{PGC 16370} 
Using the Westerbork Synthesis Radio Telescope (WRST), \citet{2016MNRAS.460..923R} also measured PGC 16370's \HI{} properties. 
Their results for \VHI{} and \Wfifty{} match those we found, but their integrated flux value is only 60\% of that 
found by our original low signal-to-noise \nan{} detection. However, our much higher signal-to-noise GBT 
follow-up observations confirm the WRST data and are consequently listed in Table~\ref{tab:avgresultsdets}. 

\subsubsection{PGC 17124} 
\begin{figure*}[ht] 
\begin{center}\includegraphics[height = 3in]{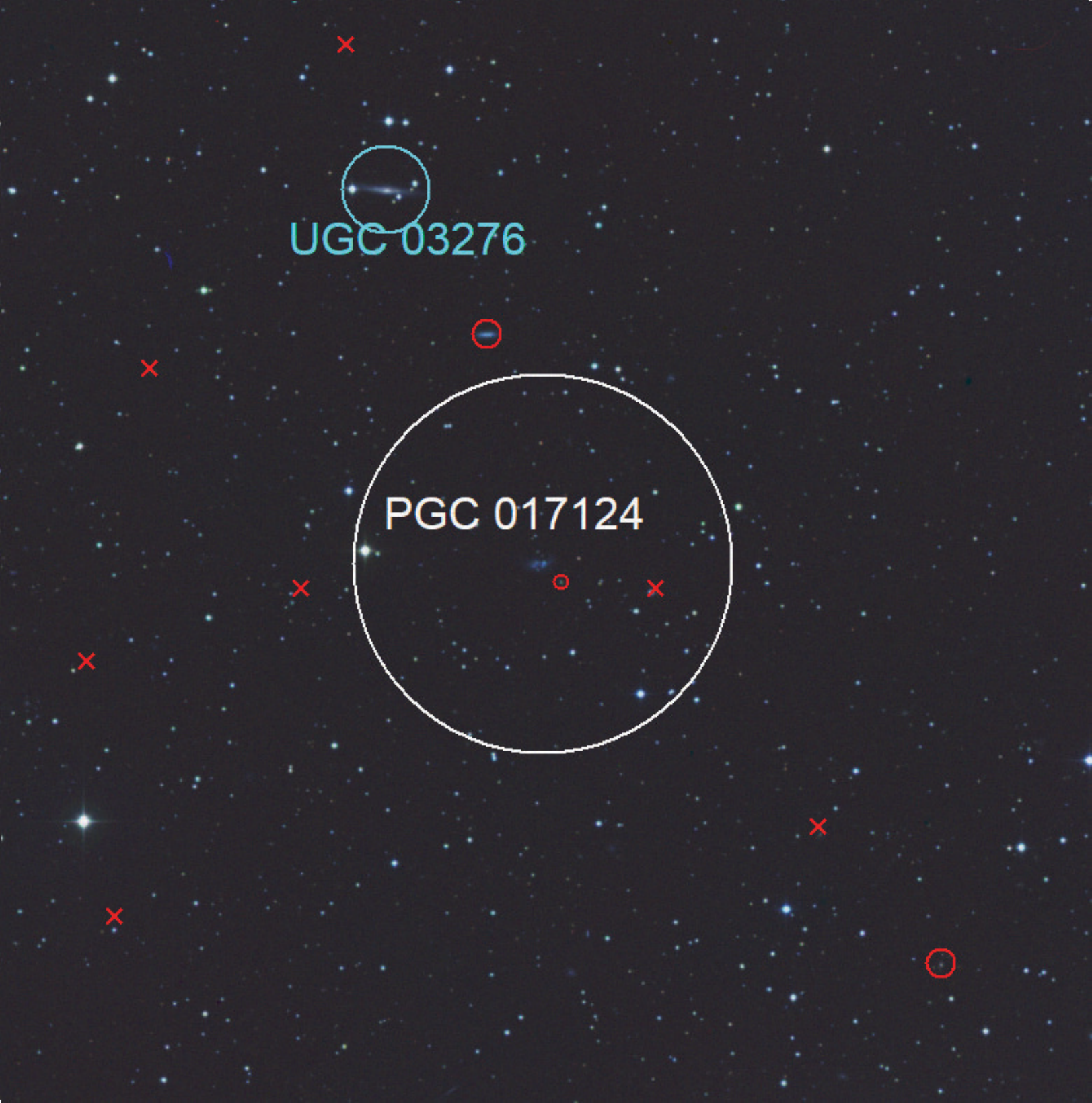}
\includegraphics[height = 3in]{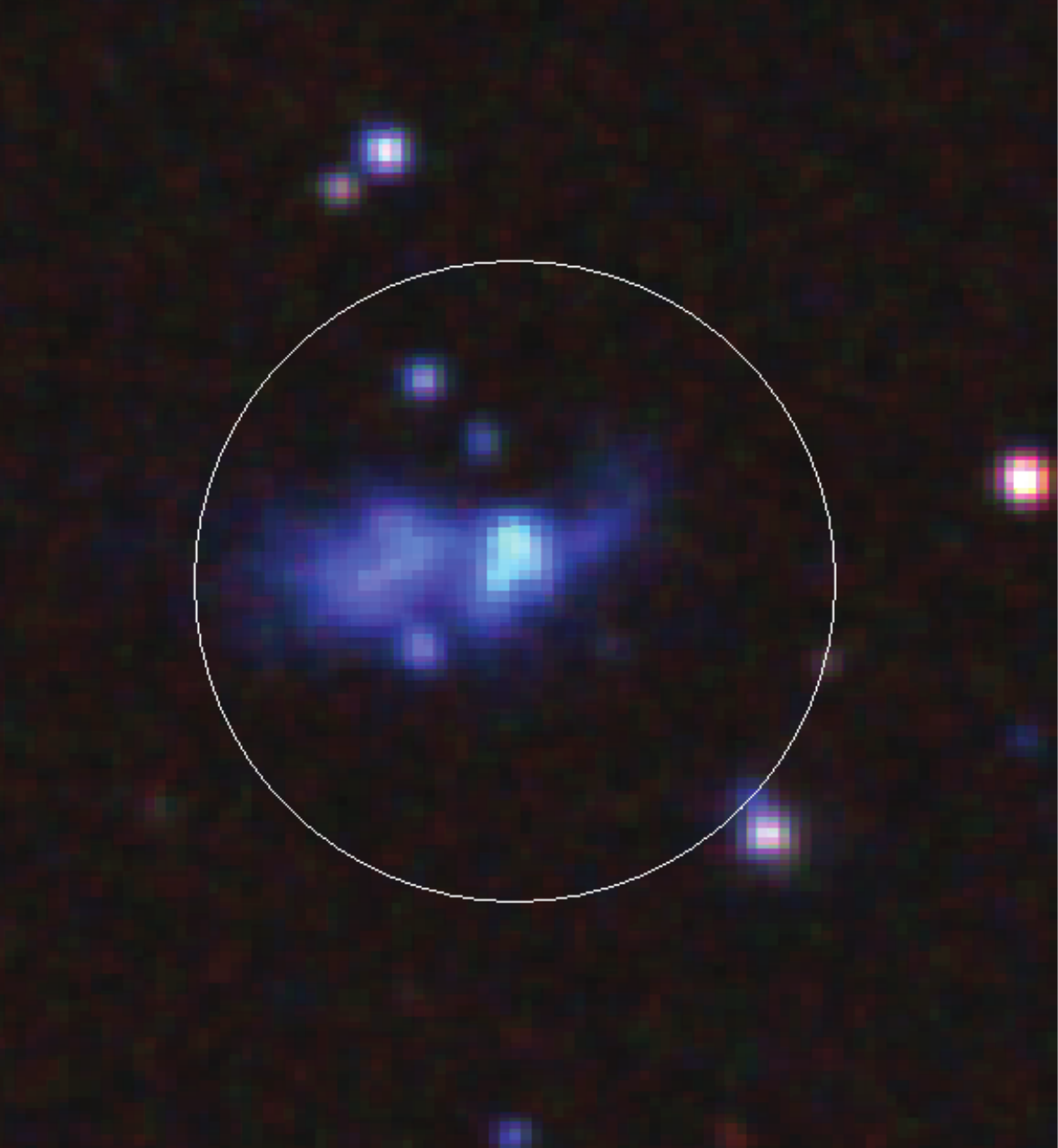}\end{center}
\caption{DSS multi-color (IR, red, and blue plates) images of PGC 17124 (center, 30\arcsec{} white circle). 
(Left:) The large white circle shows the $8\farcm7$ GBT beam, and UGC 3276 is denoted 
by the (60\arcsec{}) cyan circle. The two galaxies with known velocities more than 2,000 \kms{} from that 
of PGC 17124 are denoted by red circles, while the three previously identified galaxies 
without known velocities are denoted by red crosses. Image is $25\farcm$ across.
(Right:) Enlargement, showing the disturbed morphology of PGC 17124. \label{fig:P17124_neighbors}}
\end{figure*}

A potential source of confusion is the galaxy UGC 3276, which lies $9\farcm1$ from PGC 17124, 
i.e. inside the NRT HPBW but outside the GBT beam (see Figure~\ref{fig:P17124_neighbors}).
Therefore, our GBT profile of PGC 17124 is unlikely to be contaminated by its neighbor
and we used its line parameters in Table~\ref{tab:avgresultsdets}.
Measurements centered on UGC 3276 with the National Radio Astronomy Observatory's 300-ft telescope 
\citet{2005ApJS..160..149S} and Jodrell Bank's 76m Lovell telescope \citet{lang03}
give \HI{} profile parameters consistent with our observations: \VHI\ = 2490 \kms, \Wfifty\ = 270 \kms\ and \FHI\ = 15.3 Jy \kms. 
The beam radii of the two telescopes are more than two times smaller than the separation between the two galaxies,
so it is unlikely that these profiles are contaminated by our target galaxy, PGC 17124.
As can also be seen in Figure~\ref{fig:P17124_neighbors}, the morphology of PGC 17124 is clearly disturbed. 
It is quite possible, then, that UGC 3276 and PGC 17124 have had at least one encounter in the recent past.

\subsubsection{PGC 21133} 
There is a 90 \kms{} difference between our \nan{} and GBT \Wfifty{} line width, but the \Wtwenty{} values 
are comparable. However, this difference is not really significant given the estimated uncertainty 
in the \Wfifty{} value of the lower signal-to-noise \nan\ data. Both observations are averaged in 
Table~\ref{tab:avgresultsdets}.

\subsubsection{PGC 21529} 
At $3\farcm4$ from PGC 21529 (\Vopt\ = 4059$\pm$2 \kms{})
lies KUG 0737+323 with a 138 \kms{} lower \Vopt\ of 3921$\pm$3 \kms{} \citep{2015ApJS..219...12A} 
Both our \nan{} and GBT spectra likely include both PGC 21529 and KUG 0737+323, whereas due to its smaller 
$3\farcm6$ beam size the Arecibo data appear to be unaffected by KUG 0737+323. As a result, 
only the Arecibo result is listed in Table~\ref{tab:avgresultsdets}.


\subsubsection{PGC 21907}\label{subsec:P21907}  
\begin{figure}[ht] 
\begin{center}\includegraphics[width = 2.8in]{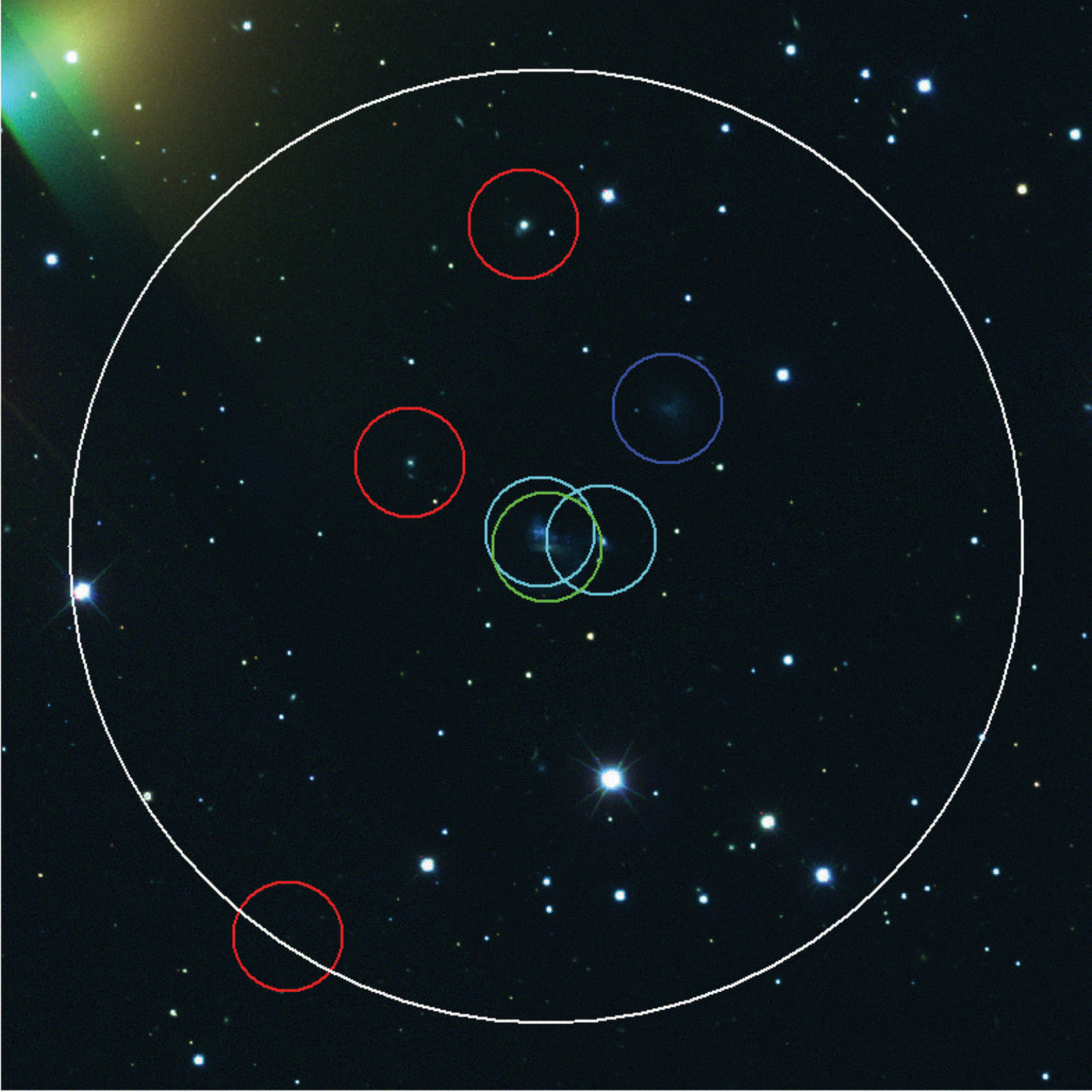}\end{center}
\caption{SDSS DR12 false color ($i$, $r$, and $g$ filter) image of the galaxies with known 
redshifts lying within the GBT beam centered on PGC 21907 (cyan circle). The large white 
circle shows the $8\farcm7$ GBT beam. The blue circle denotes the nearest galaxy, 
SDSS J074933.51+394424.3, the two cyan circles denote galaxies KUG 0746+398A (left) and 
SDSS J074939.55+394316.7 (right), the green circle PGC 21907 (aka KUG 0746+398B), and 
the red circles denote the three WISE galaxies which lie outside our redshift detection range.  
Image is $9\farcm$ across.
\label{fig:P21907_inbeam}}
\end{figure}

The intended target of this survey was PGC 21907. However observations taken after our survey, 
including optical velocity measurements, have shown it to be two distinct galaxies - KUG 0746+398A 
(the foreground LSB galaxy) and KUG 0746+398B (a.k.a. PGC 21907 - a background spiral galaxy). 
Including these two objects, there are a total of eight galaxies with known redshifts within 
the GBT beam when it was pointed toward PGC 21907, all of which have published redshifts 
(Table~\ref{tab:P21907_neighbors} \& Figure~\ref{fig:P21907_inbeam}). Four of these have 
redshifts within the frequency range of our GBT observations, whereas the others lie far outside it. 
Of these four, our GBT observations detected \HI{} gas at the redshift of SDSS J074933.51+394424.3 only. 
A possible detection was seen at 12,420 \kms{} (KUG 0746+398A), but there were significant baseline 
issues, making an the detection uncertain. No \HI{} emission was detected at the velocities 
of KUG 0746+398A, KUG 0746+398B (a.k.a. PGC 21907), or SDSS J074939.55+394316.7.


\subsubsection{PGC 23328}\label{subsec:P23328} 
\begin{figure}[ht] 
\begin{center}\includegraphics[width = 2.8in]{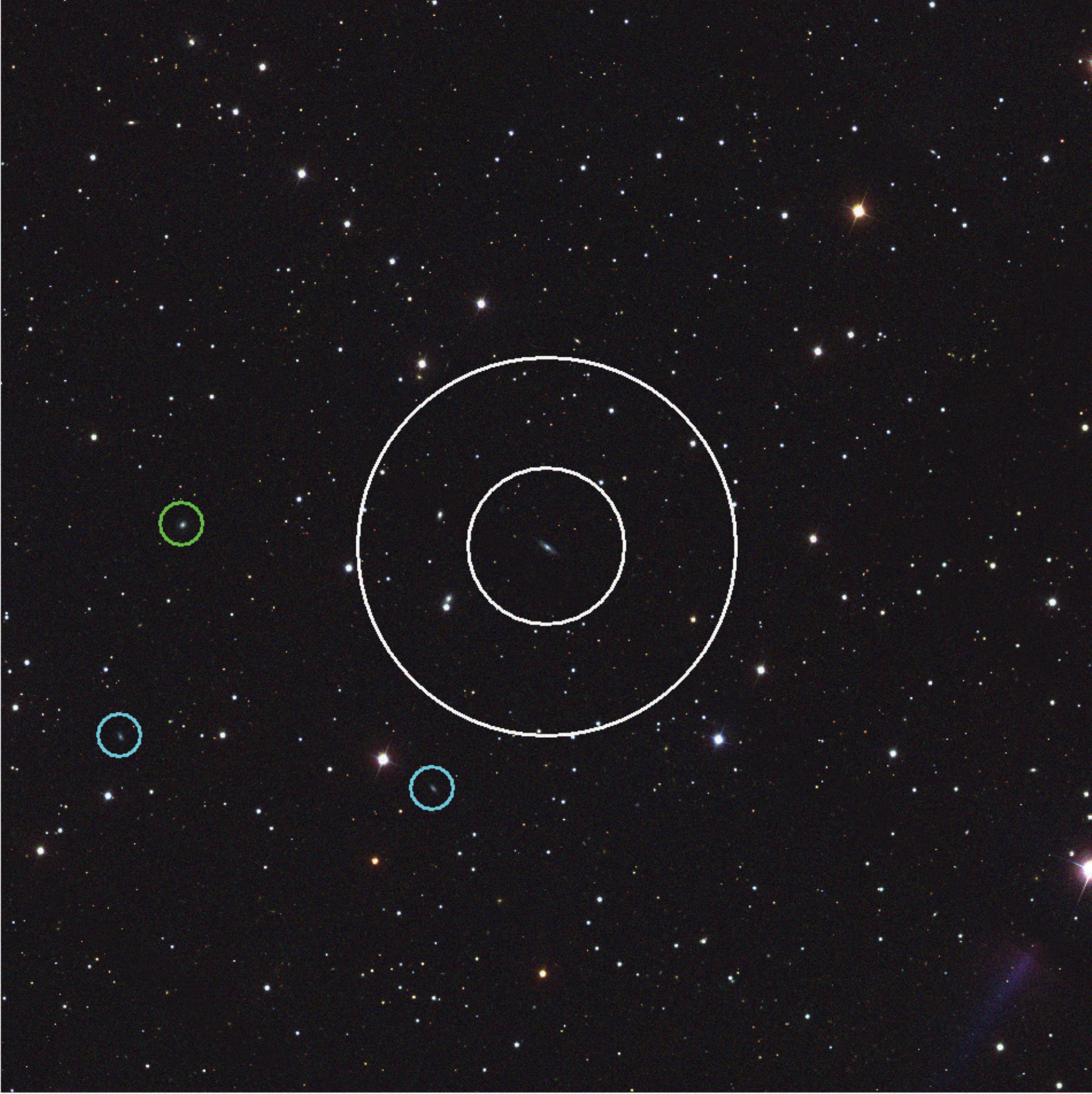}\end{center}
\caption{SDSS DR12 false color ($i$, $r$, and $g$ filter) image of the PGC 23328 galaxy group. 
The white circles show the $8\farcm7$ GBT beam and the $3\farcm6$ Arecibo beam centered on PGC 2338. 
The other two known members of the group, WISEA J081955.30+252733.0 \& WISEA J081923.48+252621.4, 
are denoted by cyan circles (listed from left to right in the image). 
The fourth potential group member, KUG 0816+256B, is denoted by the green circle.  
Image is $25\farcm$ across.
\label{fig:P23328_group}}
\end{figure}

PGC 23328 is listed as member of a small group of galaxies by both \citet{2017AA...602A.100T} and 
\citet{2012AA...540A.106T} (Table~\ref{tab:P23328_group}). This group includes, at minimum, 
PGC 23328, WISEA J081923.48+252621.4, and WISEA J081955.30+252733.0, but it is likely 
that KUG 0816+256B is also part of this group. However, as none of the galaxies in this group
were included in the telescope beams, it is likely the measured \HI{} flux belongs only to 
PGC 23328 (Figure~\ref{fig:P23328_group}).

\subsubsection{PGC 23879}
PGC 23879 has one companion, LEDA 3097429 which lies $11\farcm5$ but only 27 \kms{} away. 
There are two additional galaxies, ASK 151063.0 and ASK 151084.0, which lie 
$13\farcm1$/1675 \kms{} and $8\farcm$6/1721 \kms{} away, respectively. As the \HI{} spectrum 
of PGC 23879 was taken with the GBT, which has a beam size of $9\farcm2$ at the frequencies 
of interest, it is possible that it contains some \HI{} gas from LEDA 3097429. 
It is highly unlikely that gas was detected from the other galaxy pair, but it is worth noting 
them nonetheless, as the four galaxies could form a loose group.

 
\subsubsection{PGC 26708} 
This galaxy is part of the PGC 26708/NGC 2883 pair of galaxies, 
making it difficult to know which fraction of the measured \HI{} flux can be attributed to 
our target, PGC 26708.

\subsubsection{PGC 26936} 
The galaxy SDSS J092844.48+351641.3 is $11\farcm9$ and 48 \kms{} away from our target, PGC 26936. 
As the larger \nan{} beam does not show either a larger \Wfifty{} or line flux than we observed at Arecibo, 
it is highly unlikely that our spectra of our target are contaminated.

\subsubsection{PGC 27485} 
We observed PGC 27485 at \nan{}, Arecibo, and the GBT.
Comparing the measured \HI{} profiles it appears that the smaller Arecibo beam did not observe 
all of the extended \HI{} of this N-S oriented galaxy (the line flux measured at Arecibo is 29\% lower than at \nan;
$flux_{AO} = 4.29\pm 0.07$; $flux_{NAN} = 5.98\pm 0.26$; $flux_{GBT} = 6.29\pm 0.12$ \Jykms). 
However without a full map of the galaxy's gas distribution we cannot 
know this for certain and listed the average of both values in Table~\ref{tab:avgresultsdets}.

\subsubsection{PGC 27849} 
Although its \VHI{} values found with the three telescopes are highly consistent, the values measured 
for \Wfifty{}, \Wtwenty{}, and the \HI{} flux vary significantly. The GBT values are the largest, 
with log(\MHI{}/\Msun{}) = 9.88 and 10.05 (\Wtwenty = 393/332 and \Wfifty = 357 \& 312) while the Arecibo and \nan{} observations are similar, 
with log(\MHI{}/\Msun{}) = 9.60 and 9.73 (\Wtwenty = 247 \& 236 and \Wfifty = 221 \& 197), respectively. 
There are, however, no known galaxies within 14\arcmin{} and 2,000 \kms{} 
from our target PGC 27849, nor are there any clear, unidentified neighbors. 
It is therefore unclear what the origins are of the additional flux measured by the GBT.

Results from all four sets of observations are shown in Table~\ref{tab:indresultsdets} and
also in Figure~\ref{fig:HLdets}, with the older GBT observation shown in light blue. 
The results in Table~\ref{tab:avgresultsdets} are averaged from all four sets of observations.

      

\subsubsection{PGC 28799}\label{subsec:P28799} 
The GBT data show a clear detection of \HI{} emission at 4526 \kms{} with a second (marginal) detection at 
$\sim$6130 \kms{}. These velocities do not, however, match the 7313 \kms{} found for the \HI{} source HIPASS J0958-38 
\citep{2004MNRAS.350.1195M,2005MNRAS.361...34D}. Another galaxy, WISEA J095813.16-380926.6, lies $4\farcm5$ from our target PGC 28799 and
has an optical velocity of 7234$\pm$45 \kms{} \citep{2009MNRAS.399..683J} 
Neither of these redshifts match those found in our survey, and no \HI{} emission was found at either of these velocities.

\subsubsection{PGC 29681} 
It is possible, though not likely, that our \nan\ spectrum of our target PGC 29681 is contaminated by two nearby galaxies,
UGC 5485 and SDSS J101131.31+650524.7, which are within $8\farcm5$ and 58 \kms{} 
(SDSS optical velocities) from our target. Of the three, UGC 5485 is far the brightest, by about 2 magnitudes 
in $B$. It was observed in \HI{} at Effelsberg and \nan{} \citep{huchtmeier97,vandriel00}, and its separation 
from our target is, respectively, 1.3 and 3.5 times the beam radius for the Effelsberg and \nan{} telescopes. 
The \nan{} profiles of our target and UGC 5485 are therefore not likely to be contaminated 
by each other. The Effelsberg and NRT profiles of UGC 5485 are similar in \VHI{}, 5987 \kms{}, and \FHI{}, 15 Jy \kms{} 
(i.e., almost 4 times of that of our target), but the \nan{} \Wfifty{} of 331 \kms{} is 65 \kms{} 
broader. 

\subsubsection{PGC 30113} 
\begin{figure}[ht] 
\begin{center}\includegraphics[width = 2.8in]{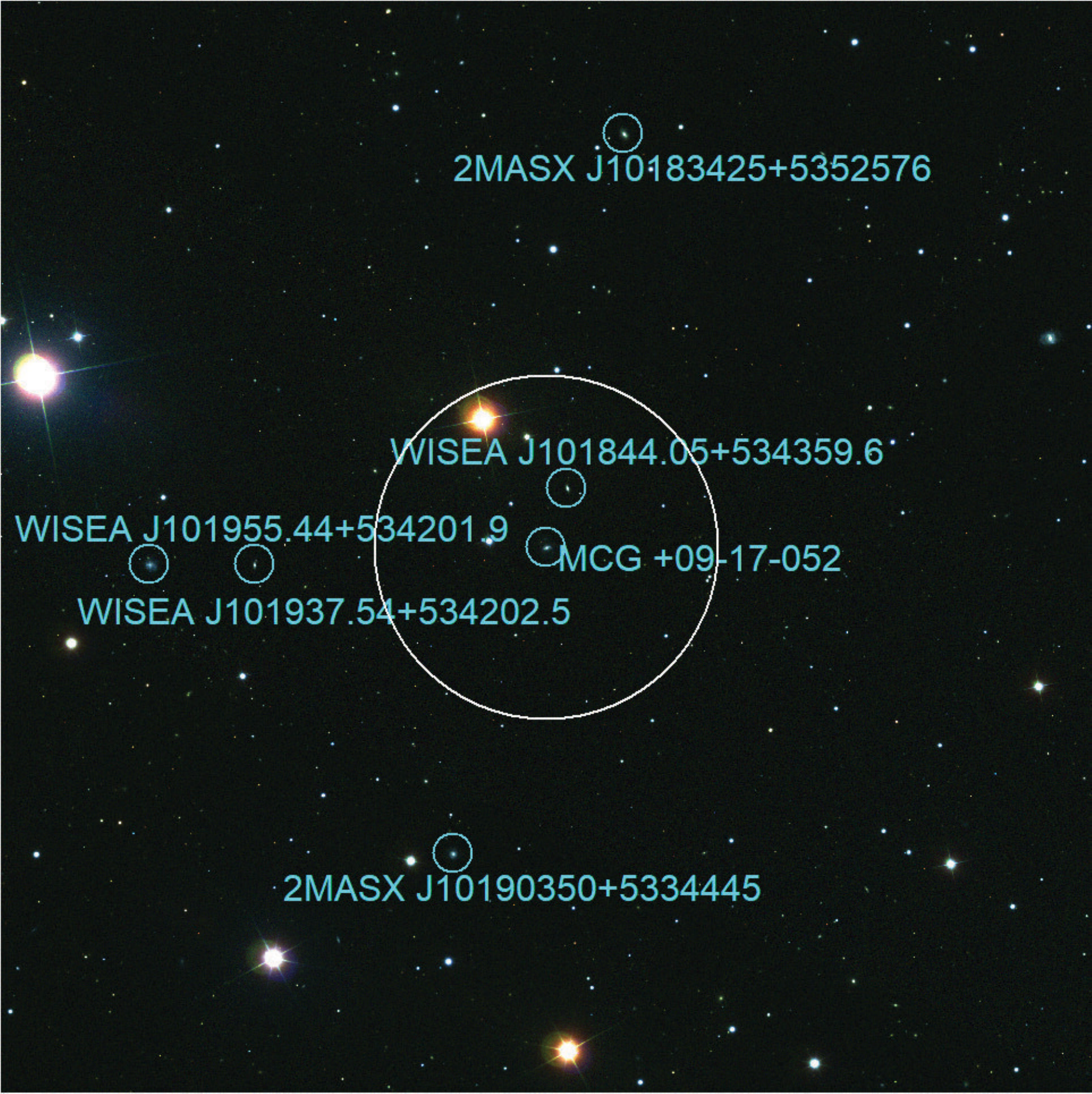}\end{center}
\caption{SDSS DR12 false color ($i$, $r$, and $g$ filter) image of the PGC 30113 galaxy group. 
The large white circle shows the $8\farcm7$ GBT beam. The six known group members are 
denoted by cyan circles. Image is $27\farcm5$ across.\label{fig:P30113_group}}
\end{figure}

It is highly likely that our measured \HI{} flux for PGC 30113 is contaminated by other galaxies.
It is part of a small group of galaxies within approximately 14\arcmin{} and 2500 \kms{} from each other 
(Figure~\ref{fig:P30113_group} and Table~\ref{tab:P30113_group}). Likely due to its group membership, 
past observations have found a large range of redshifts for PGC 30113, from 10,983 \kms{} 
\citep{2018MNRAS.473.4731K} to 21,203 \kms{} \citep{2014ApJS..210....9B}, but the majority of 
observations give 13,401$\pm$7 \kms{} 
\citep{2015ApJS..217...27A, 2004AJ....128..502A, 2015MNRAS.448.3442G, 2014ApJS..210....9B}, 
consistent with our GBT value of 13,402 \kms{}. 

\subsubsection{PGC 32862} 
\begin{figure}[ht] 
\begin{center}\includegraphics[width = 2.8in]{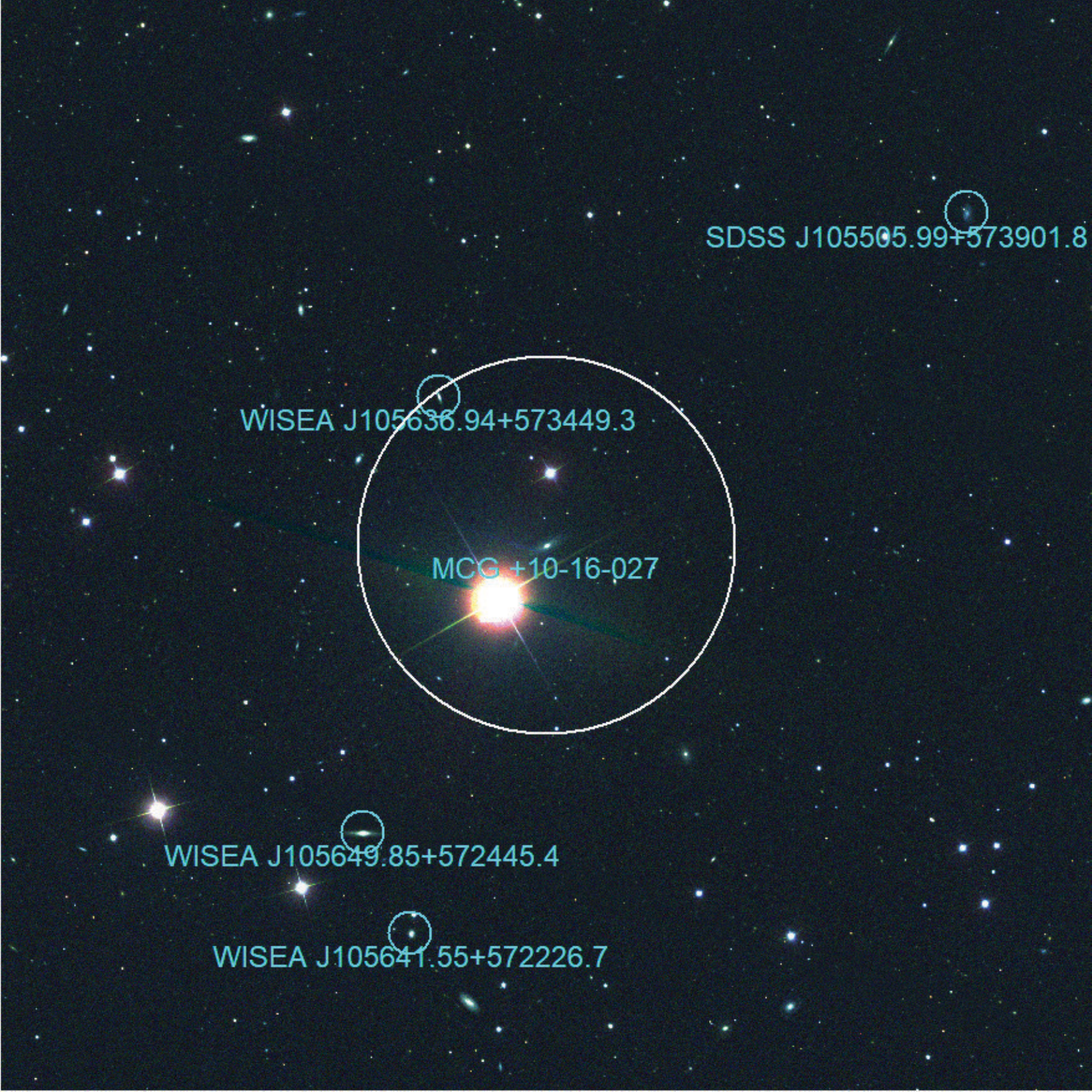}\end{center}
\caption{SDSS DR12 false color ($i$, $r$, and $g$ filter) image of the PGC 32862 galaxy group.
The large white circle shows the $8\farcm7$ GBT beam. The four known group members are 
denoted by cyan circles. Image is $25\farcm5$ across. \label{fig:P32862_group}}
\end{figure}

PGC 32862 is part of a group of spiral galaxies at approximately the same velocity 
(Figure~\ref{fig:P32862_group} and Table~\ref{tab:P32862_group}). 
Our GBT \HI{} detection at 13,969 \kms{} is therefore highly likely to be contaminated, 
even though our measurement is in agreement with its optical velocity of 13,973 \citep{2015ApJS..217...27A}.

\subsubsection{PGC 34377} 
The galaxy 2MASX J11164683+3039330 is 8$'$ (or 4.5 Arecibo beam radii) and 287 \kms{} from our target, PGC 34377. 
It is possible, but unlikely, that our measured Arecibo \HI{} profile for our target is contaminated by the galaxy. 


\subsubsection{PGC 36715} 
There are three galaxies within $4\farcm0$ (2.2 times the Arecibo beam size) and 46 \kms{} from our target, PGC 36715: 
2MASX J11462026+3543181, KUG 1143+360B, and 2MASX J11463995+3546070). It is likely that all three have
contaminated our measured Arecibo \HI{} profile.

\subsubsection{PGC 38333} 
Lying within the Coma supercluster, PGC 38333 is part of a large group of galaxies, which includes a nearby 
companion galaxy, KUG 1203+206A, as well as NGC 4090 and a score of other galaxies, all within 
15\arcmin{} and $\pm$1,000 \kms{} of PGC 38333. Only our Arecibo result is listed 
in Table~\ref{tab:avgresultsdets} since, due to the smaller beam size, it is likely 
less contaminated by the other sources than our \nan\ profile, although it is doubtful that it measured only the 
\HI{} gas in our target. 

\subsubsection{PGC 38698} 
\begin{figure}[htb] 
\begin{center}\includegraphics[width = 2.8in]{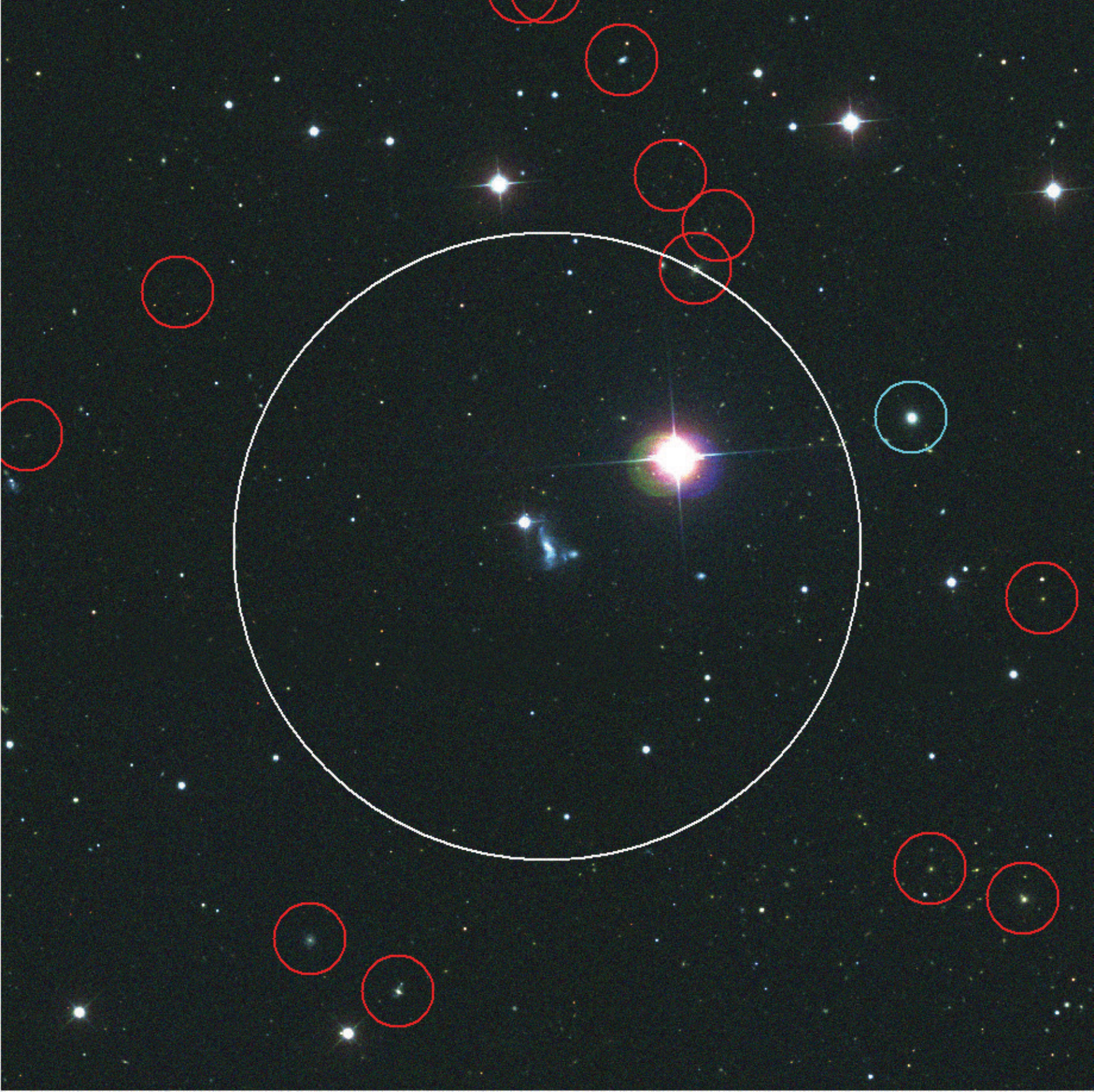}\end{center}
\caption{SDSS DR12 false color ($i$, $r$, and $g$ filter) image of PGC 38698 (center). 
The large white circle shows the $8\farcm7$ GBT beam. The galaxy I Zw 031 is denoted by the 
cyan circle on the right and the known background galaxies by red circles.  
Image is $15\farcm$ across.
\label{fig:PGC038698_neighbors}}
\end{figure}

PGC 38698's morphology indicates it has recently undergone a significant interaction, most likely with 
its nearby companion I Zw 031, at \Vopt{} = 6998 \kms{} (Figure~\ref{fig:PGC038698_neighbors}), 
as all other nearby galaxies have velocities $>$5,000 \kms{} away from I Zw 031 and PGC 38698. 
However, as the \nan\ and GBT \HI{} profile parameters are very similar, 
it is likely that all the detected \HI{} emission is associated with our target, but it is also 
likely to be greatly disturbed due to a recent interaction.

\subsubsection{PGC 38958}  
The Arecibo ALFALFA survey reported a velocity of 5909 \kms{} \citep{2018ApJ...861...49H}
for PGC 38958, significantly different from the 818$\pm$5 \kms{} we found at Arecibo. 
To further explore this discrepancy, we made very deep GBT observations of PGC 38958 
but did not detect a source at 5909 \kms{}, with an r.m.s. of 0.62 mJy, well below the $\sim$8 mJy 
($\sim$3.5$\sigma$) \HI{} peaks found by ALFALFA. It is therefore highly likely the ALFALFA 
detection is spurious. There is no other galaxy with a known redshift within 4,000 \kms{} 
and two beam-widths of our GBT detection.

\subsubsection{PGC 39701}
There is a large group of galaxies, including IC 3157, which lie $5\farcm0$ - $8\farcm9$
(1.1-2.0 times the GBT beam radius) away from PGC 39701, but their velocities (17,511-17,921 \kms{}) 
are 850-1250 \kms\ lower than that of PGC 39701, at \VHI{} = 18,768 \kms. It is possible, 
but unlikely, that our GBT detection is contaminated by that galaxy group.


\subsubsection{PGC 43880} 
\begin{figure}[ht] 
\begin{center}\includegraphics[width = 2.8in]{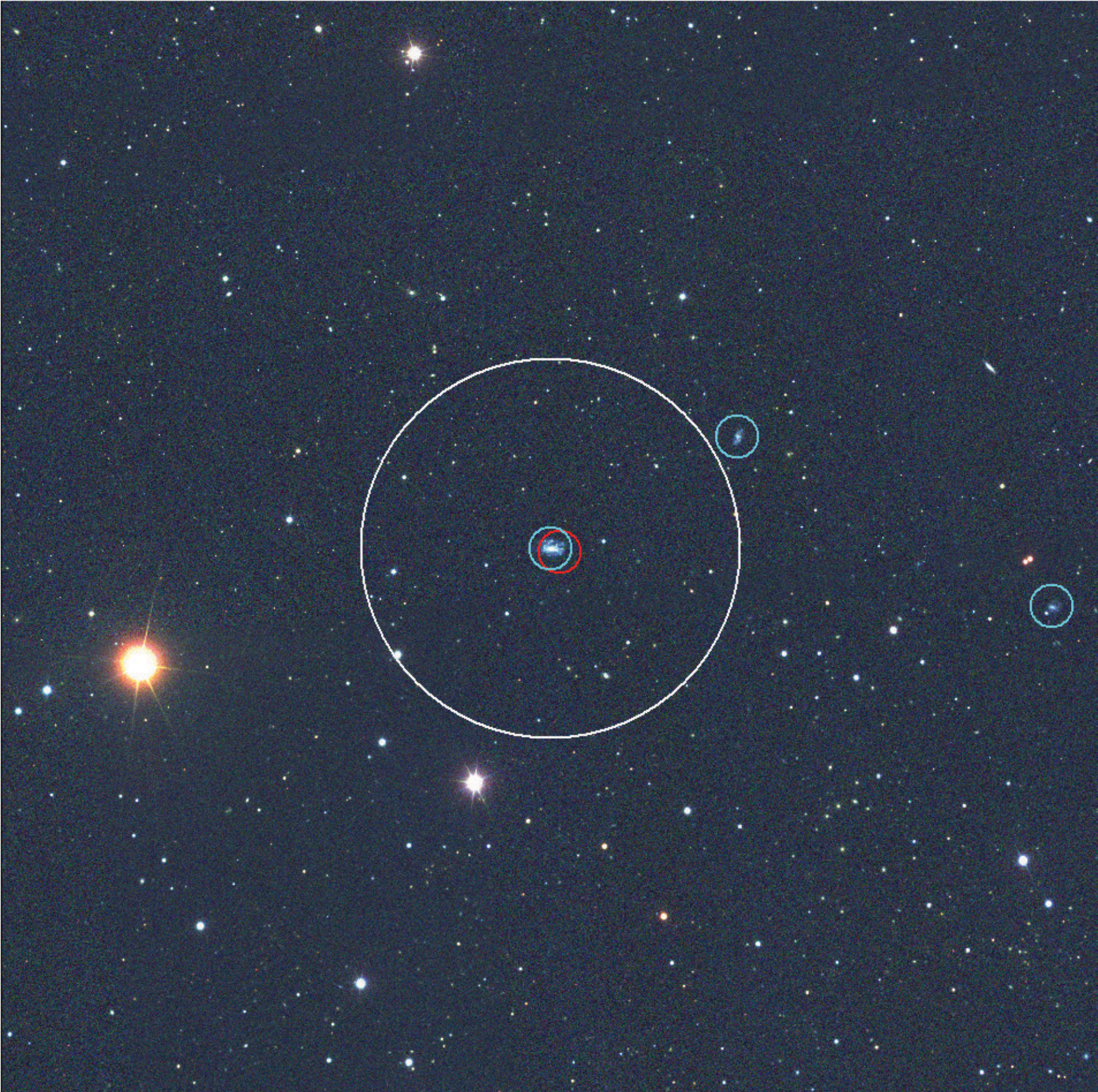}\end{center}
\caption{SDSS DR12 false color ($i$, $r$, and $g$ filter) image of our target galaxy PGC 43880 (center, cyan circle). 
The large white circle shows the $8\farcm7$ GBT beam. The position of SDSS J125405.61+481534.9 is 
denoted by the red circle, close to our target, while WISEA J125341.21+481813.1 (top center) and 
WISEA J125257.97+481417.6 (on the right) are denoted by cyan circles.  
Image is $25\farcm$ across.
\label{fig:PGC043880_group}}
\end{figure}

PGC 43880 is part of a group of at least four galaxies at similar velocities. One of these, 
SDSS J125405.61+481534.9, could in fact be considered part of our target galaxy PGC 43880 
and it is surprising the SDSS survey identified is separately. The other two galaxies, 
WISEA J125341.21+481813.1 and WISEA J125257.97+481417.6, lie outside the GBT beam but 
still could be contaminating the \HI{} profile (Figure~\ref{fig:PGC043880_group}). 
Nonetheless it is likely the measured \HI{} emission is primarily from our target.

\subsubsection{PGC 51872}  
\begin{figure}[ht] 
\centering\includegraphics[width = 2.8in]{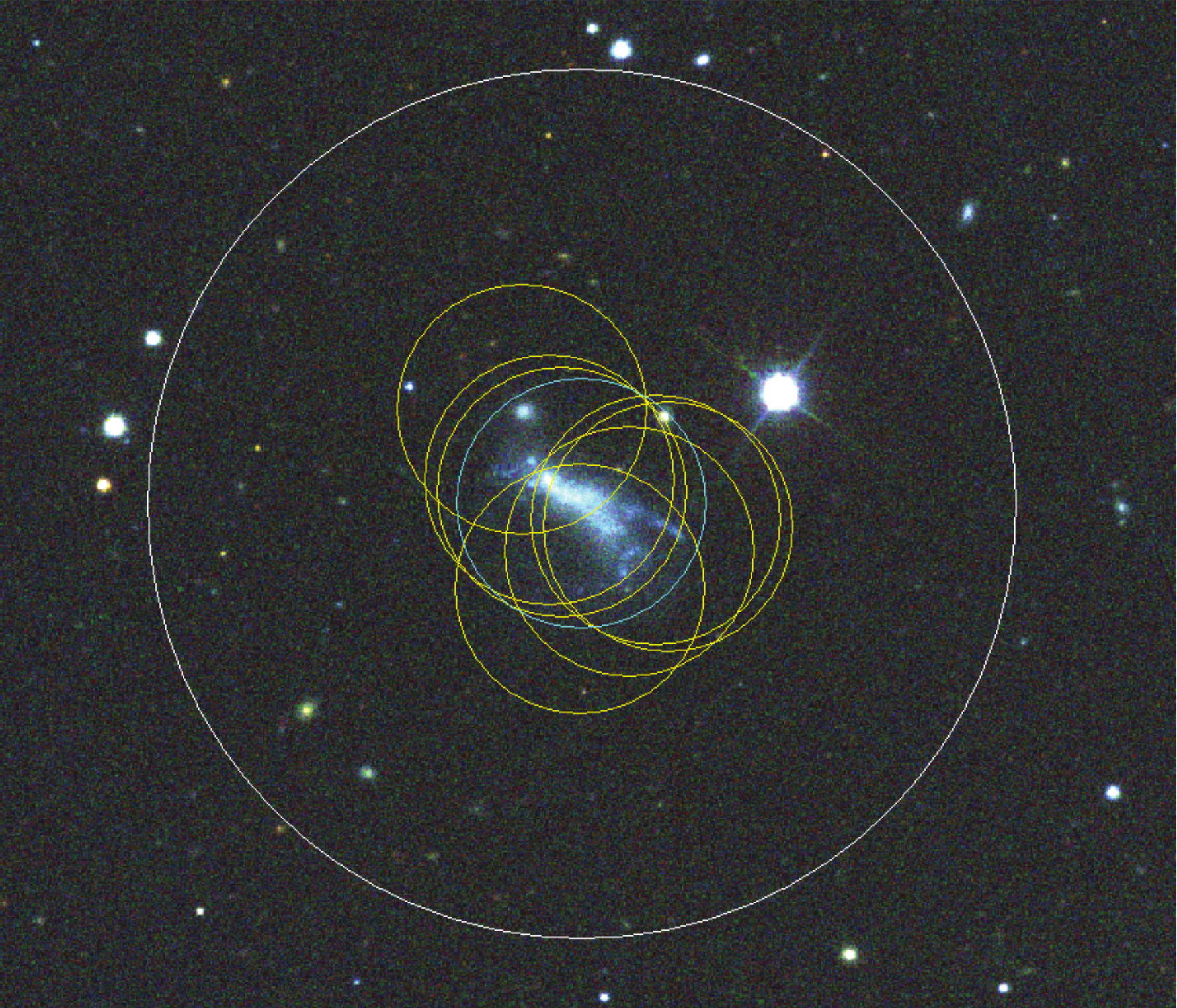} 
\caption{SDSS DR12 false color ($i$, $r$, and $g$ filter) image of PGC 51872 (center, cyan circle). 
The white circle shows the $3\farcm6$ Arecibo beam. The yellow circles denote the 
various galaxies listed in Table~\ref{tab:P51872_neighbors}, which we believe to be part of 
PGC 51872 rather than separate galaxies.  \label{fig:PGC051872_group}}
\end{figure}

PGC 51872 appears to be part of a loose group of galaxies (Figure~\ref{fig:PGC051872_group} and 
Table~\ref{tab:P51872_neighbors}). Although the various databases 
show a few objects within 1,000 \kms{} from our measured velocity and within the Arecibo beam, 
most of these appear to be parts of our target galaxy PGC 51872, which appears to be undergoing a 
significant disruptive event, rather than separate galaxies.

\subsubsection{PGC 56738} 
The galaxy UGC 10138 is 230 \kms{} and $9\farcm5$ from our target, PGC 56738. Both were observed 
at Arecibo (see also \citealt{2005ApJS..160..149S}), and their separation is 5 times the telescope 
beam radius. UGC 10138 has \VHI\ = 9645 \kms, \Wfifty\ = 685 \kms, and \FHI\ = 1.1 Jy \kms. 
It is unlikely it has contaminated our measured Arecibo \HI{} flux of our target. 
However, UGC 10138 lies within the \nan{} telescope beam and our spectrum of PGC 56738 does 
appear contamination by it. Therefore, only the Arecibo result is listed in Table~\ref{tab:avgresultsdets}.


\subsubsection{PGC 60396}
We detected an \HI{} profile centered at 7,915 \kms{} when pointing at PGC 60396 with the GBT.
The galaxy has no published redshift and no known companions within $15\farcm0$ and 
8,000 \kms{} around our GBT velocity, yet our \HI{} profile shows a clear asymmetry. 
There is, though, a nearby companion, PGC 60394, which lies only $1\farcm2$ away but 
which has an unknown velocity. 
It is highly likely that we have detected both objects - PGC 60396 at 7,915 \kms{} and 
PGC 60394 near 7,800 \kms{}.

\subsubsection{PGC 65750}
PGC 65750 has two neighbors, LEDA 888740, $13\farcm1$ and 142 \kms{} away, and 
LEDA 3320352, $11\farcm2$ and 1,172 \kms{} away. It is highly unlikely that either of these objects
are contributing to the \HI{} we detected at the GBT when pointing towards PGC 67570. As a result, 
while the optical morphology and lopsided \HI{} profile of PGC 67570 indicate a disturbed morphology, 
our gas detection appears to be wholly a part of the intended target, PGC 65750.

\subsubsection{PGC 74070} 
The galaxy NGC 798 is $9\farcm5$ and 351 \kms{} away from our target, PGC 74070. 
It is possible that our \nan{} data are slightly contaminated by NGC 798's gas,
as its distance from our target is 2.5 times the \nan{} beam radius, and as the 
smaller Arecibo beam detected a somewhat smaller flux. If that is not the case, 
then the \HI{} distribution of PGC 74070 must extend beyond the $3\farcm6$ Arecibo beam.
Both the \nan\ and Arecibo observations are included in Table~\ref{tab:avgresultsdets} 

\subsubsection{PGC 77473} 
Our \nan{} detection at 8085 \kms{} matches the Parkes' HIPASS and HIZOA \HI{} velocities 
of source HIPASS J0730-28 \citep{2004MNRAS.350.1195M, 2016AJ....151...52S}, whose optical 
counterpart is identified as 2MASX J07304535-282358 by \citet{2005MNRAS.361...34D}.

\subsubsection{PGC 82395} 
The galaxy 2MASX J09262444+3304090 is $4\farcm7$ (2.6 times the Arecibo beam radius) and 
67 \kms{} from our target, PGC 82395. It is possible, but not likely, that it contaminates our Arecibo 
\HI{} profile of our target.

\subsubsection{PGC 82550}  
As the Arecibo observations of PGC 82550 have RFI at their edge, only the \nan{} observations were used 
for our results in Table~\ref{tab:avgresultsdets}.

\subsubsection{PGC 85851} 
The galaxies SDSS J235205.07+144047.2 and KUG 2349+134A are within 
$4\farcm0$ (2.2 times the Arecibo beam radius), and 28 \kms{} from the measured \HI{} velocity 
of our target, PGC 85851. It is highly likely that the \HI{} flux we measured at Arecibo has been contaminated 
by these two other galaxies.

\subsubsection{PGC 86863} 
Only the significantly higher signal-to-noise Arecibo observations were used for line
profile parameter measurements in Table~\ref{tab:avgresultsdets}.

\subsubsection{PGC 86903} 
PGC 86903 lies within a large group of galaxies. However, none of the other galaxies
lie within a 5$'$ distance. It is therefore unlikely that the \HI{} flux we measured 
is contaminated by another source. 

\subsubsection{PGC 89535} 
2MASS \citep{2014ApJS..210....9B} lists a photometric redshift of 11,900$\pm$4500 \kms{} for 
our target PGC 89535, which is consistent with our \HI{} value of 13,387 \kms{}. Otherwise, o
ur target has no known neighbors both within the GBT beam and within 1,000 \kms{}of its \HI{} velocity. 
It is likely this is an isolated LSB galaxy.

\subsubsection{PGC 89614}
Our detection listed as PGC 89614 is more likely of KUG 1228+248, a galaxy only $0\farcm85$ from it 
and with an optical velocity of 20,099 \kms\ \citep{2017ApJS..233...25A}, only 14 \kms{} from the velocity 
we found at Arecibo.


\subsubsection{PGC 91198} 
The galaxy SDSS J122526.1+360102.5 is $8\farcm1$ (4.5 times the Arecibo beam radius) and 200 \kms{} away 
from our target, PGC 91198. It is therefore unlikely that it is significantly contaminating our Arecibo \HI{} profile 
of PGC 91198.

\subsubsection{PGC 91549} 
The galaxy SDSS J164503.80+304802.1 is $10\farcm4$ (4.4 times the Arecibo beam radius) 
and 59 \kms{} from our target, PGC 91549. It is therefore highly unlikely that it is contributing an appreciable 
amount of flux to our Arecibo \HI{} profile of our target.

\subsubsection{PGC 95598}
\begin{figure*}[ht] 
\begin{center}\includegraphics[height = 3in]{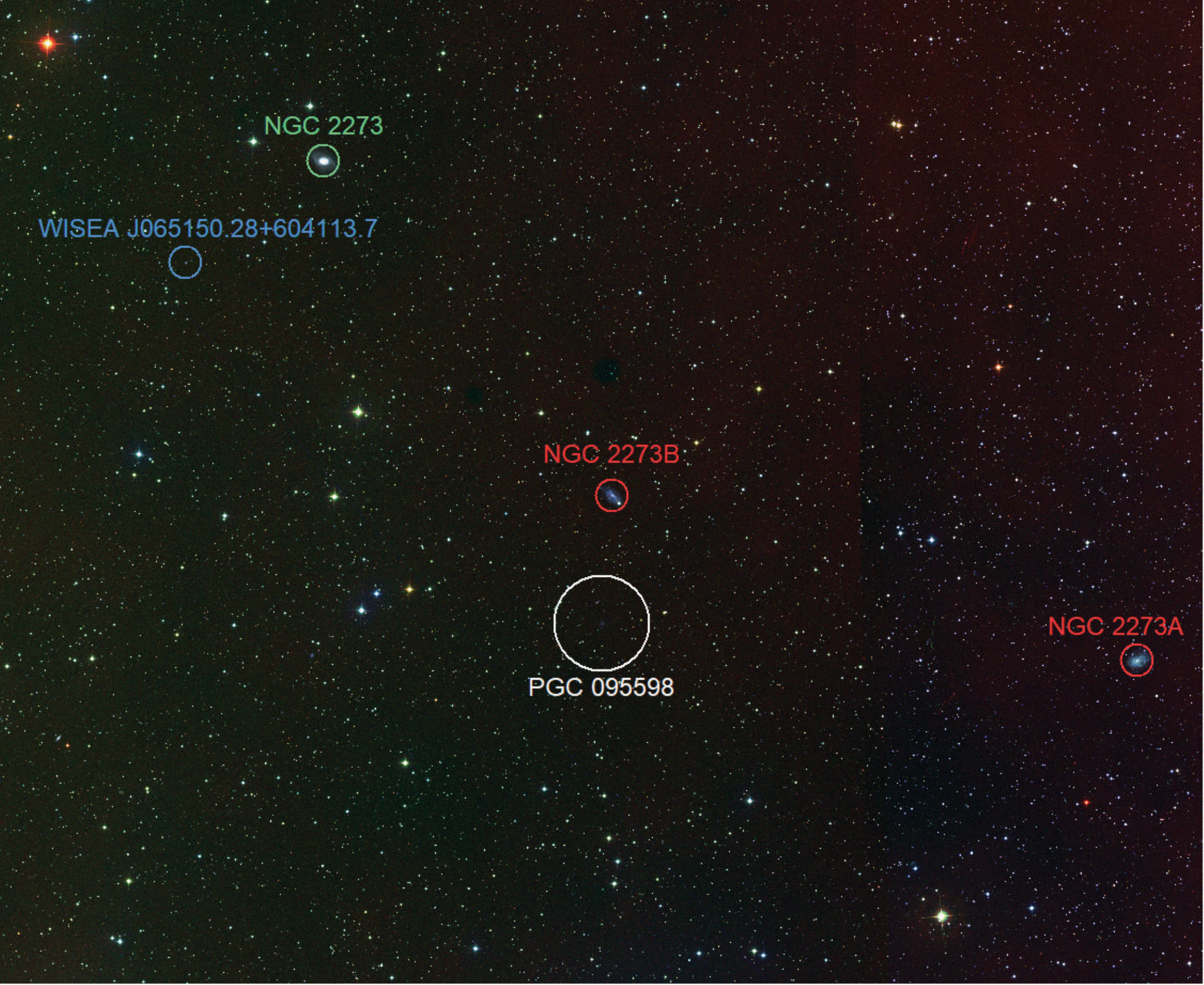}\end{center}
\caption{DSS false color (IR, red, and blue plates) image of the NGC 2273 galaxy group, which 
likely includes our target galaxy PGC 95598 -- inside the large white circle 
showing the $8\farcm7$ GBT beam. The four other circles denote the known group members, 
with their color indicating the object's relative velocity within the 
group. Image is $110\farcm \times 90\farcm$ across  \label{fig:P95598_group}}
\end{figure*}

PGC 95598 is part of the NGC 2273 group of galaxies, shown in Figure~\ref{fig:P95598_group} 
\citep[][e.g.]{1988cng..book.....T, 1992AA...255...69G, 1993AAS..100...47G}. As its nearest neighbor, 
NGC 2273B, lies $11\farcm7$ (2.7 times the GBT beam radius) and 307 \kms{} away it is possible that 
it is contaminating our GBT \HI{} profile.

\subsubsection{PGC 135624} 
Three galaxies (NGC 391, 2MASX J01072342+0053341, and 2MASX J01073154+0053232) are all 
within $7\farcm1$ and 180 \kms{} of our target PGC 135624, and within the \nan{} beam. 
All three are likely to have added to the \HI{} flux of our target we measured at \nan,
although we note that our profile is rather narrow (\Wfifty\ = 95 \kms). 
\subsubsection{PGC 135894} 
Our target PGC 135894 is in a group of at least four galaxies, three of which lie within $9\farcm0$ and 
1,000 \kms{} of it -- WISEA J233406.20+001311.8, DEEP2 33029695, and DEEP2 33029720.
Furthermore, the \HI{} flux we measured at \nan\ is 40\% (1.5$\sigma$) higher than at Arecibo, 
while the GBT flux lies between the two. 
It is therefore highly likely our \HI{} measurements of PGC 135894  are contaminated by its companions.
Please note that the \nan{} spectrum is not shown in Figure~\ref{fig:HLdets}.


\subsubsection{PGC 166523} 
There are two galaxies, WISEA J175425.16-020235.5 and WISEA J175423.40-020032.4, 
which lie within 3\arcmin{} and 1,000 \kms{} of our target, PGC 166523. It is highly likely that 
the \HI{} profile we measured at \nan\ is contaminated by these two other galaxies.

\subsubsection{PGC 2815809} 
The galaxy HIPASS J0756-26 is $6\farcm8$ and 36 \kms{} from our target, PGC 2815809. 
As HIPASS J0756-26 is an \HI{} detected galaxy only, with no optical counterpart \citep{2005MNRAS.361...34D}
it is likely that HIPASS J0756-26 and PGC 2815809 are the same source. However some
observed \HI{} properties of both sources differ significantly, as measured at Parkes by 
\citet{ 2004MNRAS.350.1195M} and at \nan\ and the GBT for this paper, with the HIPASS results having a 
50\% larger \Wfifty{} and 2$\times$ higher flux than measured by us
(PGC 2815809: \VHI{} = 6645 \kms, \Wfifty{} = 127\kms, \MHI{} = 7.5$\times \;10 ^9$\Msun{}; 
HIPASS J0756-26: \VHI{} = 6681 \kms, \Wfifty = 166 \kms, \MHI{} = 1.7 $\times \; 10^{10}$\Msun{}).
Follow-up observations to determine the true \HI{} properties of this galaxy are needed - 
to determine if PGC 2815809 and HIPASS J0756-26 are in fact the same source and, if so, 
where its center lies.

Please note that an Effelsberg \HI{} detection was reported at a much lower velocity of 
241 \kms{} by \citep{huchtmeier01}, with \Wfifty{} = 26 \kms{}, and \FHI{} = 0.7 Jy \kms{}. 
Our \nan{} and GBT data show no sign of this profile, however, which should have been detected at the 
\SNR{} = 10 level, if real. We therefore conclude the Effelsberg signal is due to 
Galactic \HI{}. The Effelsberg velocity search range ends at 4000 \kms{}, which excludes 
the profile we detected.



\subsubsection{UGC 605}
\begin{figure*}[ht] 
\centering
\includegraphics[width = 6.0in]{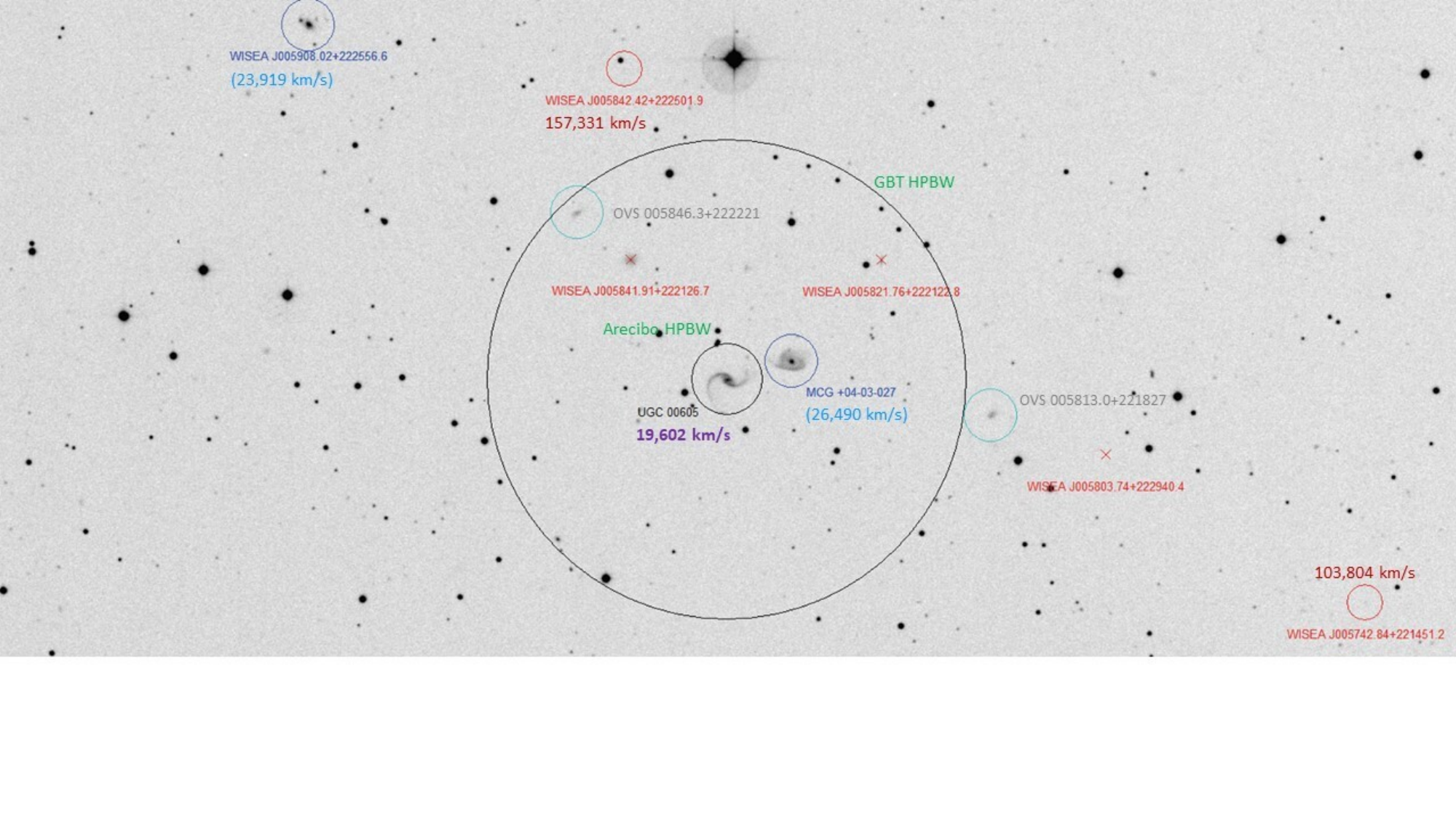}
\caption{DSS Blue image of UGC 605 showing the $3\farcm6$ Arecibo and $8\farcm7$ GBT 
beams as black circles. The galaxies possibly related to the \HI{} spectra detected 
when observing UGC 605 are denoted by the cyan circles. Other galaxies are denoted 
in dark blue and red.  \label{fig:U605_neighbors}}
\end{figure*}

UGC 605 is among our most distant detections, at 19,536$\pm$21 \kms{}, which matches the optical value 
of 19,602 \kms{} reported by \citet{wang18}.
The \HI{} velocity width and gas mass we measure towards UGC 605 are both well beyond what 
would be expected for an individual galaxy.
The mean velocities of our Arecibo and GBT profiles are the same within the uncertainties,
but the Arecibo linewidths and integrated line flux are significantly smaller than those
measured with the GBT: \Wfifty\ = 723 and 916 \kms, and \FHI\ = 2.3 and 7.5 \Jykms, respectively.
Within the GBT and Arecibo beams there is another galaxy (MCG +04-03-027) which lies only $1\farcm3$ 
away, at a photometric redshift of 26,490 \kms{} \citep{2014ApJS..210....9B}. 
As this is not a precise spectroscopic velocity, it is possible that gas from this object is also 
contained within the our \HI\ spectra. 
However, a more likely scenario is that UGC 605 is part of a loose group of interacting galaxies, 
which includes at least  MCG +04-03-027, and two other, previously unidentified LSB galaxies shown 
in Figure~\ref{fig:U605_neighbors} (OVS 005846.3+222221 \& OVS 005813.0+221827), and possibly even 
the more distant galaxies 
WISEA J005908.02+222556.6 (23,919 \kms; distance to UGC 605 $10\farcm8$), 
WISEA J005852.84+22005557.4 (17,691 \kms; at $13\farcm9$), and 
WISEA J005908.02+222556.6 (23,919 \kms; at $10\farcm8$) \citep{2014ApJS..210....9B}. 
This would also explain the notable difference in line widths and \HI{} flux found between 
the GBT and Arecibo observations. 
As the Arecibo beam is the smallest, only its values are used in Table~\ref{tab:avgresultsdets}.

\subsubsection{UGC 1127}
We first observed UGC 1127 as part of the the UGC sample and did not detect it, to an r.m.s. of 2.7 \kms{}.
Deeper follow-up observations with the GBT found a clear line signal at 20,820 \kms{}.

\subsubsection{UGC 1145} 
Our original \nan\ detection showed an \HI{} velocity at 5784$\pm$13 \kms{}, 108 \kms{} higher than 
the optical value of 5676$\pm$10 \kms{} \citep{2012ApJS..199...26H}. The \nan{} detection shows a 
single, narrow peak around 5690 \kms, somewhat off-center amidst an about 290 \kms{} wide, weaker plateau.
The velocity of the peak corresponds well to the optical value. As the noise level of
the \nan\ detection was quite high, significantly more sensitive GBT observations were taken at a later date. 
The GBT data confirmed the narrow peak, albeit at a lower peak value, which lies at the optical velocity, 
but did not confirm the wide plateau. As a result, only the GBT data were used in Table~\ref{tab:avgresultsdets},
although both results are shown in Figure~\ref{fig:HLdets} and Table~\ref{tab:indresultsdets}.

\subsubsection{UGC 1216} 
UGC 1216 was detected twice at \nan, in the velocity overlap region between our searches at 
low and high velocities. The difference in center velocity is 23 \kms, or almost 4$\sigma$. 
The line widths and line fluxes are the same within the uncertainties. 
The two results are listed separately in Table~\ref{tab:indresultsdets}.



\subsubsection{UGC 1851}
UGC 1851 was observed a total of four times -- once (using \nan) for the UGC
survey (and not detected), once at \nan\ for the HyperLeda sample, and then twice with the GBT.
The second set of GBT observations was made due to discrepancies between the earlier \nan\ detection and GBT
observations. As the new GBT values lie in between the \nan\ detection and previous GBT values, 
the thee individual detections were averaged for Table~\ref{tab:avgresultsdets}.

\subsubsection{UGC 1927} 
UGC 1927 was part of the UGC sample but not detected at \nan. Follow-up observations with the GBT
detected the object at 6280 \kms.



\subsubsection{UGC 2036}  
Our \nan\ detection (\VHI\ = 5021 \kms, \Wfifty\ = 135 \kms, \FHI\ = 3.6 \Jykms)
may be confused by CGCG 539-044, a highly inclined \BT\ = 15.2 galaxy $8\farcm2$s away from UGC 2036, 
with an optical velocity of 4920$\pm$33 \kms\ \citep{huchra99}.

\subsubsection{UGC 2068} 
Our \nan\ detection is contaminated by other galaxies within the telescope beam.
UGC 2068 was detected previously in \HI{} at \nan{} by \citet{theureau98} at 
\VHI{} = 5740 \kms; it has no other published redshift. 
Our \HI{} profile (\VHI{} = 5628 \kms, \Wfifty{} = 586 \kms, \FHI{} = 7.7 \Jykms)
is very wide and has a complex structure, due to the presence of other galaxies in the beam,
notably NGC 980 and NGC 982. NGC 980 is a \BT\ = 14.2 mag S0 galaxy at \Vopt{} = 5796$\pm$42 \kms, 
$2\farcm5$ N of our target UGC 2068, and NGC 982 is a \BT\ = 13.2 mag Sa galaxy at \Vopt{} = 5845$\pm$58 \kms, 
$1\farcm5$s from our target. The \HI{} profiles of NGC 980 and NGC 982 measured with the $3\farcm6$ 
Arecibo beam by \citet{1994AJ....108..896M} and \citet{1988AJ.....95..607H} show
\VHI{} = 5757$\pm$9 \kms, \Wfifty\ = 702 \kms{} and \FHI\ = 8.1 \Jykms, and
\VHI{} = 5737$\pm$11 \kms, \Wfifty\ = 568 \kms{} and \FHI\ = 6.9 \Jykms, respectively.
As the separation between the centers of NGC 980 and 982 is $3\farcm6$, all measured \HI{} profiles 
are bound to be confused and therefore no conclusion can be drawn on the \HI{} profile parameters 
of UGC 2068.

\subsubsection{UGC 2113} 
UGC 2113 was part of the UGC sample but not detected at \nan. Follow-up observations with the GBT
detected the object at 12,831 \kms.

\subsubsection{UGC 2139} 
UGC 2139 was part of the UGC sample but not detected at \nan. Follow-up observations with the GBT
detected the object at 5113 \kms.

\subsubsection{UGC 2235} 
UGC 2235 was detected at \nan\ in both surveys. Both sets of \HI\ profile parameters are consistent within the 
uncertainties, but it is highly likely that (most of) the \HI{} flux we detected towards UGC 2235 at 5583 \kms\ 
is associated with UGC 2234, which lies just within the \nan{} HPBW. UGC 2235 has no optical velocity, 
whereas that of UGC 2234 is 5562$\pm$25 \kms{} \citep{huchra99}. A \nan{} profile measured towards UGC 2234 by 
\citet{theureau98} shows \VHI\ = 5606 \kms, \Wfifty\ = 267 \kms, and \FHI\ = 3.5 Jy \kms. 
Follow-up observations with the GBT clearly detected UGC 2234 but not UGC 2235, to the 1.01 mJy r.m.s. level.


\subsubsection{UGC 2301 \& UGC 2305} 
Pointing towards UGC 2301, at \nan\ we detected two galaxy \HI{} profiles, centered at 1746 \kms{} and 5414 \kms{}, 
respectively. We assume the former to be of our target UGC 2301 and the latter of nearby UGC 2305, 
a \BT\ = 16.1 SBc spiral $11\farcm7$ south of our target. Taking into account the beam attenuation,
our profile parameters of UGC 2305 (\VHI{} = 5414 \kms, \Wfifty{} = 188 \kms, and 
\FHI{} = 3.0 \Jykms) match the mean of those measured at \nan{} by \citet{theureau98} 
and at Arecibo by \citet{1993AJ....105.1251W}: 
\VHI{} = 5409 \kms, \Wfifty{} = 189 \kms, \FHI{} = 4.7 \Jykms.
Please note that our NRT spectrum of UGC 2305 is not shown in Figure~\ref{fig:HLdets}


\subsubsection{UGC 2480} 
The galaxy UGC 2472 lies $7\farcm7$ (4.3 and 1.8 times the Arecibo and GBT beam radii, respectively) 
from our target UGC 2480 and has an \HI\ velocity of 10,157 \kms{} (e.g. \citealt{2018ApJ...861...49H}), 
27 \kms\ higher than what we measured for our target, UGC 2480. 
The larger \FHI{} and \Wtwenty{} values we found at the GBT, compared to Arecibo, makes it likely 
that our GBT data, at least, are affected by UGC 2472's proximity. As a result, only the Arecibo 
data are used in Table~\ref{tab:avgresultsdets}.
\subsubsection{UGC 2505} 
The galaxy 2MASX J03034116-0104249 is $10\farcm4$ and 135 \kms{} from our target, UGC 2505. 
As no difference in flux is found between the \nan{} and GBT observations, though, it is 
unlikely that flux from 2MASX J03034116-0104249 has contaminated the profile parameters 
we measured for UGC 2505. GBT follow-up observations were taken of UGC 2505 
to confirm this, with the same result as the earlier observations.

\subsubsection{UGC 2668}
\citet{2017ApJS..233...25A} gives UGC 2668 an optical velocity of 5450 \kms{}, yet we see no 
hint of a galaxy at that velocity, to an r.m.s. of 2.8 mJy. Either the object is extremely lacking
in \HI{} gas or the optical velocity is incorrect.

\subsubsection{UGC 2749} 
Our \nan\ \HI{} spectrum of UGC 2749, with \VHI{} = 4207 \kms, is very probably confused by that 
of the Sc type galaxy CGCG 541-011, $5\farcm8$ to the North, which has \Vopt{} = 4253$\pm$43 \kms.

\subsubsection{UGC 2876}  
UGC 2876 has no published optical velocity but it has a published Arecibo \HI{} detection, 
with \VHI{} = 5475 \kms, \Wfifty{} = 200 \kms{} and \FHI{} = 1.9 \Jykms{} \citep{monnier03b}.
However, upon inspection of the published spectrum we estimate its central peak velocity to be 5305 \kms{}. 
This matches the other values, which were obtained with both \nan{} and the GBT (with two separate GBT observations).


\subsubsection{UGC 3041}
\citet{2009MNRAS.399..683J} give an optical velocity of 4931 \kms{} for UGC 3041, yet we find no
\HI{} detection at that velocity to an r.m.s. of 0.81 mJy. Either this source has an extremely low
\HI{} mass or the optical velocity is incorrect.

%

\subsubsection{UGC 3797}  
It is likely that nearby galaxy UGC 3789 has contributed flux to our measured \nan\ \HI{} spectrum.
It lies $4\farcm3$ from our target UGC 3797, and has an optical velocity of 3243$\pm$70 \kms{} \citep{focardi86},
162 \kms{} lower than that of our target (3405$\pm$15 \kms{} \citealt{2000AJ....120.2338R}).
We observed UGC 3797 at \nan{} and measured \VHI\ = 3399 \kms, \Wfifty\ = 163 \kms, 
and \FHI\ = 2.9 Jy \kms, whereas UGC 3798 was observed at \nan{} by \citet{theureau98}, 
who reported \VHI\ = 3325 \kms, \Wfifty\ = 389 \kms, and \FHI\ = 1.7 Jy \kms. 
The separation between both objects is 2.5 times the beam radius. 



\subsubsection{UGC 3963}
UGC 3963 was originally observed at \nan\ as part of the UGC sample. Subsequently an optical 
redshift of 20,236 \kms\ was reported \citep{huc2012} which places it outside our \nan\ search range. 
However, follow-up observations with the GBT which included the optical velocity still did not 
detect the galaxy. Only the GBT r.m.s. is listed in Table~\ref{tab:resultsnondets}.

\subsubsection{UGC 5071} 
Although noted as a POSS blue plate defect and as not visible on the red 
POSS plate in \citet{UGC}, UGC 5071 is listed as a galaxy in \citet{2000AAS..144..475P} and HyperLeda. 
We did not detect \HI{} emission in its direction.

\subsubsection{UGC 5127} 
Our Arecibo spectrum shows two detections, at 4521 \kms{} and 5892 \kms{}. The former is of our target 
UGC 5127, with \Vopt{} = 4508 \kms{} \citep{2015ApJS..217...27A}, whereas the latter detection appears to be of 
SDSS J093743.77+370631.3, $1\farcm2$ from our target with \Vopt{} = 5894 \kms{} \citep{2015ApJS..217...27A}. 
The data we used are for the lower velocity detection. 

\subsubsection{UGC 5293}
We detected UGC 5293 at \nan\ and twice with the GBT, and found 
\VHI{} = 4986 \kms, \Wfifty{} = 198 \kms, and \FHI{} = 2.4 Jy \kms.
The galaxy WISEA J095234.27+412207.1 lies $7\farcm5$ (1.6 times the GBT beam size) away 
and has a velocity of 5019 \kms\ based on spectroscopic measurements by \citet{2000AJ....120.1579Y}, 
33 \kms\ higher than our \HI\ detections. It is possible, then, that UGC 5293 and 095234.27+412207.1
are interacting companions.

\subsubsection{UGC 5412} 
The intent of this survey was to detect our target, UGC 5410. A photometric redshift of 35,120\kms{} 
was reported subsequently \citep[e.g.,][]{2017ApJS..233...25A} for our target, well outside our \HI\ search range.
However, the GBT and \nan\ beams also readily covered nearby UGC 5412, whose optical spectroscopic velocity of
9610 \kms\ \citep{2015ApJS..219...12A} corresponds to our \HI\ detection, and the results 
of this spurious detection are given in Tables~\ref{tab:avgresultsdets} and \ref{tab:indresultsdets}. 



\subsubsection{UGC 5613} 
Due to residual RFI centered on its optical velocity of 9656 \kms{} \citep{2015ApJS..217...27A} we are unable 
to draw any conclusions regarding the \HI{} line signal of this galaxy.

\subsubsection{UGC 5983} 
It is highly likely that most, or all, of our measured \HI\ flux is from nearby NGC 3432, 
a much larger galaxy whose center lies $3\farcm4$ from the position of our target, UGC 5983.
The \HI\ profiles we measured towards UGC 5983 with the GBT and NRT are significantly 
different. Both telescope beams include a significant portion of NGC 3432, especially at the GBT.
An Arecibo measurement of NGC 3432 by \citet{hewitt83} shows \VHI{} = 608 \kms, \Wtwenty{} = 248 \kms, 
and \FHI{} = 138 Jy \kms. We measured an \FHI{} of 78 and 32 Jy \kms\ at the GBT and NRT, respectively.
Please note that the \nan{} spectrum of this galaxy is not shown in Figure~\ref{fig:HLdets}

\subsubsection{UGC 6005} 
UGC 6005 was not detected in \HI{} by \citet{schneider90} with an r.m.s. noise level
of 8.7 mJy, which is consistent with the average line flux density level of our detection
of \FHI/\Wfifty{} = 15.5 mJy.


\subsubsection{UGC 6179} 
\begin{figure*}[hbtp] 
\begin{center}
\gridline{\fig{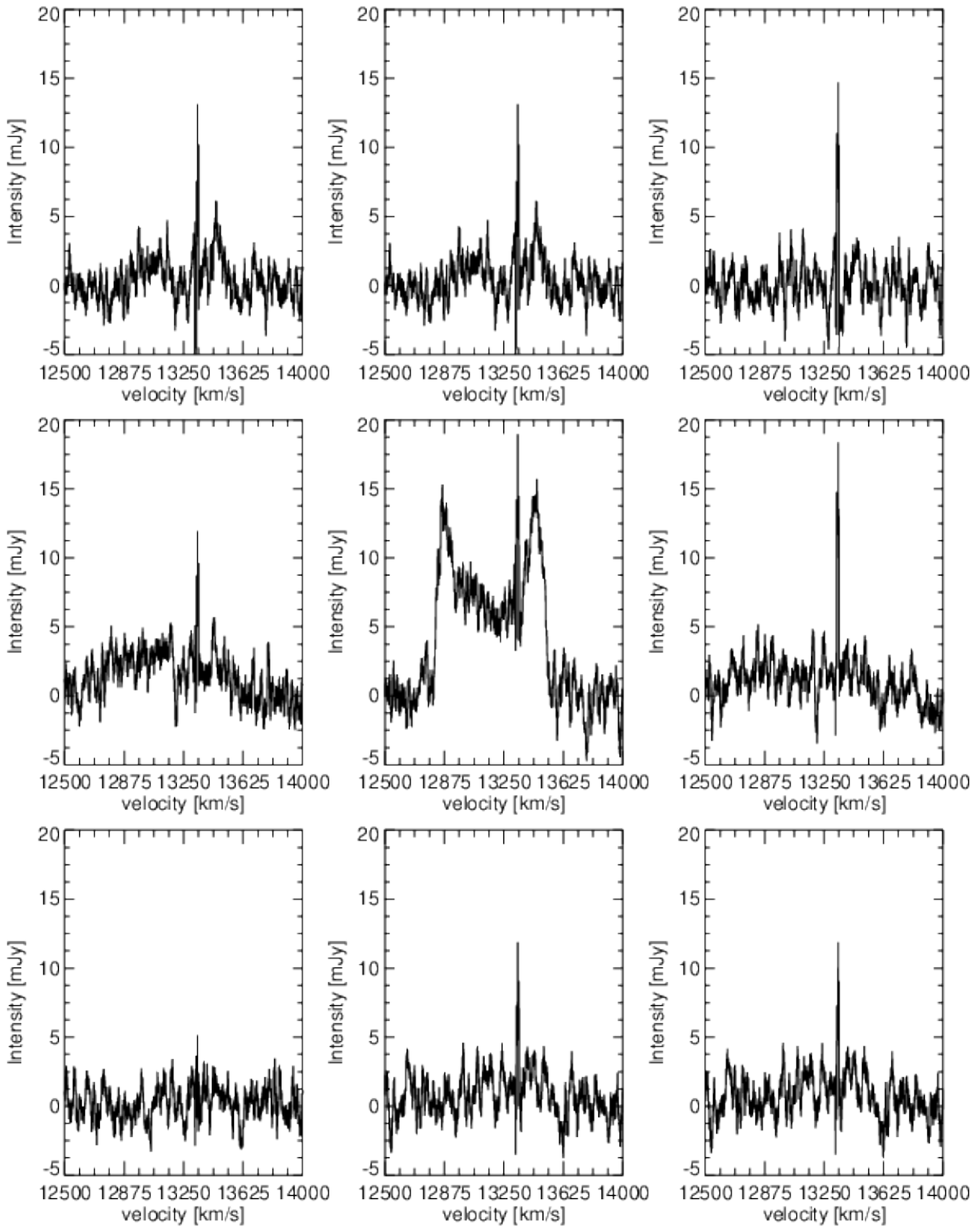}{0.3\textwidth}{(a)}
         \fig{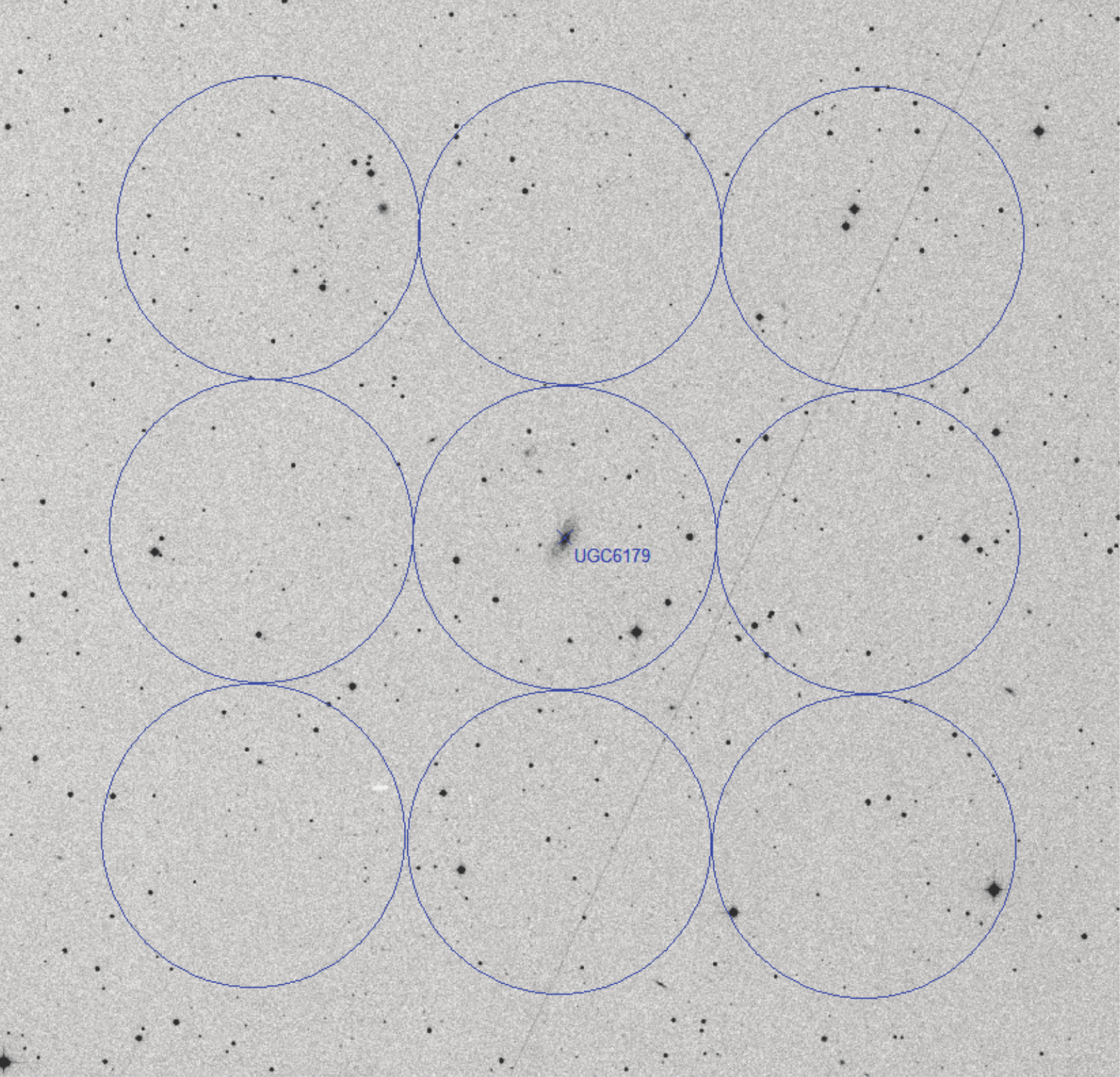}{0.4\textwidth}{(b)}}
\gridline{\fig{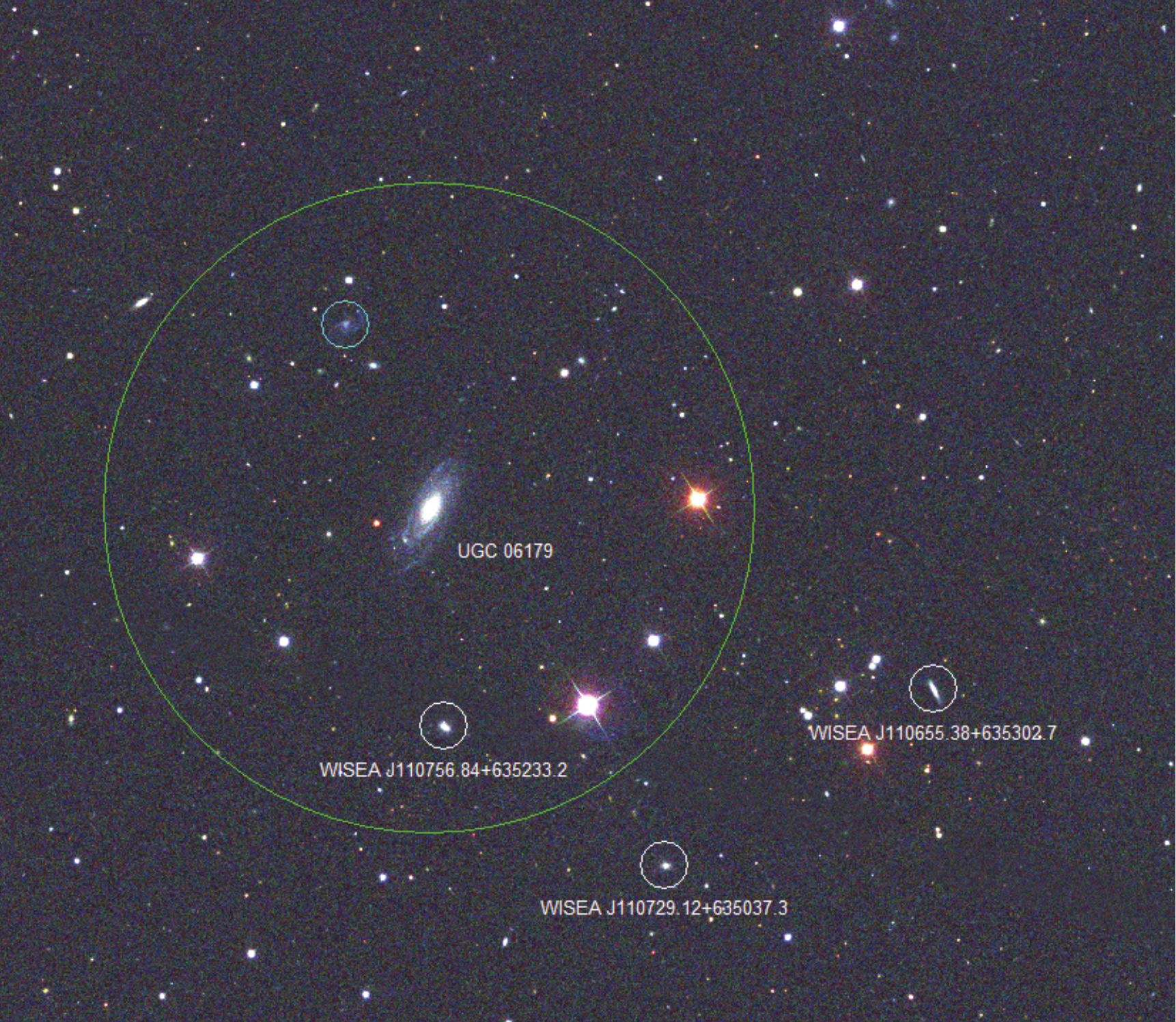}{0.7\textwidth}{(c)}}
\end{center}
\caption{GBT observations made in a grid around UGC 6179, which indicates that the 
measured \HI{} gas lies within the central beam. 
(a): \HI{} spectra measured in each of the the nine $8\farcm7$ GBT beams. Please note 
that the ubiquitous strong, narrow line at $\sim$13,340 \kms{} is due to RFI;
(b): Optical DSS2 Blue \citep{2004AJ....128..502A} image with the nine GBT beams overlaid; 
(c): SDSS DR12 \citep{2015ApJS..219...12A} false color ($i$, $r$, and $g$ filter) image of 
UGC 6179 showing the $8\farcm7$ GBT beam (green circle) and four known neighbors. 
The three galaxies with velocities near that of IC 06170 are denoted by white, 
labelled circles, and the previously unidentified low surface brightness galaxy, 
[OVS23] 110809.3+635806, is denoted by the cyan circle.  \label{fig:U6179_GBT}}
\end{figure*}

The original GBT observation, pointed at the galaxy center, showed a huge profile width (\Wtwenty{} = 913 \kms) 
and total \HI{} mass (log(\MHIMsun{}) = 10.81), and \VHI{} = 13,155 \kms\ in agreement with its optical spectroscopic
velocity of 13,145 \kms\ \citet{2017ApJS..233...25A}. Confirmation observations with the GBT on a grid surrounding the 
galaxy indicate the detected \HI{} mass is confined within the original GBT beam (Figure~\ref{fig:U6179_GBT}). 
A literature search shows only one other galaxy, WISEA J110756.84+635233.2, which lies within the central GBT beam 
and has a velocity similar to that of UGC 6179 (\Vopt{} = 13,173 \kms). However, two more galaxies,
WISEA J110655.38+635302.7 and WISEA J110729.12+635037.3, lie just outside the GBT beam at very similar velocities 
(\Vopt{} = 13,061 \& 13,117 \kms, respectively). Additionally, though, inspection of images from the SDSS 
\citep{2015ApJS..219...12A} indicates there are a number of previously unidentified galaxies within the beam. 
One of these, shown in Figure~\ref{fig:U6179_GBT}, appears to be a blue LSB galaxy which is likely also part
of the galaxy group surrounding UGC 6179. 
As the \nan{} spectrum for this galaxy has RFI in the middle of the measured spectrum, only the GBT data are used 
for determining the line parameters of UGC 6179. 

\subsubsection{UGC 6369} 
There are at least five other galaxies within $11\farcm0$ and 1,000 \kms{} of UGC 6369,
see Table~\ref{tab:UGC6369_neighbors} for a list of its known neighbors. 
It is therefore highly likely that our \nan\ \HI{} profile of UGC 6369 is confused.

\subsubsection{UGC 6489}
UGC 6489 has two nearby neighbors -- LEDA 2198942 at $6\farcm9$ and 335 \kms{} away and 
ASK 347484.0 at $8\farcm4$ and 653 \kms{} away \citep{2004AJ....128..502A, 2014ApJS..210....9B}.
However based both on the \HI{} profile and morphology of UGC 6489 it is likely
that the spectrem is not contaminated.
\subsubsection{UGC 6812}
\citet{2015ApJS..219...12A} gives an optical spectroscopic velocity for UGC 6812 
of 6785 \kms{}, yet there is no \HI{} detected at that velocity to a 1$\sigma$ limit of 1.7 mJy. 
Either the reported velocity is incorrect or UGC 6182 has a very low \HI{} mass.

\subsubsection{UGC 7146} 
The SDSS DR12 \citep{2015ApJS..217...27A} lists an optical velocity of 19,3916 \kms{} for UGC 7146, 
significantly different from earlier SDSS values, e.g. 1060 \kms\ (DR5), and the 1060 \kms\ we found 
in \HI\ at the GBT and NRT. 
However numerous other \HI{} studies have found velocities similar to ours, giving significant
confidence in our results \citep[e.g.][]{2014MNRAS.445..630P, 2013MNRAS.428.1790W, vandriel16}.
Our target UGC 7146, is listed as a member of galaxy group [RPG97] 186 \citep{ramella97}, whose center position lies 
11\arcmin{} and 216 \kms{} from our target, but all 18 objects within 12\arcmin{} distance with 
known redshifts are much more distant (at 20,000-50,000 \kms) than our nearby (1060 \kms) target. 
It therefore seems unlikely that other group members have contaminated our GBT or \nan{} spectra 
of UGC 7146. 

\subsubsection{UGC 7553} 
UGC 7553 has an optical spectroscopic velocity of 8733 \kms{} \citep{2001MNRAS.328.1039C}.
Our \HI{} detection at \nan{}, and another by \citet{matthews00}, both measured 
a 90 \kms{} higher \HI{} velocity of 8823 \kms{}. 
Both \HI{} spectra may  be confused by nearby CGCG 014-041, a \BT\ = 14.8 mag 
galaxy $9\farcm8$ to the North, at \Vopt{} = 8806$\pm$35 \kms, although 
it is not expected to be \HI-rich given its S0 classification.

\subsubsection{UGC 7953} 
We detected UGC 7953 at 17,211 \kms{} with the GBT, whereas the SDSS DR12 \citep{2015ApJS..219...12A}
reported an optical velocity of 1093$\pm$14 \kms. However, as the SDSS classified the spectrum as
"star F9" and it shows no emission lines, it is likely that the reported velocity
is erroneous. Furthermore, our GBT and Arecibo observations did not detect \HI{} at the 1093 \kms{}velocity.


\subsubsection{UGC 8107} 
Using the GBT, \citet{2014MNRAS.443.1044M} measured \Wfifty{} = 747 \kms{} towards UGC 8107, 
significantly larger than the 602$\pm$34 \kms{} we found at \nan{}. Our follow-up GBT observations, 
though, gave an even higher value, of \Wfifty{} = 988\kms{}. The measured \HI{} line fluxes of all three 
measurement are consistent. 
As the signal-to-noise ratios for the two GBT measurements are significantly 
higher than for the \nan{} measurement, the GBT value is more reliable, but the measured error is high.
The average of the measured values if given here.

\subsubsection{UGC 8222}
The detection of UGC 8222 is marginal, with our confidence in the detection due to \cite{2015ApJS..219...12A}'s
earlier determination of the galaxy's spectroscopic velocity.


\subsubsection{UGC 8659} 
Given its much higher \HI{} velocity of 5001 \kms, UGC 8659 is clearly not a member of the M101 
group (mean velocity $\sim$240 \kms), as was considered a possibility by \citet{bremnes99}.
Surface photometry in the $B$ and $R$ bands \citep{bremnes99} shows it has an Im morphology, 
an extrapolated \BT{} = 16.16 mag, a central blue disk surface brightness 23.2 \masq, 
a blue disk scalelength of \as{11}{2} and an increasingly blue color with radius. 


\subsubsection{UGC 8802} 
UGC 8802 has two nearby galaxies, WISEA J135337.14+354117.7 at $6\farcm0$ and 400 \kms{} away from it,
and CGCG 191-005 at $8\farcm1$ and 329 \kms\ away.
However, as both the \nan{} and Arecibo spectra look similar it is 
unlikely that either galaxy has contributed to the measured \HI\ flux of UGC 8802.



\subsubsection{UGC 9783} 
The SDSS optical value of 11,168$\pm$6 \kms{} \citep{2017ApJS..233...25A} of UGC 9783 corresponds exactly 
to the 11,168$\pm$10 \kms{} we measured in our GBT follow-up observations. 
Our \nan\ profile is quite different (see Table~\ref{tab:indresultsdets}), but as it was detected near the edge 
of the NRT bandpass and the rms of our GBT spectrum is five times smaller, only the GBT results 
are reported in Table~\ref{tab:avgresultsdets}.


\subsubsection{UGC 10666, UGC 10668} 
Our Arecibo data show two separated  profiles, centered at 9809 and 10,101 \kms{} respectively. 
Our target galaxy UGC 10666, has a 2MASS photometric redshift of 16,047 \kms{} 
\citep{2014ApJS..210....9B} whereas UGC 10668, at $2\farcm2$ separation, has an optical 
spectroscopic velocity of 10,162$\pm$45 \kms{} \citep{2012ApJS..199...26H}. 
UGC 10668's optical velocity corresponds to the higher-velocity \HI{} detection, although 
as UGC 10668 lies on the edge of the telescope beam its flux is likely underestimated by 
a factor of 2 or so.
It is also likely that the lower measured \HI{} velocity corresponds to UGC 10666.
Our \nan\ observations were not sensitive enough (r.m.s. 3.2 mJy) to detect the profiles.
 


\subsubsection{UGC 11900}
UGC 11900 has one neighbor, LEDA 167572, which lies $4\farcm5$ and 405 
\kms{} away. The standard, double-horned shape of the UGC 11900 \HI{} profile combined with the 
difference in velocities of the two galaxies makes it highly unlikely that its 
\HI{} profile is contaminated by its neighbor.


\clearpage 


\defcitealias{2009MNRAS.399..683J}{6dF}
\defcitealias{2011AA...527A..58B}{Bel11a}
\defcitealias{2011AA...533A..37B}{Bel11b}
\defcitealias{2002AJ....123.2261B}{Ber02}
\defcitealias{2014ApJS..210....9B}{Bil14}
\defcitealias{2009AJ....137.3038B}{Bou09}
\defcitealias{2003AA...406..829B}{Bra03}
\defcitealias{2017ApJ...848L..10B}{Bra17}
\defcitealias{2016MNRAS.460.4492D}{Dab16}
\defcitealias{2001MNRAS.328.1039C}{Col01}
\defcitealias{courtois09}{Cou09}
\defcitealias{2004AJ....128..502A}{DR4} 
\defcitealias{2007ApJS..172..634A}{DR5}
\defcitealias{2015ApJS..217...27A}{DR7}
\defcitealias{2012ApJS..203...21A}{DR9}
\defcitealias{2015ApJS..219...12A}{DR12}
\defcitealias{2017ApJS..233...25A}{DR13} 
\defcitealias{1999PASP..111..438F}{Fal99}
\defcitealias{2017ApJ...847....4G}{Geh17}
\defcitealias{2018ApJ...861...49H}{Hay18}
\defcitealias{2001AA...377..801H}{Huc01}
\defcitealias{2012ApJS..199...26H}{Huc12}
\defcitealias{huchra83}{Huc83}
\defcitealias{huchra99}{Huc99}
\defcitealias{1996ApJS..105..209I}{Imp96}
\defcitealias{2001AJ....122.2341I}{Imp01}
\defcitealias{jones05}{Jon05}
\defcitealias{jones09}{Jon09}
\defcitealias{2013AJ....145..101K}{Kar13}
\defcitealias{1984Afz....21..641K}{Kar84}
\defcitealias{2002Ap.....45..458K}{Kaz02}
\defcitealias{2017ApJ...843...16K}{Kou17}
\defcitealias{2001AstL...27..213M}{Mak01}
\defcitealias{2003AA...405..951M}{Mak03}
\defcitealias{1997AstL...23..445M}{Mak97}
\defcitealias{1996AJ....112.1803M}{Mar96}
\defcitealias{2014MNRAS.443.1044M}{Mas14}
\defcitealias{matthews00}{Mat00}
\defcitealias{1996AJ....111.1098M}{Mat96}
\defcitealias{2004MNRAS.350.1195M}{Mey04} 
\defcitealias{monnier03b}{Mon03}
\defcitealias{1992NED11.R......1N}{NED}
\defcitealias{2003AA...412...57P}{Pat03}
\defcitealias{2008MNRAS.385.1709P}{Pro08}
\defcitealias{2016MNRAS.460..923R}{Ram16}
\defcitealias{2000AJ....120.2338R}{Rin00}
\defcitealias{rines03}{Rin03}
\defcitealias{2016ApJ...819...63R}{Rin16}
\defcitealias{2002AJ....124.1954R}{Rya02}
\defcitealias{2002AA...384..371S}{Sim02}
\defcitealias{2005ApJS..160..149S}{Spr05}
\defcitealias{1984PASP...96..128S}{Ste84}
\defcitealias{2016AJ....151...52S}{Sta16}
\defcitealias{1988LNP...297..361S}{Str88}
\defcitealias{2005AA...430..373T}{The05}
\defcitealias{theureau98}{The98}
\defcitealias{vandriel16}{Van16}         
\defcitealias{2006AA...451..851V}{Ver06}
\defcitealias{2007AJ....134.1849W}{War07}
\defcitealias{1998ApJ...496...39Z}{Zab98}

\startlongtable
\begin{deluxetable*}{lllllllllll}
\tabletypesize{\small}
\tablecaption{Literature values -- optical properties and known velocities for all galaxies.
In the case more than one type of velocity is given, the references are listed in order. 
In the case more than one velocity of a given type is found in the literature, the most
recent value is listed.  \label{tab:opticalprops}}
\tablehead{
\colhead{Name} & \colhead{RA} & \colhead{Dec} & \colhead{Morph.} & \colhead{\BT{}} & \colhead{D25} & \colhead{b/a} & \colhead{\Vopt} & \colhead{Opt. ref.} & \colhead{\VHI} & \colhead{\HI\ ref.}\\ 
\colhead{} & \colhead{(J2000.0)} & \colhead{(J2000.0)} & \colhead{} & \colhead{(mag)} & \colhead{(\arcmin{})} & \colhead{} & \colhead{(\kms)} & \colhead{} & \colhead{(\kms)} & \colhead{} \\
}
\startdata
\multicolumn{11}{l}{\bf \HIbf{} Detections} \\
ESO 300-020 & 03 14 07.0 & -37 59 29 &  (dIAB(s)m?) & 17.27 &  0.50 &  0.54 &         &                                  &         &                                         \\
ESO 338-020 & 19 41 17.0 & -38 12 49 &          Sbc & 17.08 &  0.62 &  0.45 &         &                                  &         &                                         \\
ESO 346-034 & 23 10 00.0 & -37 20 01 &      (dwarf) & 16.42 &  0.69 &  0.49 &         &                                  &         &                                         \\
ESO 359-009 & 03 54 59.0 & -35 47 21 &          Sbc & 16.52 &  0.79 &  0.48 &         &                                  &         &                                         \\
ESO 368-004 & 07 32 54.0 & -35 29 19 &            I & 16.49 &  0.91 &  0.63 &         &                                  &    1397 & \citetalias{2016AJ....151...52S}        \\
ESO 371-004 & 08 40 14.0 & -32 55 49 &          Sbc & 16.18 &  0.56 &  0.49 &         &                                  &         &                                         \\
ESO 371-007 & 08 41 16.0 & -33 15 53 &          Sbc & 16.09 &  0.42 &  0.56 &         &                                  &         &                                         \\
\enddata
\end{deluxetable*}

\clearpage

\startlongtable
\begin{deluxetable*}{lcccllllllllcll}
\tabletypesize{\footnotesize}
\tablecaption{\HI{} results: Average \HI{} values for all detections \label{tab:avgresultsdets}}
\tablehead{
\colhead{Name} &  \colhead{Sample} & \colhead{Notes} & \colhead{Conf.} & \colhead{\VHI{}} & \colhead{error} & \colhead{\Wfifty{}} & \colhead{\Wfiftycor{}} &
\colhead{\Wtwenty{}} & 
\colhead{r.m.s.} & \colhead{flux} & \colhead{error} & \colhead{log(\MHI{}/\Msun)} & \colhead{error} & \colhead{Tel.}\\ 
\colhead{}     & \colhead{}      &\colhead{}      & \colhead{}     & \multicolumn{3}{c}{(\kms{})}&  \multicolumn{2}{c}{(\kms{})} & 
\colhead{mJy}   & \multicolumn{2}{c}{(Jy \kms{})}  & \colhead{}  & \colhead{} & \colhead{} 
}
\startdata
        CGMW1-0409 &    HLS &  Y &  1 &   2805 &      1 &    132 &      177 &    147 &      1.70 &      8.59 &      0.19 &      9.51 &      0.01 &        G\\
       ESO 300-020 &     HL &  N &  1 &  13256 &      5 &     81 &      110 &     85 &      1.46 &      0.22 &      0.10 &      9.27 &      0.20 &        G\\
       ESO 338-020 &     HL &  Y &  3 &   6377 &      6 &    379 &      438 &    396 &      4.11 &      3.88 &      0.76 &      9.88 &      0.09 &        N\\
       ESO 346-034 &     HL &  N &  1 &   1805 &     14 &     58 &       80 &    107 &      1.60 &      0.89 &      0.16 &      8.12 &      0.08 &        G\\
       ESO 359-009 &     HL &  N &  1 &   5850 &      7 &    221 &      265 &    247 &      4.08 &      3.24 &      0.44 &      9.72 &      0.06 &        N\\
       ESO 368-004 &     HL &  Y &  1 &   1397 &      2 &     77 &      114 &     95 &      4.08 &     13.98 &      0.12 &      9.07 &      0.00 &      NGG\\
       ESO 371-004 &     HL &  N &  1 &   1393 &      4 &     74 &       98 &    102 &      4.11 &      3.18 &      0.29 &      8.43 &      0.04 &        N\\
       ESO 371-007 &     HL &  N &  1 &   2454 &      3 &    132 &      173 &    150 &      4.16 &      3.28 &      0.35 &      8.98 &      0.05 &        N\\
\enddata
\end{deluxetable*}

\newpage 

\startlongtable
\begin{deluxetable*}{lcccccccccccccc}
\tabletypesize{\footnotesize}
\tablecaption{Results: \HI{} detections, values for individual telescopes when more than one telescope observed a galaxy (HyperLeda sample) \label{tab:indresultsdets}}
\tablehead{
\colhead{Name} & \colhead{Notes} & \colhead{Conf.} & \colhead{\VHI{}} & \colhead{error} & \colhead{\Wfifty{}} & \colhead{error} &
\colhead{\Wtwenty{}} & \colhead{error} &
\colhead{r.m.s.} & \colhead{flux} & \colhead{error} & \colhead{\MHI{}/\Msun{}} & \colhead{error} & \colhead{Tel.}\\ 
\colhead{}     & \colhead{}      & \colhead{}     & \multicolumn{2}{c}{(\kms{})}&  \multicolumn{2}{c}{(\kms{})} & 
 \multicolumn{2}{c}{(\kms{})} &  \colhead{Jy}   & \multicolumn{2}{c}{(Jy \kms{})}  & \colhead{}  & \colhead{} & \colhead{} 
}
\startdata
ESO 368-004 &  Y &  1 &   1396 &      1 &     58 &      2 &     78 &      3 &      3.78 &     12.91 &      0.23 &      9.04 &      0.01 &        N\\
ESO 368-004 &  Y &  1 &   1397 &      1 &     86 &      1 &    101 &      1 &      1.50 &     14.51 &      0.14 &      9.09 &      0.00 &        G\\
ESO 368-004 &  Y &  1 &   1397 &      5 &     89 &     10 &    108 &     15 &      1.30 &     13.45 &      0.11 &      9.06 &      0.00 &        G\\
ESO 424-009 &  N &  1 &   4057 &      6 &    140 &     12 &    174 &     18 &      3.70 &      3.27 &      0.34 &      9.40 &      0.05 &        N\\
ESO 424-009 &  N &  1 &   4052 &      1 &    153 &      2 &    165 &      3 &      1.00 &      3.22 &      0.13 &      9.39 &      0.02 &        G\\
ESO 463-005 &  Y &  2 &   2809 &      3 &    519 &      6 &    552 &      9 &      2.80 &     11.70 &      0.46 &      9.64 &      0.02 &        N\\
ESO 463-005 &  Y &  2 &   3014 &     10 &    159 &     15 &    191 &     15 &      1.41 &      3.26 &      0.17 &      9.15 &      0.02 &        G\\
ESO 491-003 &  Y &  1 &   7166 &     11 &    154 &     22 &    221 &     34 &      2.77 &      2.22 &      0.26 &      9.73 &      0.05 &        N\\
\enddata
\end{deluxetable*}
\clearpage 
\clearpage

\startlongtable
\begin{deluxetable*}{lcccccc}
\tabletypesize{\footnotesize}
\tablecaption{\HI{} r.m.s. values in mJy for all non-detections.  \label{tab:resultsnondets}}
\tablehead{
\colhead{Name} & \colhead{Sample} &\colhead{Notes} & \colhead{\nan} & \colhead{GBT} & \multicolumn{2}{c}{Arecibo}\\
\colhead{}     & \colhead{}      &\colhead{}      &  \colhead{}& \colhead{} & \colhead{low} & \colhead{high} 
}
\startdata
ESO 366-014 &     HL &  N &      2.61 &      1.11 &           &         \\
ESO 435-001 &     HL &  N &           &      0.95 &           &         \\
ESO 491-002 &     HL &  Y &      1.69 &           &           &         \\
ESO 527-015 &     HL &  N &      3.18 &           &           &         \\
ESO 540-030 &     HL &  Y &      2.58 &      1.84 &           &         \\
ESO 582-010 &     HL &  N &      1.62 &           &           &         \\
ESO 598-019 &     HL &  N &      1.45 &           &           &         \\
PGC 4620    &     HL &  N &      2.08 &           &           &         \\
\enddata
\end{deluxetable*}
\clearpage

\begin{deluxetable}{lcccccl} 
\tablecaption{Galaxies with large differences between \HI{} properties measured by more than one telescope.\label{tab:outliers}}
\tablewidth{0pt}
\tablehead{
\colhead{Object} & \colhead{Confusion} &
\colhead{$\Delta$\Vhel{}} & \colhead{$\Delta$\Wfifty{}} & 
\colhead{$\Delta$\FHI{}} & \colhead{Telescopes} & \colhead{Notes}\\
\colhead{} &\colhead{(1-3)} &
\colhead{(\kms))}  & \colhead{(\kms)} & \colhead{(Jy \kms)} & \colhead{} &\colhead{}
}
\startdata
ESO 463-005 & 2 & {\bf -205} & {\bf 360}& {\bf 8.44}  & \nan{}-GBT & \nan{} obs. confused\\
ESO 492-001 & 1 &  9, 2 & {\bf -84, -82} & -1.0, 0.0 & \nan{}-GBT, \nan{}-GBT$_2$ & \nan{} obs. low S/N\\ 
IC 3852     & 1  &  14, 18  &   {\bf 3, 108} & {\bf 0.6, 0.5}  & GBT-AO, GBT$_2$-AO & \\
PGC 16370   & 1 & {\bf -198}& -49 & 2.0  & \nan{}-GBT$_2$ &\nan{} obs.low S/N\\ 
PGC 17124   & 2 & 3 & -25 & {\bf 7.4} & \nan{}-GBT$_2$ &\nan{} obs. confused\\ 
PGC 21133   & 1 & -30 & {\bf -92} & -0.5 & \nan{}-GBT & \nan{} obs have high error\\ 
PGC 21529   & 2 & 8, {\bf -60, -68} & {\bf -290, -102, 188} & -0.9, {\bf 2.8, 3.7} & \nan{}-GBT, \nan{}-AO, GBT-AO &  \nan{} \& GBT obs. confused\\
PGC 27849   & 1 & 16, 4, -12  &{\bf -160, -115}, -24 & {\bf -1.4}, 0.2, -0.5 & \nan{}-GBT$_2$, \nan{}-GBT, \nan{}-AO  &\nan{} obs. confused \\
\enddata
\end{deluxetable}

\begin{deluxetable}{lccl}  
\tablecaption{Galaxies with $|$\VHI{} - \Vopt{}$| >$100 \kms{}.
\label{tab:voptdif}}
\tablewidth{0pt}
\tablehead{
\colhead{Object} & \colhead{\VHI} & \colhead{\Vopt} & \colhead{Note}\\
\colhead{}v& \colhead{(\kms))}  & \colhead{(\kms)} & \colhead{}
}
\startdata
PGC 7225  &   5193 &   5071$\pm$80  & 1 \\ 
UGC 263   & 16,853 & 16,718$\pm$105 & 2  \\ 
UGC 1145  &   5809 &   5676$\pm$13  & 3 \\ 
\enddata
\tablecomments{
1. Our \HI{} results are confirmed by two telescopes - see Appendix~\ref{Sec:IndGals}.
2. \Vopt{} has a $\pm$100 \kms{} error, putting it close to our measured value.
3. Peak \VHI{} value confirmed by two telescopes - see Appendix~\ref{Sec:IndGals}.
}
\end{deluxetable}


\begin{deluxetable*}{lcccc} 
\tablecaption{All galaxies with known redshifts within the GBT beam when pointed at PGC 21907.  \label{tab:P21907_neighbors}}
\tablewidth{0pt}
\tablehead{
\colhead{Object} & \colhead{Velocity} & \colhead{GBT Det?} & \colhead{GBT r.m.s.} & \colhead{Vel. Ref.} \\
\colhead{} &\colhead{(km s$^{-1}$)} & \colhead{(Y/N)}  & \colhead{(mJy)} & \colhead{}
}
\startdata
SDSS J074933.51+394424.3  &   2943           & Y & 3.2 & 5,7,8,9,10\\
KUG 0746+398A             &  12,420          & Y & 2.2 & 2 \\
PGC 21907                 &  22,210$\pm$4500 & N & 1.8 & 1 \\
SDSS J074939.55+394316.7  &  12,536          & N & 2.2 & 3,4,5,6\\
WISEA J074940.39+394604.7 &  28,603          & N & N/A & 2,5,6,10\\
WISEA J074945.78+394354.4 &  46,750          & N & N/A & 5,6,10\\
WISEA J074949.97+393945.9 & 151,577          & N & N/A & 10\\
WISEA J074951.56+393934.9 & 150,496          & N & N/A & 10\\
\enddata
\tablecomments{Velocity references:
$^1$\citet{2006ApJS..167....1B}; $^2$\citet{2014ApJS..210....9B}; $^3$\citet{2016ApJS..224...38A}; $^4$\citet{2015MNRAS.448.3442G}; $^5$\citet{2004AJ....128..502A}; $^6$\citet{2003AJ....126.2081A}; $^7$\citet{2017ApJ...843...16K}; $^8$\citet{2015ApJS..217...27A}; 
$^9$\citet{vandriel16}; $^{10}$\citet{2015ApJS..217...27A} 
}
\end{deluxetable*}

\begin{deluxetable}{lcc} 
\tablecaption{Galaxies in the PGC 23328 group.  \label{tab:P23328_group}}
\tablewidth{0pt}
\tablehead{
\colhead{Object} & \colhead{Velocity} & \colhead{Vel. Ref.} \\
\colhead{} &  \colhead{(km s$^{-1}$)} & \colhead{}
}
\startdata
PGC 23328                 & 5829 & 1,2,3\\
WISEA J081923.48+252621.4 & 5770 & 1,2,3 \\
WISEA J081955.30+252733.0 & 6006 & 1 \\
KUG 0816+256B             & 7316 & 1,2,3 \\
\enddata
\tablecomments{Velocity references:
$^1$\citet{2015ApJS..217...27A}; $^2$\citet{2018ApJ...861...49H};
$^3$\citet{2011AJ....142..170H}
}
\end{deluxetable}

\begin{deluxetable}{lcc} 
\tablecaption{Galaxies in the PGC 30113 group.  \label{tab:P30113_group}}
\tablewidth{0pt}
\tablehead{
\colhead{Object} & \colhead{Velocity} & \colhead{Vel. Ref.} \\
\colhead{} &  \colhead{(km s$^{-1}$)} & \colhead{}
}
\startdata
PGC 30113                 & 13,491 & 1,2,3,4, 5\\
2MASX J10183425+5352576   & 13,607 & 1,2,4,5 \\
WISEA J101844.05+534359.6 & 14,159 & 1,4,5 \\
WISEA J101937.54+534202.5 & 14,169 &  1,4,5 \\
WISEA J101955.44+534201.9 & 14,269 & 1,2,5 \\
2MASX J10190350+5334445   & 15,137 & 1,4,5 \\
\enddata
\tablecomments{\raggedright Velocity references:\\
$^1$\citet{2015ApJS..217...27A}; $^2$\citet{2015MNRAS.448.3442G};\\
$^3$\citet{2018MNRAS.473.4731K}; $^4$\citet{2014ApJS..210....9B}; \\
$^5$\citet{2004AJ....128..502A}
}
\end{deluxetable}

\begin{deluxetable}{lc} 
\tablecaption{Galaxies in the PGC 32862 group.  \label{tab:P32862_group}}
\tablewidth{0pt}
\tablehead{
\colhead{Object} & \colhead{Velocity}  \\
\colhead{} &  \colhead{(km s$^{-1}$)}  
}
\startdata
PGC 32862                 & 13,973 \\ 
SDSS J105505.99+573901.8  & 13,839  \\  
WISEA J105636.94+573449.3 & 13,968  \\ 
WISEA J105649.85+572445.4 & 14,031  \\ 
\enddata
\tablecomments{Velocity reference: \citet{2017ApJS..233...25A}}
\end{deluxetable}

\begin{deluxetable}{lcc} 
\tablecaption{Galaxies likely misidentified as separate from PGC 51872.  \label{tab:P51872_neighbors}}
\tablewidth{0pt}
\tablehead{
\colhead{Object} & \colhead{Velocity} & \colhead{Vel. Ref.} \\
\colhead{} &  \colhead{(km s$^{-1}$)} & \colhead{}
}
\startdata
GALEXMSC J143103.29+353112.0 & 3897 & 1\\
WISEA J143104.15+353132.1    & 3897 & 2\\
SDSS J143103.66+353115.8     & 4067 & 3\\
SDSS J143102.07+353058.6     & 4117 & 4 \\
SSTSL2 J143101.52+353105.9   & 3298 & 5\\
SDSS J143103.06+353049.7     & 4179 & 5\\
SDWFS J143101.34+353104.7    & 4197 & 1\\
\enddata
\tablecomments{Velocity references:
$^1$\citet{2010ApJ...716..530K}; $^2$\citet{2007ApJS..169...21C};
$^3$\citet{2007ApJ...671..153Y}; $^4$\citet{2008ApJ...675.1459K}; 
$^5$\citet{2012ApJS..200....8K}
}
\end{deluxetable}

\begin{deluxetable}{lcc} 
\tablecaption{Galaxies likely misidentified as separate from UGC 6369.  \label{tab:UGC6369_neighbors}}
\tablewidth{0pt}
\tablehead{
\colhead{Object} & \colhead{Velocity} & \colhead{Vel. Ref.} \\
\colhead{} &  \colhead{(km s$^{-1}$)} & \colhead{}
}
\startdata 
MCG +10-16-124 &  1774 & 1 \\
NGC 3625       &  1944 & 2 \\
UGC 6344       &  1952 & 3 \\
PGC1 0034566   &  2045 & 1 \\
SBS 1118+578B  &  2173 & 4 \\
\enddata
\tablecomments{Velocity references:
$^1$\citet{2017ApJS..233...25A}; $^2$\citet{2005ApJS..160..149S};
$^3$\citet{2015MNRAS.447.1531C}; $^4$\citet{2014ApJS..210....9B}
}
\end{deluxetable}
%

\clearpage 


\figsetstart
\figsetnum{1}
\figsettitle{\HI{} Spectra}
\figsetgrpstart
\figsetgrpnum{HISpectra.1}
\figsetgrptitle{HI Spectra, Image 1}
\figsetplot{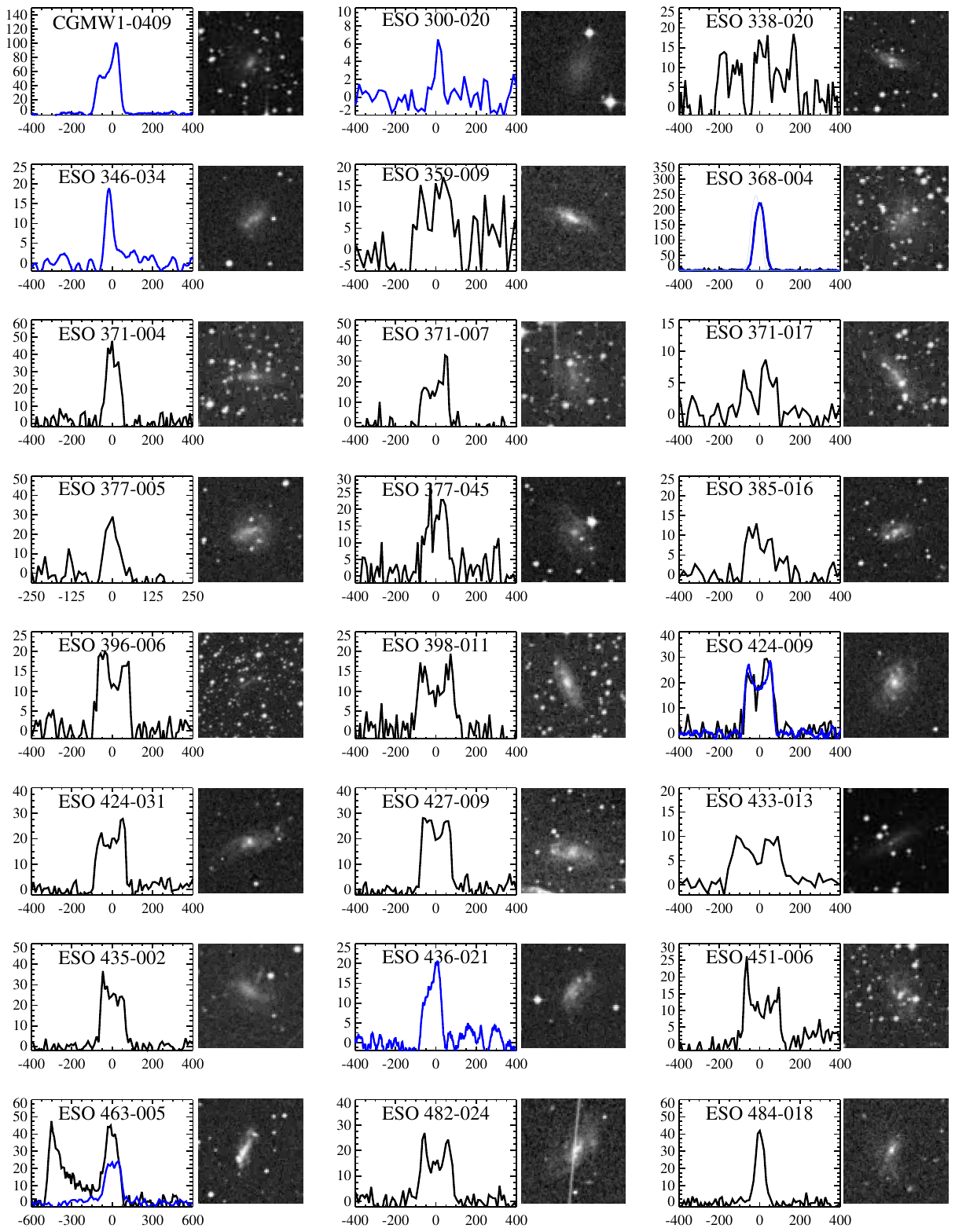}
\figsetgrpnote{21 cm \HI{} line spectra and optical images of the detected galaxies. 
Black lines in the spectra represent \nan{} data, blue lines GBT data, and red 
lines Arecibo data. In the case where an object was observed twice by the same telescope, 
the earliest observation is shown in by a dashed line.
Optical images are 2$\times$2 arcmin in size. False color ($i$, $r$, and $g$ filter) images are 
from the SDSS DR12 \citep{2015ApJS..219...12A}, and black and white images are from the 
2$^{nd}$ Digitized Sky Survey (DSS2) Blue plates \citep{2004AJ....128..502A}, 
used when SDSS images are not available. Objects are arranged in alphabetical order.}
\figsetgrpend
\figsetgrpstart
\figsetgrpnum{HISpectra.2}
\figsetgrptitle{HI Spectra, Image 2}
\figsetplot{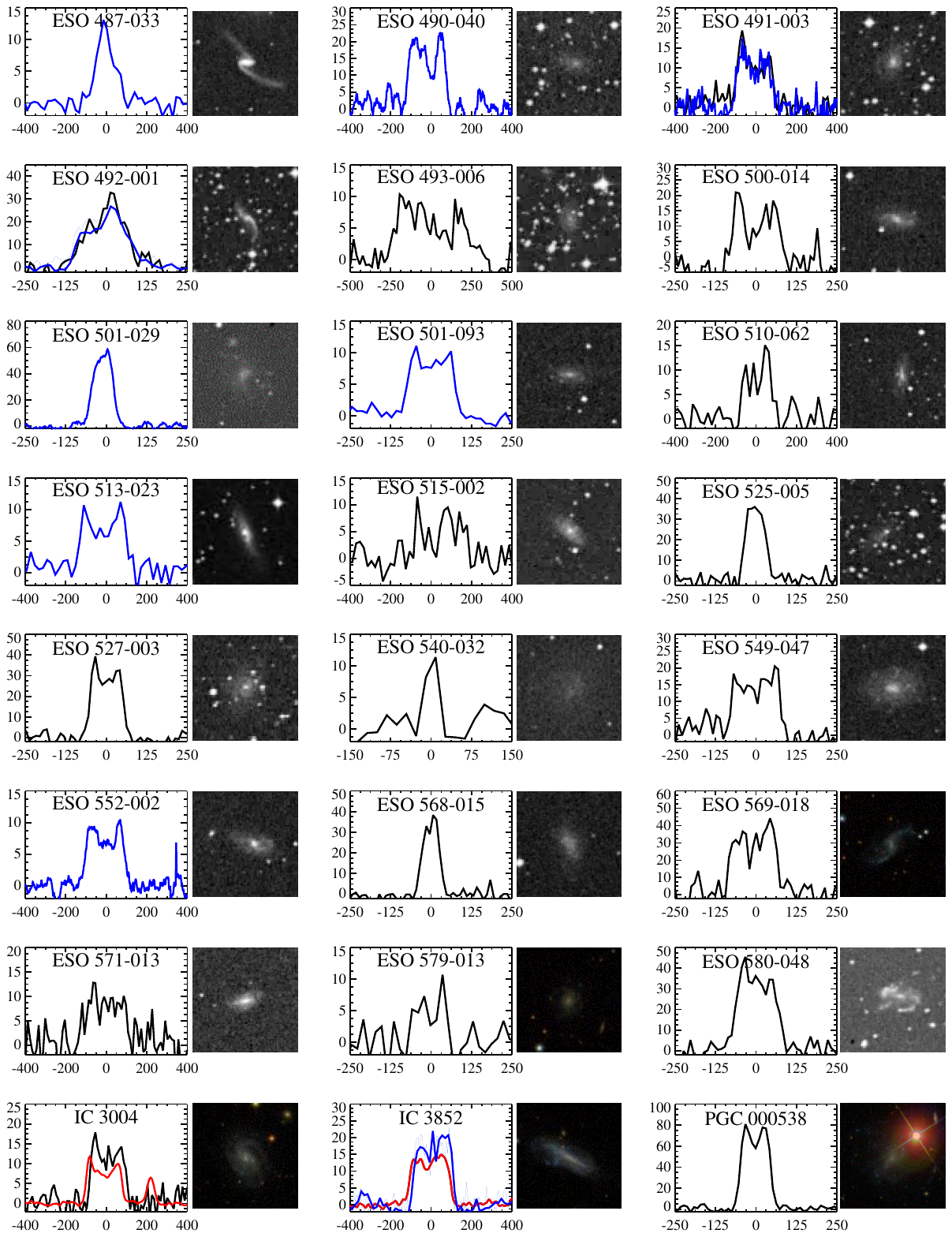}
\figsetgrpnote{21 cm \HI{} line spectra and optical images of the detected galaxies. 
Black lines in the spectra represent \nan{} data, blue lines GBT data, and red 
lines Arecibo data. In the case where an object was observed twice by the same telescope, 
the earliest observation is shown in by a dashed line.
Optical images are 2$\times$2 arcmin in size. False color ($i$, $r$, and $g$ filter) images are 
from the SDSS DR12 \citep{2015ApJS..219...12A}, and black and white images are from the 
2$^{nd}$ Digitized Sky Survey (DSS2) Blue plates \citep{2004AJ....128..502A}, 
used when SDSS images are not available. Objects are arranged in alphabetical order.}
\figsetgrpend
\figsetgrpstart
\figsetgrpnum{HISpectra.3}
\figsetgrptitle{HI Spectra, Image 3}
\figsetplot{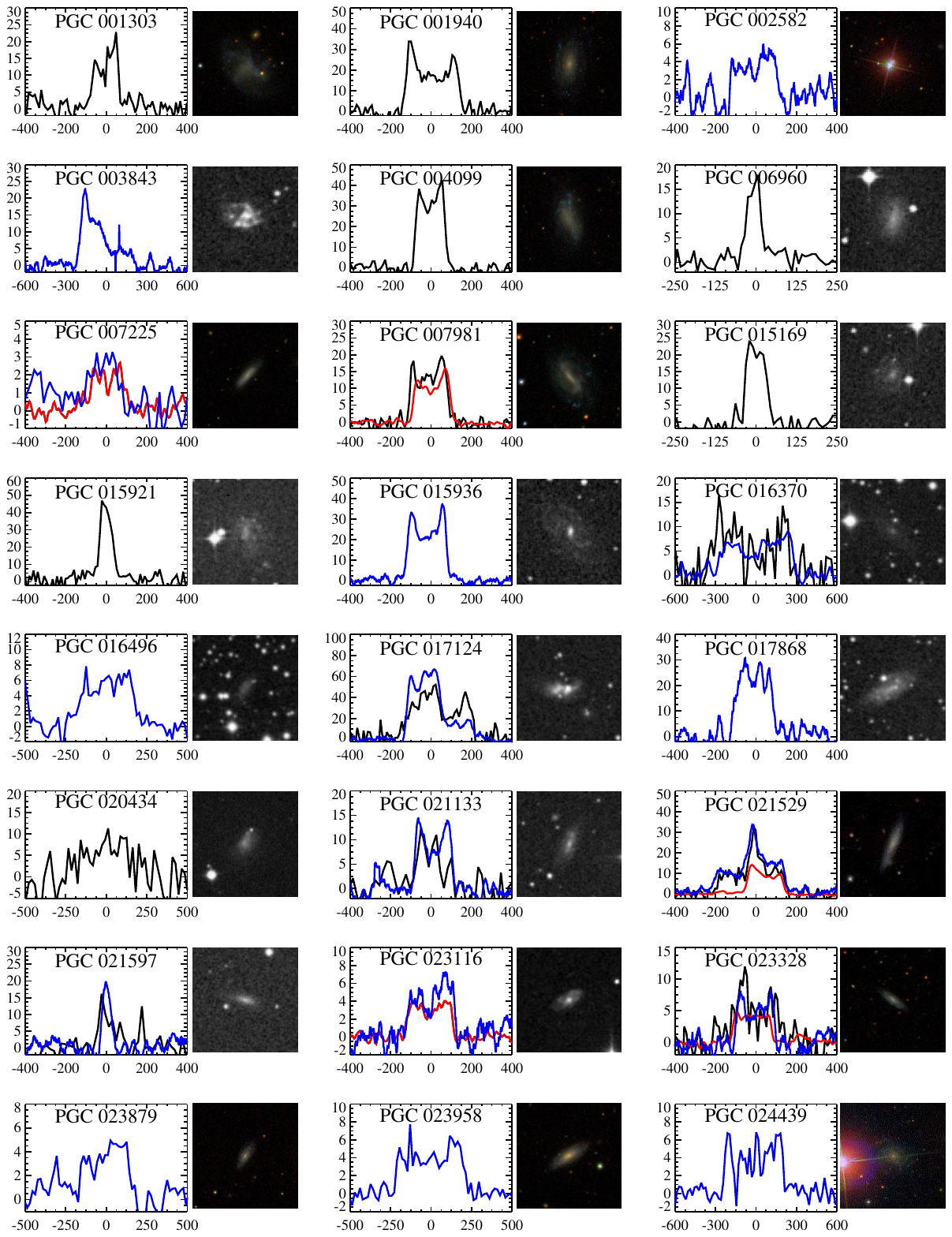}
\figsetgrpnote{21 cm \HI{} line spectra and optical images of the detected galaxies. 
Black lines in the spectra represent \nan{} data, blue lines GBT data, and red 
lines Arecibo data. In the case where an object was observed twice by the same telescope, 
the earliest observation is shown in by a dashed line.
Optical images are 2$\times$2 arcmin in size. False color ($i$, $r$, and $g$ filter) images are 
from the SDSS DR12 \citep{2015ApJS..219...12A}, and black and white images are from the 
2$^{nd}$ Digitized Sky Survey (DSS2) Blue plates \citep{2004AJ....128..502A}, 
used when SDSS images are not available. Objects are arranged in alphabetical order.}
\figsetgrpend
\figsetgrpstart
\figsetgrpnum{HISpectra.4}
\figsetgrptitle{HI Spectra, Image 4}
\figsetplot{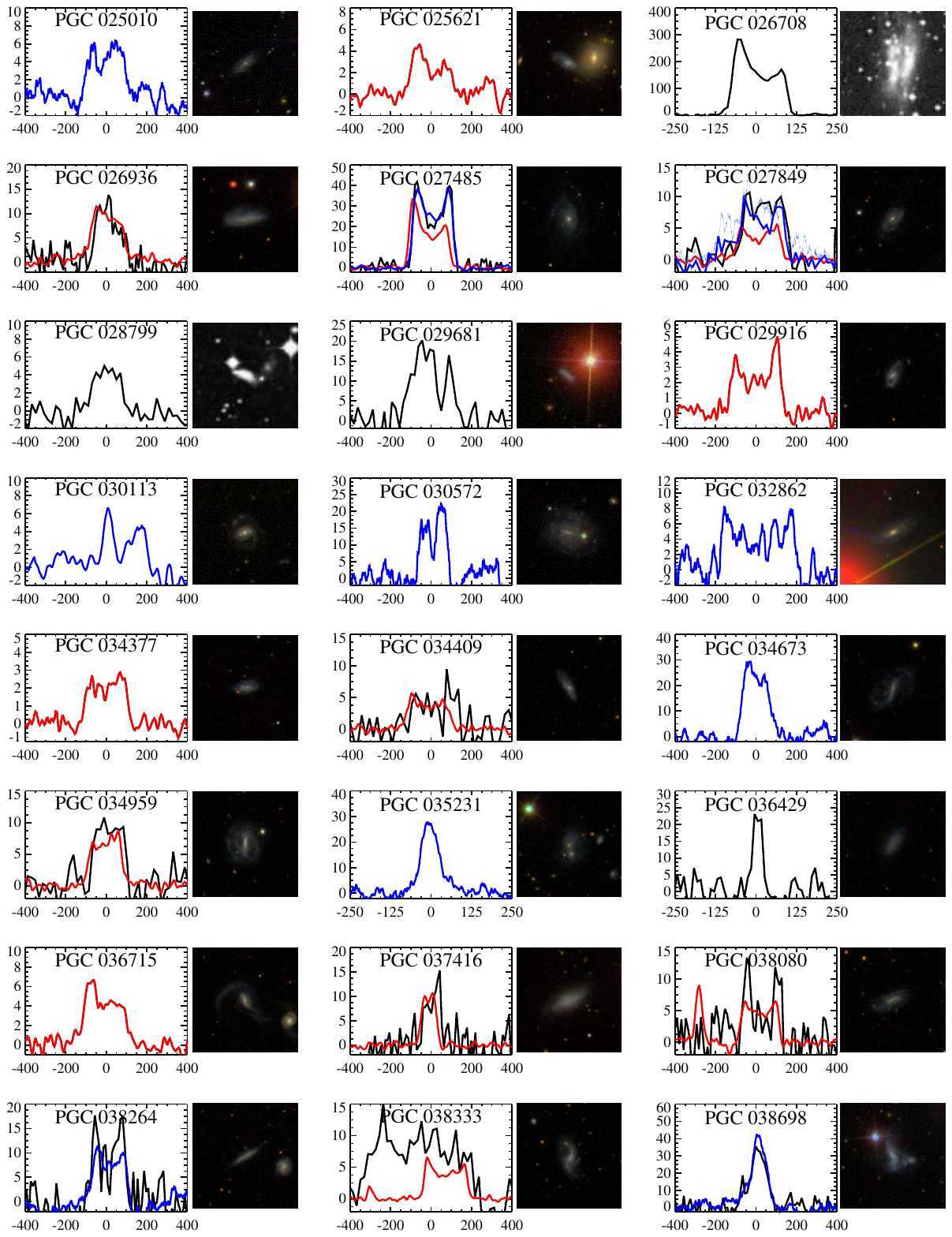}
\figsetgrpnote{21 cm \HI{} line spectra and optical images of the detected galaxies. 
Black lines in the spectra represent \nan{} data, blue lines GBT data, and red 
lines Arecibo data. In the case where an object was observed twice by the same telescope, 
the earliest observation is shown in by a dashed line.
Optical images are 2$\times$2 arcmin in size. False color ($i$, $r$, and $g$ filter) images are 
from the SDSS DR12 \citep{2015ApJS..219...12A}, and black and white images are from the 
2$^{nd}$ Digitized Sky Survey (DSS2) Blue plates \citep{2004AJ....128..502A}, 
used when SDSS images are not available. Objects are arranged in alphabetical order.}
\figsetgrpend
\figsetgrpstart
\figsetgrpnum{HISpectra.5}
\figsetgrptitle{HI Spectra, Image 5}
\figsetplot{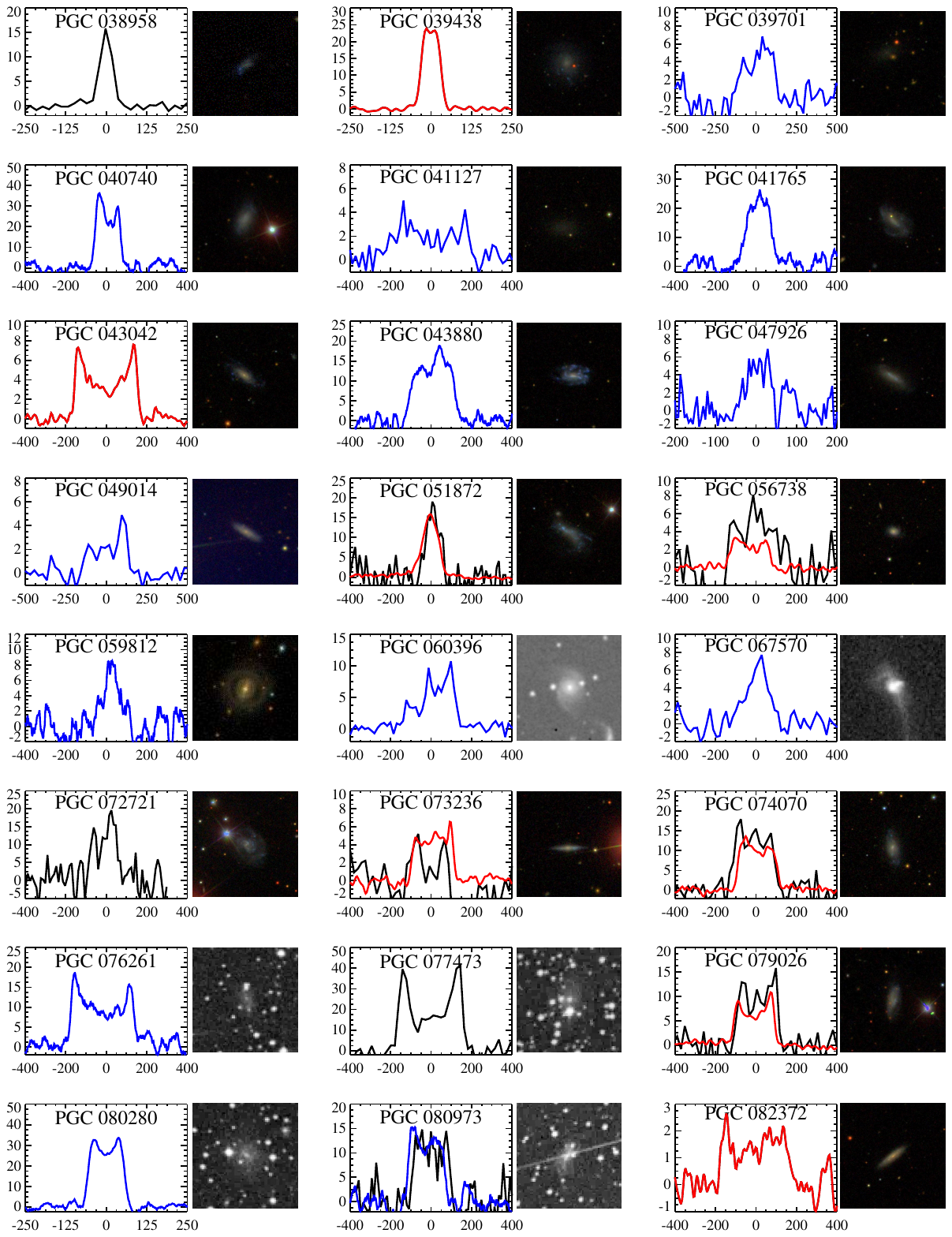}
\figsetgrpnote{21 cm \HI{} line spectra and optical images of the detected galaxies. 
Black lines in the spectra represent \nan{} data, blue lines GBT data, and red 
lines Arecibo data. In the case where an object was observed twice by the same telescope, 
the earliest observation is shown in by a dashed line.
Optical images are 2$\times$2 arcmin in size. False color ($i$, $r$, and $g$ filter) images are 
from the SDSS DR12 \citep{2015ApJS..219...12A}, and black and white images are from the 
2$^{nd}$ Digitized Sky Survey (DSS2) Blue plates \citep{2004AJ....128..502A}, 
used when SDSS images are not available. Objects are arranged in alphabetical order.}
\figsetgrpend
\figsetgrpstart
\figsetgrpnum{HISpectra.6}
\figsetgrptitle{HI Spectra, Image 6}
\figsetplot{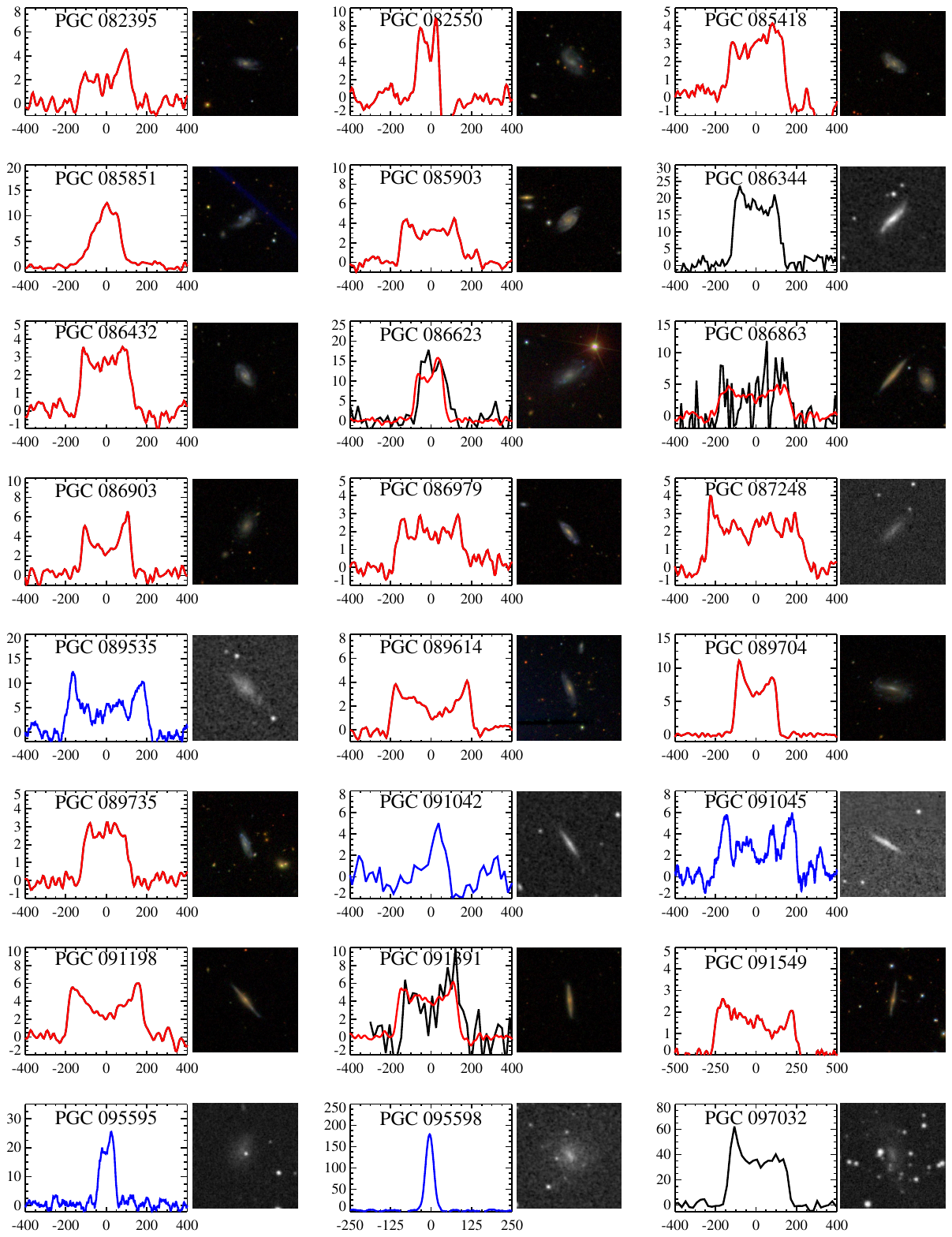}
\figsetgrpnote{21 cm \HI{} line spectra and optical images of the detected galaxies. 
Black lines in the spectra represent \nan{} data, blue lines GBT data, and red 
lines Arecibo data. In the case where an object was observed twice by the same telescope, 
the earliest observation is shown in by a dashed line.
Optical images are 2$\times$2 arcmin in size. False color ($i$, $r$, and $g$ filter) images are 
from the SDSS DR12 \citep{2015ApJS..219...12A}, and black and white images are from the 
2$^{nd}$ Digitized Sky Survey (DSS2) Blue plates \citep{2004AJ....128..502A}, 
used when SDSS images are not available. Objects are arranged in alphabetical order.}
\figsetgrpend
\figsetgrpstart
\figsetgrpnum{HISpectra.7}
\figsetgrptitle{HI Spectra, Image 7}
\figsetplot{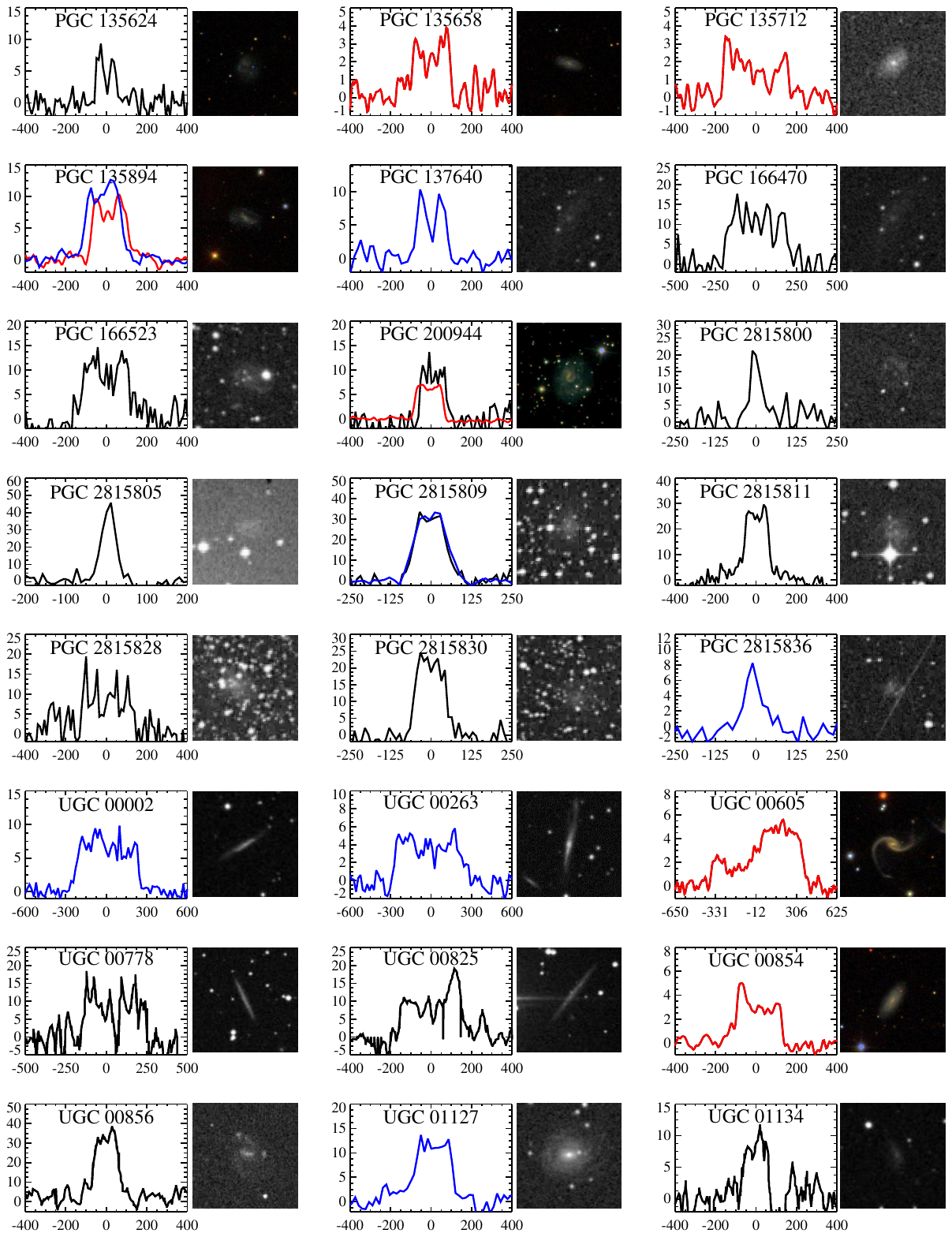}
\figsetgrpnote{21 cm \HI{} line spectra and optical images of the detected galaxies. 
Black lines in the spectra represent \nan{} data, blue lines GBT data, and red 
lines Arecibo data. In the case where an object was observed twice by the same telescope, 
the earliest observation is shown in by a dashed line.
Optical images are 2$\times$2 arcmin in size. False color ($i$, $r$, and $g$ filter) images are 
from the SDSS DR12 \citep{2015ApJS..219...12A}, and black and white images are from the 
2$^{nd}$ Digitized Sky Survey (DSS2) Blue plates \citep{2004AJ....128..502A}, 
used when SDSS images are not available. Objects are arranged in alphabetical order.}
\figsetgrpend
\figsetgrpstart
\figsetgrpnum{HISpectra.8}
\figsetgrptitle{HI Spectra, Image 8}
\figsetplot{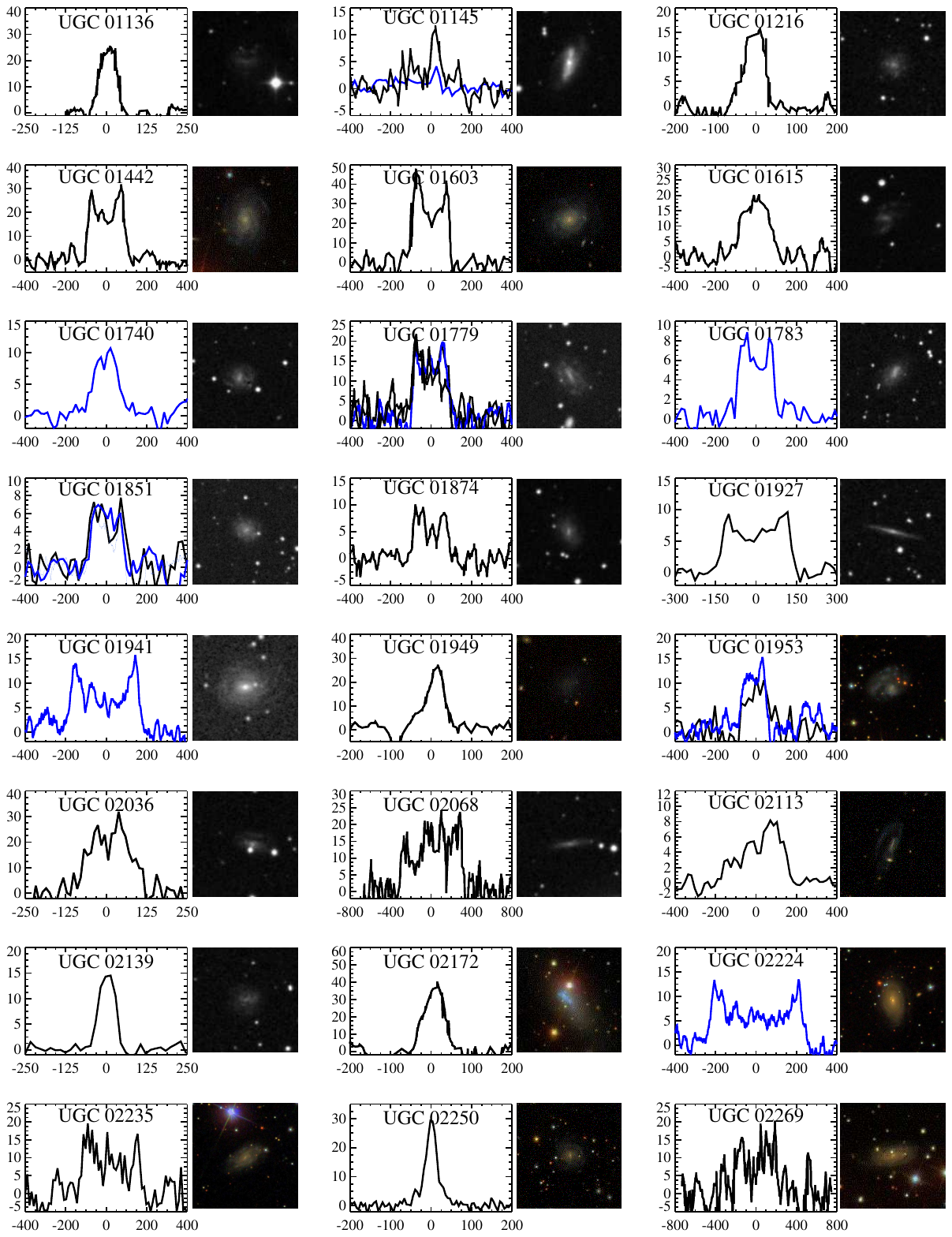}
\figsetgrpnote{21 cm \HI{} line spectra and optical images of the detected galaxies. 
Black lines in the spectra represent \nan{} data, blue lines GBT data, and red 
lines Arecibo data. In the case where an object was observed twice by the same telescope, 
the earliest observation is shown in by a dashed line.
Optical images are 2$\times$2 arcmin in size. False color ($i$, $r$, and $g$ filter) images are 
from the SDSS DR12 \citep{2015ApJS..219...12A}, and black and white images are from the 
2$^{nd}$ Digitized Sky Survey (DSS2) Blue plates \citep{2004AJ....128..502A}, 
used when SDSS images are not available. Objects are arranged in alphabetical order.}
\figsetgrpend
\figsetgrpstart
\figsetgrpnum{HISpectra.9}
\figsetgrptitle{HI Spectra, Image 9}
\figsetplot{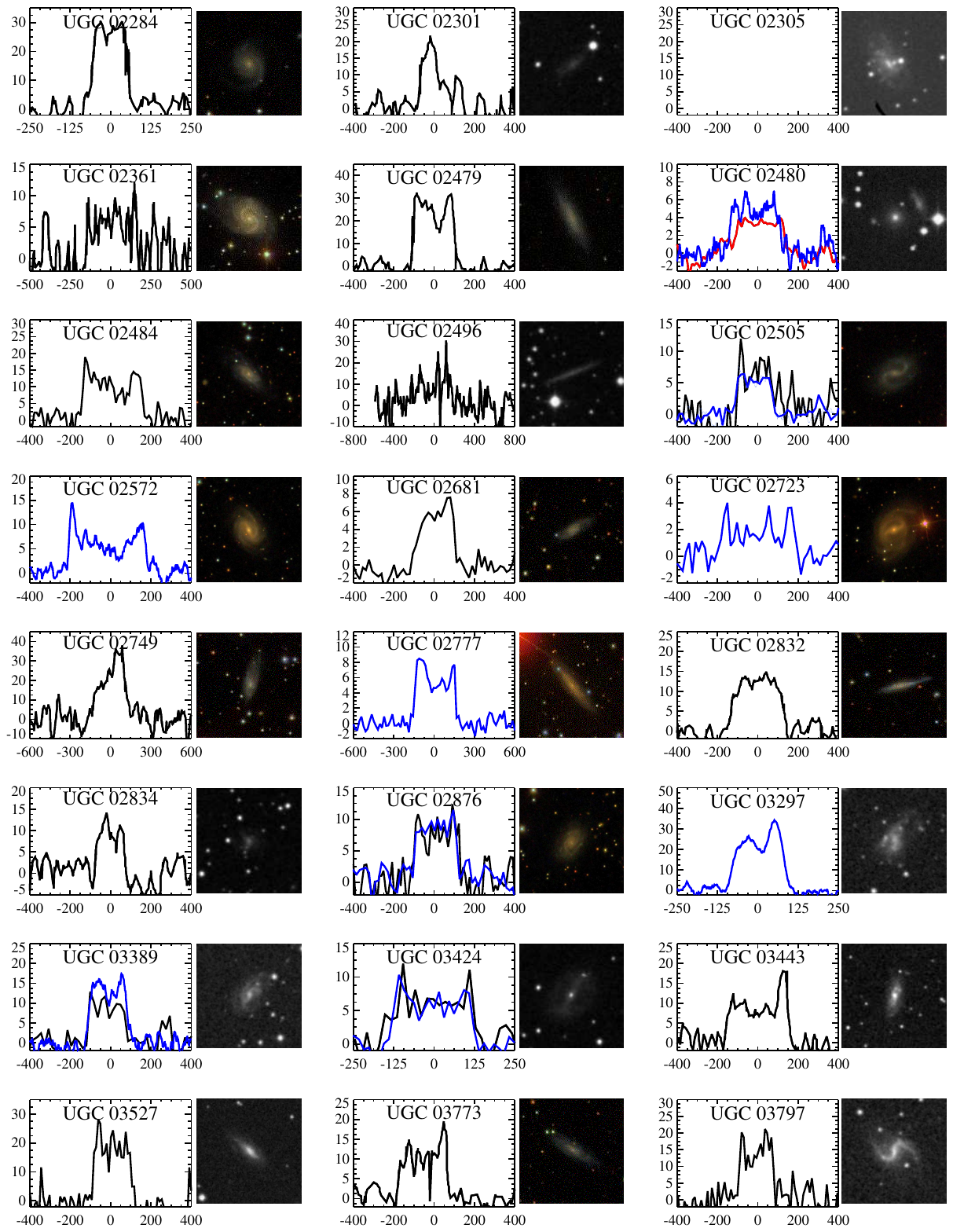}
\figsetgrpnote{21 cm \HI{} line spectra and optical images of the detected galaxies. 
Black lines in the spectra represent \nan{} data, blue lines GBT data, and red 
lines Arecibo data. In the case where an object was observed twice by the same telescope, 
the earliest observation is shown in by a dashed line.
Optical images are 2$\times$2 arcmin in size. False color ($i$, $r$, and $g$ filter) images are 
from the SDSS DR12 \citep{2015ApJS..219...12A}, and black and white images are from the 
2$^{nd}$ Digitized Sky Survey (DSS2) Blue plates \citep{2004AJ....128..502A}, 
used when SDSS images are not available. Objects are arranged in alphabetical order.}
\figsetgrpend
\figsetgrpstart
\figsetgrpnum{HISpectra.10}
\figsetgrptitle{HI Spectra, Image 10}
\figsetplot{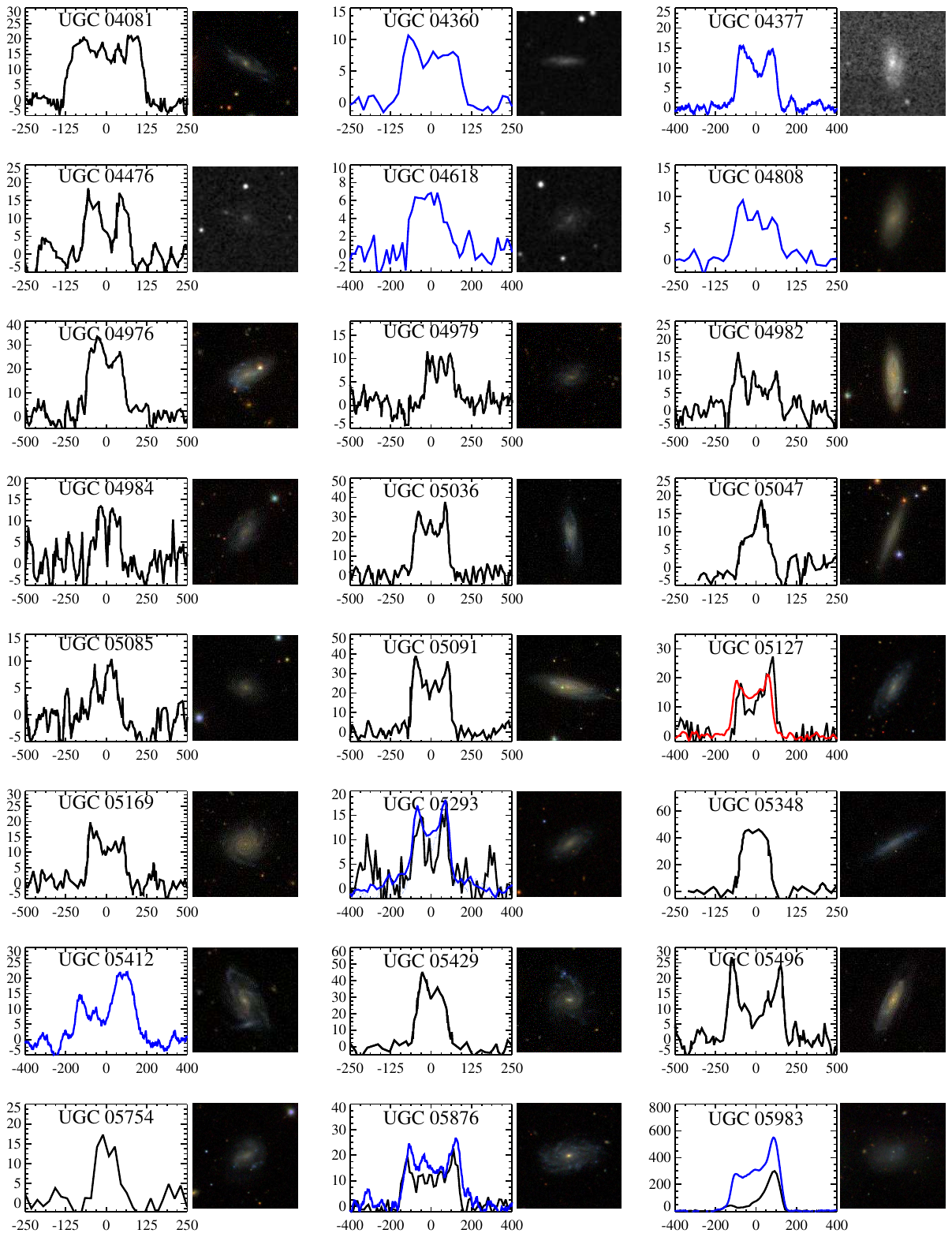}
\figsetgrpnote{21 cm \HI{} line spectra and optical images of the detected galaxies. 
Black lines in the spectra represent \nan{} data, blue lines GBT data, and red 
lines Arecibo data. In the case where an object was observed twice by the same telescope, 
the earliest observation is shown in by a dashed line.
Optical images are 2$\times$2 arcmin in size. False color ($i$, $r$, and $g$ filter) images are 
from the SDSS DR12 \citep{2015ApJS..219...12A}, and black and white images are from the 
2$^{nd}$ Digitized Sky Survey (DSS2) Blue plates \citep{2004AJ....128..502A}, 
used when SDSS images are not available. Objects are arranged in alphabetical order.}
\figsetgrpend
\figsetgrpstart
\figsetgrpnum{HISpectra.11}
\figsetgrptitle{HI Spectra, Image 11}
\figsetplot{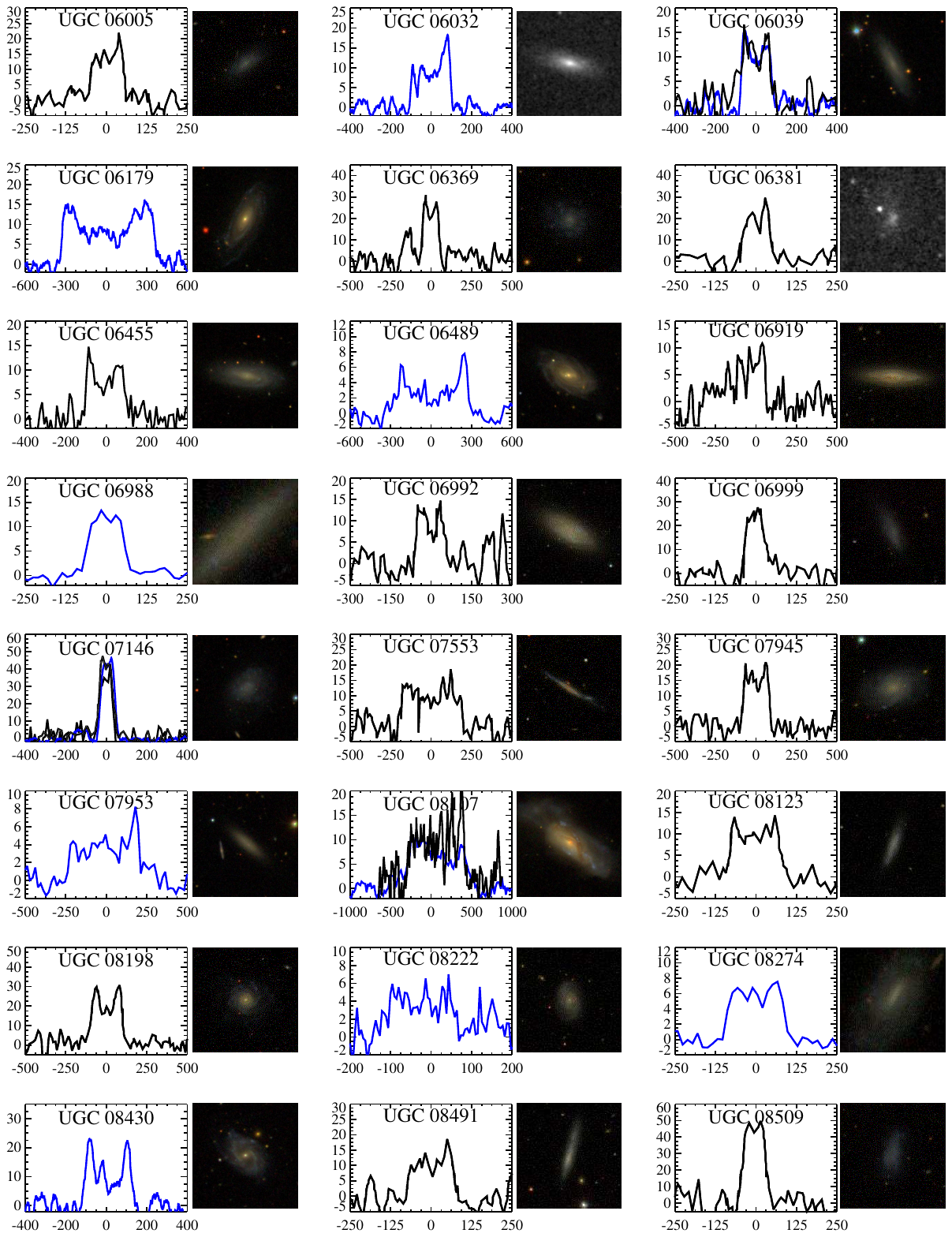}
\figsetgrpnote{21 cm \HI{} line spectra and optical images of the detected galaxies. 
Black lines in the spectra represent \nan{} data, blue lines GBT data, and red 
lines Arecibo data. In the case where an object was observed twice by the same telescope, 
the earliest observation is shown in by a dashed line.
Optical images are 2$\times$2 arcmin in size. False color ($i$, $r$, and $g$ filter) images are 
from the SDSS DR12 \citep{2015ApJS..219...12A}, and black and white images are from the 
2$^{nd}$ Digitized Sky Survey (DSS2) Blue plates \citep{2004AJ....128..502A}, 
used when SDSS images are not available. Objects are arranged in alphabetical order.}
\figsetgrpend
\figsetgrpstart
\figsetgrpnum{HISpectra.12}
\figsetgrptitle{HI Spectra, Image 12}
\figsetplot{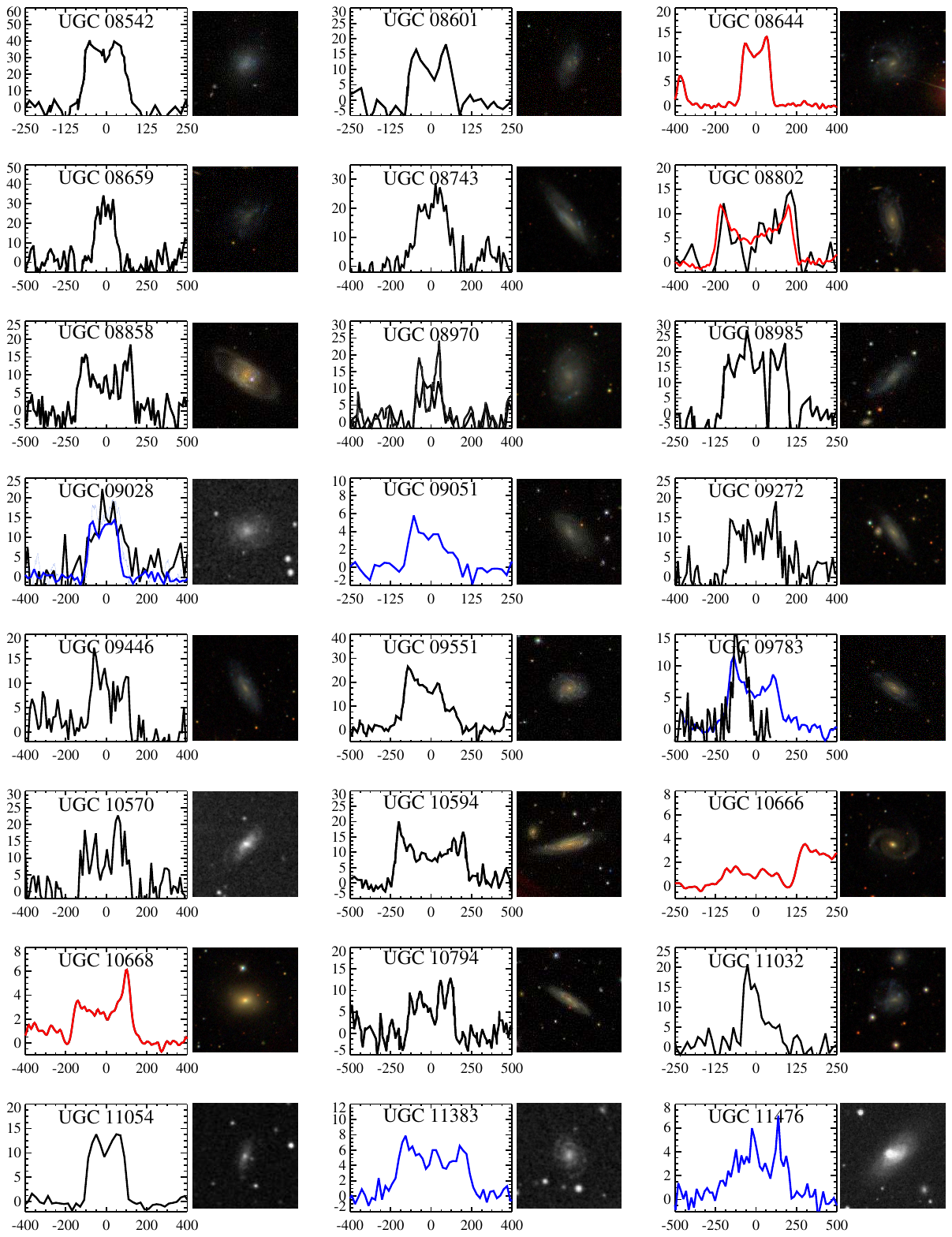}
\figsetgrpnote{21 cm \HI{} line spectra and optical images of the detected galaxies. 
Black lines in the spectra represent \nan{} data, blue lines GBT data, and red 
lines Arecibo data. In the case where an object was observed twice by the same telescope, 
the earliest observation is shown in by a dashed line.
Optical images are 2$\times$2 arcmin in size. False color ($i$, $r$, and $g$ filter) images are 
from the SDSS DR12 \citep{2015ApJS..219...12A}, and black and white images are from the 
2$^{nd}$ Digitized Sky Survey (DSS2) Blue plates \citep{2004AJ....128..502A}, 
used when SDSS images are not available. Objects are arranged in alphabetical order.}
\figsetgrpend
\figsetgrpstart
\figsetgrpnum{HISpectra.13}
\figsetgrptitle{HI Spectra, Image 13}
\figsetplot{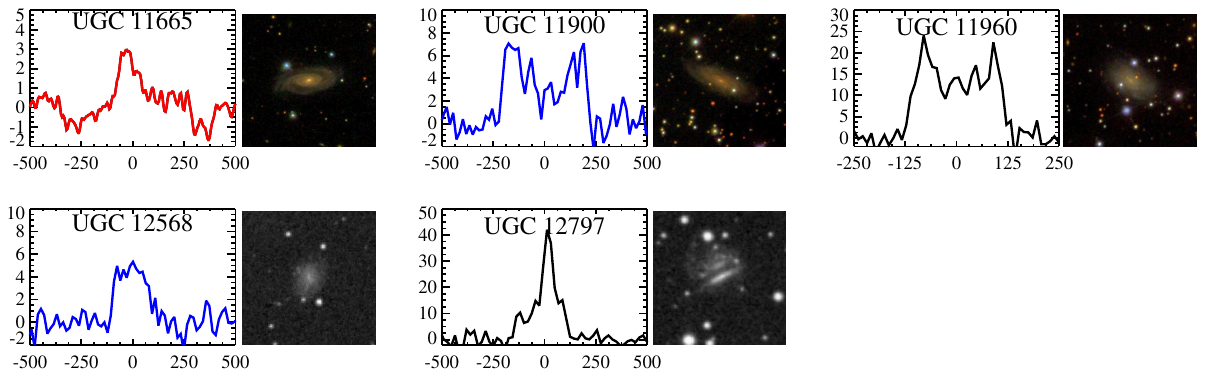}
\figsetgrpnote{21 cm \HI{} line spectra and optical images of the detected galaxies. 
Black lines in the spectra represent \nan{} data, blue lines GBT data, and red 
lines Arecibo data. In the case where an object was observed twice by the same telescope, 
the earliest observation is shown in by a dashed line.
Optical images are 2$\times$2 arcmin in size. False color ($i$, $r$, and $g$ filter) images are 
from the SDSS DR12 \citep{2015ApJS..219...12A}, and black and white images are from the 
2$^{nd}$ Digitized Sky Survey (DSS2) Blue plates \citep{2004AJ....128..502A}, 
used when SDSS images are not available. Objects are arranged in alphabetical order.}
\figsetgrpend
\figsetend

\clearpage


\clearpage

\clearpage

\clearpage

\clearpage

\clearpage

\clearpage

\end{document}